\documentclass[12pt]{iopart}

\usepackage{iopams}  
\usepackage[pdftex]{graphicx}
\usepackage[pdftex]{color}
\usepackage[pdftex]{hyperref}
\usepackage{cite} 
\usepackage{relsize}
\def\babar{\mbox{\slshape B\kern-0.1em{\smaller A}\kern-0.1em
    B\kern-0.1em{\smaller A\kern-0.2em R}}}

\newcommand{\be}{\begin{eqnarray}}
\newcommand{\ee}{\end{eqnarray}}

\newcommand{\ket}{\rangle}
\newcommand{\bra}{\langle}
\newcommand{\del}{\partial}

\newcommand{\vslash}{{v\hspace{-5.4pt}/}}

\newcommand{\dsz}{\mbox{$\bar D^{\ast 0}$}}

\newcommand{\ubar}{\bar{{u}}}
\newcommand{\dbar}{\bar{{d}}}
\newcommand{\cbar}{\bar{{c}}}

\newcommand{\qbar}{\bar{{q}}}
\newcommand{\Dbar}{\bar{{D}}}
\newcommand{\Jpsi}{J\!/\!\psi}

\newcommand{\up}{{\uparrow}}
\newcommand{\dw}{{\downarrow}}

\newcommand{\footref}[1]{\textsuperscript{\ref{#1}}}

\usepackage[normalem]{ulem}  
\renewcommand\sout{\bgroup \color{red} \ULdepth=-.5ex \ULset}

\begin{document}

\title[Hadronic molecules with pion exchange and quark core]{Heavy hadronic molecules with pion exchange and quark core couplings: a guide for practitioners}

\author{
Yasuhiro Yamaguchi$^{1}$, 
Atsushi Hosaka$^{2,3,1}$, 
Sachiko Takeuchi$^{4,1,2}$, 
and 
Makoto Takizawa$^{5,1,6}$
}

\address{$^1$Theoretical  Research  Division,  Nishina  Center,RIKEN,  Hirosawa,  Wako,  Saitama  351-0198,  Japan}
\address{$^2$Research  Center  for  Nuclear  Physics  (RCNP),  Ibaraki,  Osaka  567-0047,  Japan}
\address{$^3$Advanced Science Research Center,
Japan Atomic Energy Agency, Tokai, Ibaraki 319-1195, Japan}
\address{$^4$Japan  College  of  Social  Work,  Kiyose,  Tokyo  204-8555,  Japan}
\address{$^5$Showa  Pharmaceutical  University,  Machida,  Tokyo  194-8543,  Japan}
\address{$^6$J-PARC Branch, KEK Theory Center, IPNS, KEK, Tokai, Ibaraki, 319-1106, Japan}
\ead{yasuhiro.yamaguchi@riken.jp}
\ead{hosaka@rcnp.osaka-u.ac.jp}
\ead{s.takeuchi@jcsw.ac.jp}
\ead{takizawa@ac.shoyaku.ac.jp}
\vspace{10pt}

\begin{abstract}
We discuss selected and important features of hadronic molecules as one of several promising forms of exotic hadrons near thresholds.  
Using examples of $D \bar D^*$ systems such as $X(3872)$ and $Z_c$, emphasis is put on the roles of the one pion exchange interaction between them 
and their coupling to intrinsic quark states.  
Thus hadronic molecules emerge as admixtures of the dominant long-range hadron structure and short-range quark structure.  
For the pion exchange interaction, properties of the tensor force are analyzed 
in detail.  
More coupled channels supply more attractions, and heavier constituents suppress kinetic energies, 
providing more chances to form hadronic molecules of heavy hadrons.  
Throughout this article, we show details of basic ideas and methods.   
\end{abstract}

 \vspace{2pc}
 \noindent{\it Keywords}: hadronic molecule, $X(3872)$, pion, tensor force, quark core


\section{Introduction}

\subsection{Exotic phenomena}

Since the discovery of the $X(3872)$ in 2003 at Belle/KEK and BaBar/SLAC, 
many candidates of new hadrons have been observed~\cite{Choi:2003ue,Aubert:2004zr}.~\footnote{
More complete references are given in \sref{sec:X3872_review} for the $X(3872)$ and in \sref{sec:charged_overview} for $Z_{c}$ and $Z_{b}$. 
}
Their observed properties such as masses and life times are not easily explained 
by conventional methods and models of QCD.
Thus they have been called exotic hadrons or simply exotics.  
Historically, exotic hadrons of multiquarks are already predicted by Gell-Mann 
in his original work of the quark model~\cite{GellMann:1964nj}.
The states with quantum numbers  that are not accessed by the standard quark model, 
mesons as quark and antiquark ($q\bar q$) 
and baryons as three quarks ($qqq$), are often referred to as manifest or genuine exotics.  
In this regard, the $X(3872)$ is not manifestly exotic, but it shows up with many unusual properties.  
By now the $X(3872)$ has been observed by many experimental facilities, and is well established
with its quantum numbers determined by LHCb, $J^{PC} = 1^{++}$~\cite{Aaij:2013zoa}.  
In the latest PDG data base, more than thirty particles are listed as candidates of exotic hadrons.
Many of them are considered to contain charm quarks as constituents, while some of them 
only  light quarks~\cite{Tanabashi:2018oca}.  

Those exotic candidates are observed near thresholds.
For charmonia ($c \bar c$ pairs), the thresholds are the energies of $D^{(*)} \bar D^{(*)}$ above which 
an excited charmonium may decay into a $D^{(*)} \bar D^{(*)}$ 
pair~\footnote{Here $D^{(*)}$ stands for either $D$ or $D^*$ meson.  
In this article we do not consider systems 
containing strange quarks, 
and therefore the notation $q$ is used for light $u, d$ quarks.}. 
Therefore, near the threshold region systems may contain an extra light $q \bar q\ (q = u, d)$ in addition to the heavy quark-antiquark pair ($c \bar c$ or $b \bar b$).  
The nature of hadrons near thresholds and of those well below thresholds are qualitatively different from each other.  
Quarkonium-like states of $c\bar c$ or $b \bar b$ well below the threshold are essentially 
non-relativistic systems of a slowly moving heavy quark pair~\cite{Brambilla:2010cs}.  
In contrast,  exotics containing both heavy and light quarks may show up with various configurations
such as compact multiquarks~\cite{Jaffe:1976ig,Jaffe:1976ih,Maiani:2004vq,Terasaki:2007uv}, 
hadronic molecules~\cite{Dalitz:1959dn,Tornqvist:1993ng} and hybrids or complicated structure of quarks and gluons~\cite{Close:2005iz,Kou:2005gt,Voloshin:2013dpa}.  
The question of how and where 
these different structures show up is an important issue in hadron physics and has been 
discussed in references~\cite{Karliner:2017qhf,Karliner:2018qnp}.

In multi-quark systems, the quarks may rearrange into a set of colorless clusters.  
For instance, a hidden charm four quarks rearrange as 
$c \bar c q \bar q \to (c\bar c)(q \bar q) \sim J/\psi \pi$, or $(c\bar q)(q \bar c) \sim D \bar D^*$.
$J/\psi \pi$ dominantly appears in decays because the pion is  light and unlikely to be a constituent 
of hadrons.  
In the chiral limit massless pions behave just as chiral radiations.  
In contrast,  $D \bar D^*$ may form quasi-stable states if suitable interactions 
are provided via light meson exchanges, in particular pion exchanges between light quarks.  
This is the crude but basic idea of how hadronic molecules are formed.  
The idea of hadronic molecule is dated back to the discussion of 
$\Lambda(1405)$ as a $\bar KN$ molecule~\cite{Dalitz:1959dn}, 
and more were conjectured  in the context of $c \bar c$ productions after the discovery of 
$J/\psi$~\cite{DeRujula:1976qd}.  

\subsection{Clusterization}

The rearrangement of multi-quarks shares a general feature of clustering phenomena by neutralizing 
the original strong force among the constituents.  
Then among the neutralized clusters only relatively weak forces act.  
In the present case the color force is strong, while the meson exchange force is weak.
In this clustering process hierarchies of matter, or separation of the energy scale occurs.  
Strong color force is of order hundred MeV while the weak meson exchange force is of order ten MeV.  
This qualitatively explains how hadronic molecules are bound with a binding energy of order ten MeV.  
In table~\ref{Table_exoticcandidates}, several  candidates of hadronic molecules are shown.  
From these small binding energies can verify that the spatial sizes of these systems are 
of order one fm or larger.  
With  this inter-distance, the constituent hadrons in molecules can maintain their identity.  

\begin{table}[htp]
\caption{Candidates of exotic hadrons near thresholds, where
$\Delta E = {\rm Mass} - {\rm Threshold\ mass} $.  Data are taken from PDG~\cite{Tanabashi:2018oca}
for $\Lambda$, $X$ and $Z_c$, and from~\cite{Aaij:2019vzc} for $P_c$'s.  
The lower raw of $\Lambda(1405)$ is for the higher pole of the two-pole scenario.}
\begin{center}
\begin{tabular}{c c c c c }
\hline
State  & Mass (MeV) & Width (MeV) & Threshold & $\Delta E$  \\
\hline
$\Lambda(1405)$ & 1405 & $\sim 50$ & $\bar K N$ & $ -30$ \\
    & 1421-1434 &  & $\bar K N$ & $-15$ to $-1$\\
$X(3872)$ & 3872 & $< 1$ & $D^+ \bar D^{*-}$ & $-8$ \\
  &   &   & $D^0 \bar D^{*0}$ &  $ < 1$ \\
 $Z_c(3900)$ & 3887  & 26-31  & $D \bar D^{*}$ &  $ +15$ \\
 $Z_c(4020)$ & 4024  & 9-18  & $D^* \bar D^{*}$ &  $ +4$ \\
 $P_c(4312)$ & 4312  & 7-12  & $\bar{D}\Sigma_c$ &  $-8$ \\
 $P_c(4440)$ & 4440  & 15-25  & $\bar{D}^* \Sigma_c$ &  $  - 26$ \\
$P_c(4457)$  & 4457  & 4-8  & $\bar{D}^* \Sigma_c$ &  $  - 7$ \\
$Z _b(10610)$  & 10610  & 16-21  & $B \bar B^*$ &  $ +7$ \\
$Z _b(10650)$  & 10650  & 9-14  & $B^* \bar B^*$ &  $ +2$ \\
\hline
\end{tabular}
\end{center}
\label{Table_exoticcandidates}
\end{table}%

Analogous phenomena are found in nuclear excited states in which alpha cluster correlations are strongly developed.  
A well known example is the Hoyle state, the first $0^+$ excited state of $^{12}$C~\cite{Hoyle:1954zz}.
The formation of alpha clusters near the threshold of alpha decays
is known as the Ikeda rule that predicts the dominance of alpha cluster 
components in nuclear structure in the threshold region of $4N$ ($N=$ integer) nuclei~\cite{Ikeda:1968kk}.
Threshold phenomena are now regarded as universal phenomena and are discussed 
in the context of universality that covers various  systems 
from quarks to atoms and molecules~\cite{Braaten:2006vd,Pascal:2017xxx}. 

By now there are many articles that discuss hadronic molecules including comprehensive 
reviews~\cite{Guo:2017jvc}.
Here in this article we do not intend to list all of the previous works, but rather 
focus on limited subjects that we believe important  for the discussions of hadronic molecules.  
To elucidate the points we discuss $D^{(*)}\bar{D}^{(*)}$ systems, especially for the $X(3872)$ and some related states.  
We do not discuss baryons; 
for $\Lambda(1405)$, there are many discussions including the summary one in 
PDG~\cite{Tanabashi:2018oca}; 
for $P_c$'s, discussions have just started and we need more studies to make conclusive statements.  
In this way, this article is not inclusive.  
However, we try to emphasize general features by using a few specific examples.  
We also try to show  some details of how basic ingredients are derived.
Sometimes, we discuss items that are by now taken for granted.
We think that this strategy is important because many current discussions seem to be based on 
ad hoc assumptions, and many explanations and predictions depend very much on them.

\subsection{Pions and interactions}

\begin{figure}[h]
\begin{center}
\includegraphics[width=0.9\linewidth, clip]{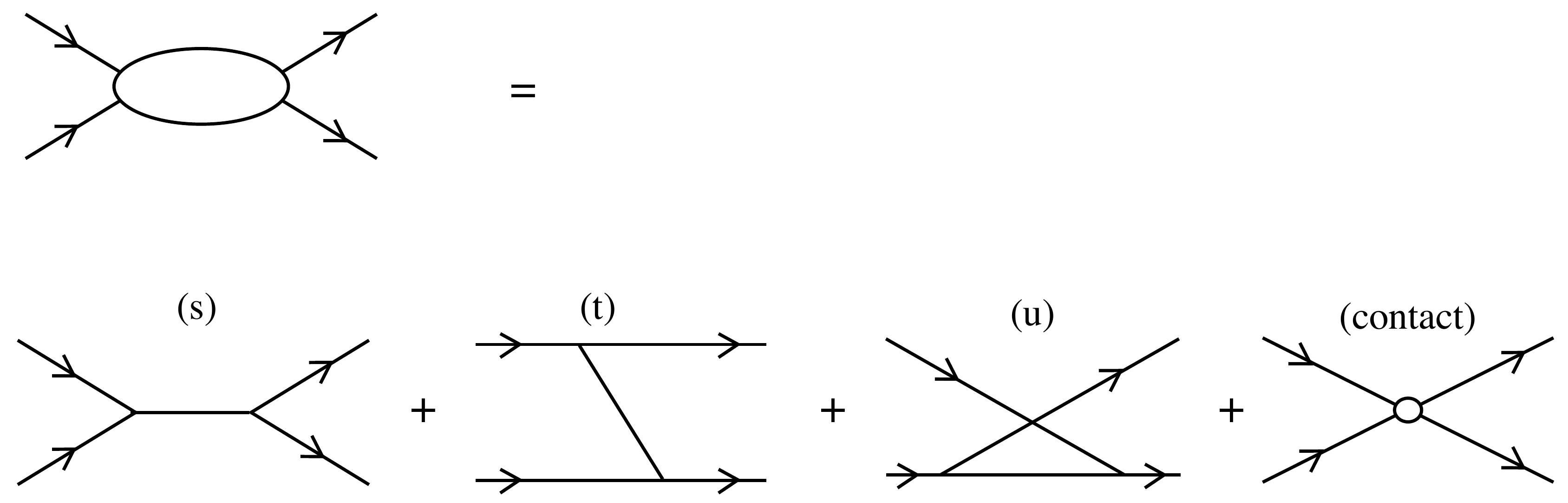}
\end{center}
\vspace{-5mm}
\caption{Decomposition of two-body interaction into s, t, u and c (contact) channels.  }
\label{fig_V_stuc}
\end{figure}

Now the most important ingredient is the interaction that 
is provided by light meson exchanges at long and medium distances.  
Among them best established is the one-pion exchange potential (OPEP). 
The pion is  the Nambu-Goldstone boson 
of the spontaneous symmetry breaking (SSB) of chiral symmetry~\cite{Nambu:1961tp,Nambu:1961fr}.
Its interaction with  hadrons is dictated by low-energy theorems.  
The leading term is the Yukawa term of $\pi hh^\prime $ ($h, h^\prime$: hadrons).  
By repeating this twice, the OPEP emerges in the $t$-channel as shown in (t) of
\fref{fig_V_stuc}, where 
general structure of two-body amplitudes is shown.  
Hence, in hadronic molecules, pions play a role of the mediator of the force 
between constituent hadrons.  

\begin{figure}[h]
\begin{center}
\includegraphics[width=0.7\linewidth, clip]{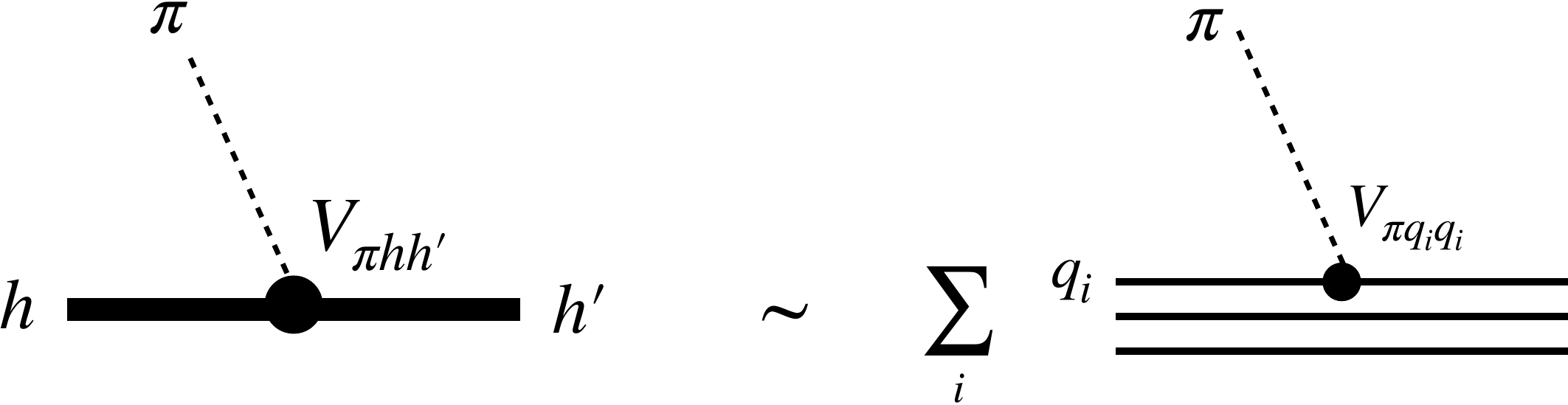}
\end{center}
\vspace{-5mm}
\caption{Schematic  view of a pion hadron (nucleon in this figure) coupling. }
\label{fig_piqq}
\end{figure}

Microscopically, the pion couples to the constituent quarks
that are dynamically generated by SSB.  
Combined with the quark model wave functions of hadrons, the coupling strengths as well as 
form factors are estimated, schematically by 
\be
V_{\pi hh^\prime} = \sum_i \bra h^\prime | V_{\pi q_iq_i} | h\ket \, , 
\ee
where the sum is taken over the light quarks ($i$) in the hadrons as shown in \fref{fig_piqq}.  
This method works qualitatively well for nuclear interactions and is now extended to other 
hadrons for the study of hadronic molecules.  
Other meson interactions  such as $\sigma$, $\omega$ and $\rho$ mesons are also employed 
but then more  parameters are needed.  
In fact, the masses of these mesons are of the same order of the inverse of hadron size, 
and their contributions may be masked by the form factors.  
Thus, the pion interaction is the best known and under control.
In most part of this paper, we test models of hadronic molecules 
with the pion interaction.  

Another feature of the pion interaction is in its tensor structure.  
This is the consequence of SSB of chiral symmetry which leads to pseudoscalar nature of the pion with 
spin-parity $J^P = 0^-$.  
Therefore, the coupling structure of the pion to hadrons is of $\bsigma \cdot \bi r$ type.  
This leads to the tensor force causing mixing of 
orbital motions of different angular momenta by two units.
This provides extra attraction which contributes significantly to
the formation of molecules.  
Although the importance of the tensor force has long been  recognized in nuclear 
physics~\cite{Nagata:1959tfp,Terasawa:1960sos}, 
quantitative understanding has progressed by  developments 
in the microscopic treatment of many-body systems and in computer 
power~\cite{Pieper:2001mp,Carlson:2014vla,Otsuka:2005zz,Myo:2007vm}.  

In addition to the pion exchange interaction at long distances, 
we also discuss $s$-channel 
interactions at short distances where the incoming hadrons merge into a single hadron (one-particle) 
as an intermediate state (see 
\fref{fig_V_stuc}).  
Hence this process leads to the mixing of configurations, an extended molecular structure of two particles and a compact one-particle state.  
The problem is also related to the question of the so-called compositeness~\cite{Weinberg:1962hj,Weinberg:1965zz,Nagahiro:2013hba,Nagahiro:2014mba,Sekihara:2014kya}. 
We emphasize the importance of such  mixing for $X(3872)$; 
a molecular component of $D\bar D^*$ at long distances and a $c \bar c$ component at short distances~\cite{Takeuchi:2014rsa}.  

\subsection{Contents of this paper}

This paper is organized as follows.  
In section 2, we show how the Yukawa vertex of a pion to heavy hadrons are derived. 
Coupling constants in different schemes are discussed in some detail.
Estimation in the quark model is also discussed.  
In section 3, OPEP is derived with emphasis on general features of the potential.
Special attention is 
paid
to the tensor structure and  form factors.  
A non-static feature is also discussed when the mass of the interacting hadrons changes, which is  
taken into account by an effective mass of the pion. 
In section 4, we discuss the structure of $X(3872)$.  
After briefly reviewing experimental status, we discuss the molecular nature made by the one-pion exchange.  
An important role of the short distance dynamics is also discussed, and consider 
a mixing structure of hadronic molecule coupled by a compact quark $c \bar c$ component.  
In section 5, a brief review for $Z_c$ with some discussions are given. 
Section 6 is for a few subjects for pentaquarks, where we quickly overview for a few candidates including the most recent ones
from LHCb, $P_c$ baryons.  
We summarize the paper with some remarks and prospects in section 7.  

\section{Heavy hadron interactions}

Hadronic molecules are composite systems of hadrons which are loosely bound or resonate.
``Loosely" means that the binding or resonant energies are small as compared 
to the QCD scale of  $\Lambda_{QCD} \sim$ some hundreds MeV, 
which is relevant to intrinsic structure of hadrons by quarks.  
In such a situation, the constituent hadrons can 
retain their intrinsic structure in the molecules.
The interaction among the hadrons is colorless and its dominant part is expected to be 
dictated by meson exchanges.  
Among them, pion exchange interaction is the best under control. 
The pion couples to the light $u, d$ quarks, 
and their dynamics is determined by 
the nature of Nambu-Goldstone bosons of spontaneous breaking of chiral symmetry.  
This is the case if hadrons contain light $u, d$ quarks as constituents such as protons, neutrons 
and also heavy open flavor hadrons such as $D$ mesons ($c \bar u$)  and  $\Sigma_c$ baryons ($cuu$).  

In this section, we discuss basic interactions of heavy mesons,  that is the Yukawa vertices 
for $P$ and $P^*$ with the pion, 
where $P$ stands for $D$ or $\bar B$ meson, and $P^*$ for  $D^*$ or $\bar B^*$.  
We also employ the notation $P^{(*)}$ for either $P$ or $P^*$.  
In addition to chiral symmetry features associated with light quarks, 
heavy quark spin symmetry also applies in the presence of heavy quarks 
(either charm $c$ or bottom $b$ quark)~\cite{Casalbuoni:1996pg,Manohar:2000dt}.  
In particular, heavy quark spin symmetry relates 
the mesons with different spins under spin transformations.
For example, $P$-meson of spin-parity $J^P = 0^-$ is a spin partner of 
$P^*$ meson of $J^P = 1^-$; 
they are the same particles under heavy quark spin symmetry.  

To implement the aspects of heavy quark spin and chiral  symmetries in the effective Lagrangian,
we shall quickly overview several issues such as 
a convention for  heavy quark normalization, 
representations of heavy fields for $D$ and $D^*$ mesons, 
and their  properties 
under the heavy quark spin and chiral symmetry transformations.  
We also discuss how the relevant coupling constants are determined.  
We see that the constituent picture of the light quark coupled by the pion 
consistently describes the decay properties of the $D^{(*)}$ mesons 
as well as axial properties of the nucleon.

\subsection{Heavy fields}

When considering quantum fields of heavy particles of mass $m_H$, 
it is convenient to redefine the effective heavy fields 
in which the rapidly oscillating component in time, $\exp(-i m_H t)$, is factored out.  
For QCD  ``heavy" means that  $m_H$ is sufficiently larger than the QCD scale, 
$\Lambda_{\rm QCD} \ll m_H$, 
and the heavy quarks almost stay on mass-shell with quantum fluctuations being suppressed.  
In accordance with the redefinition of the field, 
the normalization of the effective heavy fields are naturally modified from 
the familiar one of quantum fields by the factor $\sqrt{m_H}$.  

To show this point 
let us consider the standard Lagrangian for a complex scalar meson field 
of heavy mass $m_H$, $\phi(x) = \frac{1}{\sqrt{2}}(\phi_1(x) + i \phi_2(x))$, 
\be
{{{\cal L}}} = (\del_\mu \phi^\dagger)(\del^\mu \phi) - m_H^2 \phi^\dagger \phi\, .  
\label{eq_L_heavyscalar}
\ee
The factor 1/2 is recovered when using the real components $\phi_{1,2}$.
From this Lagrangian, the current is given by
\be
j^\mu = 
i 
\left(
\phi^\dagger \del^\mu \phi - \phi \del^\mu \phi ^\dagger
\right)\, .
\label{eq_def_current}
\ee
The field expansion may be done as 
\be
\phi(x) &=& \int\frac{\rmd^3p}{2E(2\pi)^3}
\left(
e^{-ipx} a_{\bi p}+ e^{+ipx} b_{\bi p}^\dagger
\right), \; \; \; E = \sqrt{m_H^2 + \bi p^2} 
\ee
in the standard conventions.  

In the heavy mass limit $m_H \to \infty$, the particle is almost on-mass shell, and it is convenient to define the velocity
$v_\mu, v^2 = 1$ which defines the on-shell momentum $m_H v_\mu$.  
Thus 
the momentum fluctuation $k_\mu$ around it is considered to be small, $k_\mu \ll m_H$
\be
p_\mu = m_H v_\mu + k_\mu\, .
\ee
Moreover the Hilbert space of different heavy particle numbers decouple because
particle-antiparticle creation is suppressed in the considering energy scale.  

Hence we define the heavy field by 
\be
\phi(x) = \exp(-im_H vx) \tilde \phi(x)\, .
\label{eq_Hshift}
\ee
This means that the energy of $\tilde \phi(x)$ is measured from $m_H$.  
Inserting the relation
\be
\del_\mu \phi(x) =  (-i m_H v_\mu \tilde \phi(x) + \del_\mu \tilde \phi(x)) \exp(-im_H vx) 
\ee
into (\ref{eq_L_heavyscalar}), we find
\be
{{{\cal L}}}
= 2 i m_H (\tilde \phi^\dagger v \del \tilde \phi) + {\cal O}(1)
\ee
and for the current, 
\be
    j^\mu = 2 m_H \tilde \phi^\dagger v^\mu \tilde \phi + {\cal O}(1) \, , 
\ee
where in both equations we have only shown the leading term of order ${\cal O}(m_H)$.  
Note that the mass term in (\ref{eq_L_heavyscalar}) disappears in the Lagrangian as expected
because the energy is measured from $m_H$.  
By absorbing the factor $m_H$ into the field as
\be
\phi_H \equiv \sqrt{m_H} \tilde \phi  
\label{eq_re_standard_heavy}
\ee
then we have
\be
{{{\cal L}}}_H &=& 2 i (\phi_H^\dagger v \del \phi_H)  \, , 
\nonumber \\
j_H^\mu &=& 
2 \phi_H^\dagger v^\mu \phi_H\, .
\label{L_and_j_HQ}
\ee

In this convention, the heavy (boson) field $\phi_H$ 
carries dimension 3/2 in units of mass, unlike 1 in the standard boson theory.  
In this paper, as in many references, we follow this convention, while we also 
come back to the ordinary convention of dimension 1.  
In terms of one-particle states, these two conventions correspond 
to different normalizations~\cite{Manohar:2000dt}
\be
\bra v | v \ket = \sqrt{2v_0}\ , \; \; \; \bra p | p \ket = \sqrt{2p_0}\ , \; (p_0 = E) \ .
\ee
Moreover the one-particle to vacuum matrix element of the field is given by
\be
\bra 0 | \phi_H(x) | p\ket = \sqrt{m_H} e^{-ipx} \ .
\ee

\subsection{Interaction Lagrangian}

Let us consider the pseudoscalar $P$  and vector $P^*$ mesons as 
a pair of heavy quark $Q$ and light antiquark $\bar q$ in the lowest S-wave orbit. 
Having overviewed the features of heavy particles in the previous subsection, 
the heavy meson field is defined in the frame of a fixed velocity $v_\mu$ 
and contains multiplet of spin 0 pseudoscalar and spin 1 vector mesons.  
They are $(D, D^*)$ in the charm sector, 
and $(\bar B, \bar B^*)$ for the bottom sector.
For convenience, namings and quark contents of various mesons are given in table~\ref{Tab_defKDB}.
In this article throughout, we place symbols without bar on the left of those with bar.  
This is a convention that is consistent with quark model calculations.  

\begin{table}[htp]
\caption{Various heavy mesons and quark contents, where $q = u, d$ quarks. }
\begin{center}
\begin{tabular}{l c c c c c c }
\hline
Mesons $P^{(*)}$ & $K^{(*)}$ & $\bar K^{(*)}$ & $D^{(*)}$ & $\bar D^{(*)}$ & $B^{(*)}$ & $\bar B^{(*)}$ \\
Quark contents & $q \bar s$ & $s \bar q $ & $c \bar q$ & $q \bar c$ & $q\bar b$ & $b\bar q$\\
\hline
\end{tabular}
\end{center}
\label{Tab_defKDB}
\end{table}%

A convenient way to express such heavy meson fields (including antiparticles) is 
\be
H^a &=&  \frac{1+ \vslash}{2}\left[ -P^{*a}_\mu \gamma^\mu + P^a \gamma_5 \right] 
\sim Q\bar q \ , 
\nonumber \\
\bar H^a &=& \gamma_0 H^\dagger_a \gamma_0 
= \left[ - (P^{*\dagger})^a_\mu \gamma^\mu - (P^{\dagger})^a \gamma_5 \right] \frac{1+ \vslash}{2} 
\sim q \bar Q\ , 
\label{eq_def_PPstar}
\ee
where $P^a$ and $P^{*a}_\mu$  carry an index of isospin 1/2, $a = 1, 2 \sim u, d$.
The factor $(1+ \vslash)/{2}$ is a projector to constrain the heavy quark velocity at $v$.
We employ the convention for $\gamma$-matrices
\be
\gamma^0 = 
\left(
\begin{array} {c c}
1 & 0 \\
0 & -1 
\end{array}
\right)\, , \; \; \; 
\gamma^i = 
\left(
\begin{array} {c c}
0 & \sigma_i \\
- \sigma_i &0
\end{array}
\right)\, , \; \; \; 
\gamma_5 = 
\left(
\begin{array} {c c}
0 & 1 \\
1 &0
\end{array}
\right)\, ,
\ee
and contractions $A^\mu B_\mu = A^0B_0 + A^iB_i = A^0B_0 - \bi A \cdot \bi B$. 
For later convenience, we express the meson fields explicitly in terms of the quark fields
\be
P^a = \bar q^a  i \gamma_5 Q, \; \; \; P^{*a}_\mu = \bar q^a \gamma_\mu Q \ .
\ee

Let us consider 
$D^+ = \bar d  i \gamma_5 c, D^{*+}_\mu = \bar d  \gamma_\mu c$, 
where $d, c$ express the Dirac fields for the down and charm quarks.  
Under charge conjugation transformations, 
\be
c \to -i (\bar c \gamma^0 \gamma^2)^T, \; \; \; 
\bar d \to -i ( \gamma^0 \gamma^2 d)^T \ , 
\ee
we can verify that
\be 
D^+ \to D^- = (D^+)^\dagger  , \; \; \; D^{*+}_\mu \to - D^{*-}_\mu = - (D^{*+}_\mu)^\dagger \ .
\label{eq_cconjugate}
\ee
In this convention, again using the notation $P$'s, the operators for the charge conjugated anti-particles are 
$P^\dagger$ for pseudoscalars and $- P^{*\dagger}_\mu$ for vectors.  
The corresponding states are defined by 
\be
|P\ket = P^\dagger |0\ket, \; \; \; |P^*_\mu \ket = P^{*\dagger}_\mu |0\ket \ , 
\nonumber\\
|\bar P\ket = P |0\ket, \; \; \; |\bar P^*_\mu \ket = - P^{*}_\mu |0\ket \ .
\label{eq_cc_state}
\ee
These relations will be used when forming eigenstates of charge conjugation 
of molecules formed by $P^{(*)}$ and $\bar P^{(*)}$ mesons.  

The spin multiplet nature of $P$ and $P^*$ is verified by writing (\ref{eq_def_PPstar}) in the rest frame 
$v^\mu = (1, 0, 0, 0)$, where only the spatial components remain for the vector meson as its degrees of freedom
\be
H = 
\left(
\begin{array}{ c c} 
0 &  P + {\bsigma}\cdot \bi{P}^{*} \\
0 & 0
\end{array}
\right) \ .
\ee
Here the isospin label $a$ is suppressed, since it is irrelevant under spin transformations.  
The combination $P + \bi{\bsigma}\cdot \bi{P}^*$ indicates that 
$P$ and $\bi P^*$ 
are the spin multiplet of $(1/2, 1/2)$ representation of the heavy and light spin
group $SU(2)_Q \times SU(2)_q$.  
They transform under the heavy and light spin transformations as
\be
H \to \exp(i \bi{\bSigma} \cdot \bi \btheta_Q) H \exp(-i \bi{\bSigma} \cdot \bi \btheta_q)
\ , 
\ee
where $\bi{\bSigma}$ are the four component spin matrices 
defined by $\Sigma_k = \frac{i}{2} \epsilon_{ijk} [ \gamma^i, \gamma^j]$, 
and $\bi \btheta_Q$ and $\bi \btheta_q$ are 
the rotation angles of heavy and light quark spins, respectively.  
The diagonal part of $\bi \btheta_Q = \bi \btheta_q$
corresponds to the total spin rotation.  

Chiral symmetry property of the heavy meson field (\ref{eq_def_PPstar}) is 
inferred by the constituent nature of the light quark $q$.  
It is subject to nonlinear transformations of chiral 
symmetry~\cite{Weinberg:1968de,Coleman:1969sm,Callan:1969sn,Hosaka:2001ux}.  
In this article, we consider two light flavors and  therefore $SU(2)_L \times SU(2)_R$ is the relevant chiral symmetry group, 
where the left ($L$) and right ($R$) transformations act on the two isospin groups. 
Explicitly, the quark field $q$ of isospin 1/2 are transformed as
\be
q \to h(g_R, g_L, \bpi(x)) q \ , 
\label{eq_NL_chi_trans}
\ee
where the isospin $SU(2)$ matrix function $h(g_R, g_L, \bpi(x)) $ characterizes nonlinear chiral transformations 
determined by global chiral transformations of $g_{L,R }\in SU(2)_{L,R}$  at the pion field
$\bpi(x)$.     
Therefore, the heavy meson fields of isospinor transforms
under chiral symmetry transformations as
\be
H \to H h^\dagger, \; \; \; \bar H \to h \bar H \ .
\label{eq_H_chi_trans}
\ee

The isovector pion field parametrizes unitary matrices as 
\be
U(x) = \exp\left(i \frac{\btau \cdot \bpi (x)}{f_\pi}\right) \ , 
\ee
which linearly transforms as 
\be
U(x) \to g_L U(x) g_R^\dagger \ .
\ee
The nonlinear transformation for the pion field is then conveniently expressed in terms of the square root
\be
\xi(x) = \exp\left(i \frac{\btau \cdot \bpi (x)}{ 2f_\pi}\right) \ , 
\label{eq_def_xi}
\ee
which is subject to
\be
\xi \to h \xi g_R^\dagger = g_L \xi h^\dagger \ .
\ee
Here $f_\pi$ is the pion decay constant for which our convention is $f_\pi \sim 93$ MeV.  
The $\xi$-field defines the vector and axial-vector currents 
\be
V_\mu &=& \frac{1}{2} ( \xi^\dagger \del_\mu \xi + \xi \del_\mu \xi^\dagger) \ , 
\nonumber \\
A_\mu &=& \frac{1}{2} ( \xi^\dagger \del_\mu \xi - \xi \del_\mu \xi^\dagger) \ , 
\label{def_v_av_currents}
\ee
which are transformed as 
\be
V_\mu &\to& h (V_\mu + \del_\mu )h^\dagger \ , 
\nonumber \\
A_\mu &\to& h A_\mu h^\dagger \ .
\label{eq_v_a_trans}
\ee
Note that the currents (\ref{def_v_av_currents}) are anti-Hermitian.  
Moreover, the vector and axial-vector currents are of even and odd power 
with respect to the pion field (see (\ref{eq_VA_exp})), while properly satisfying the 
parity of the currents, $1^-$ and $1^+$, respectively.  

With the  heavy quark spin and chiral transformation properties established,
we can write down the invariant Lagrangian.  
To the leading order of derivative expansion, we find
\be
{{{\cal L}}}&=& 
-i \tr \left[ H v_\mu (\del^\mu + i V^\mu) \bar H \right]
+
ig_A \tr \left[ H \gamma_\mu \gamma_5 A^\mu \bar H \right] \ . 
\label{eq_L_DandDstar}
\ee
By expanding the vector and axial-vector currents, $V_\mu$ and $A_\mu$ 
with respect to the pion field, 
\be
V_\mu &=& \frac{i}{4f_\pi^2}\btau \cdot \bpi \times \del_\mu \bpi + \cdots 
\nonumber \\
A_\mu &=& \frac{i}{2f_\pi} \btau \cdot \del_\mu \bpi + \cdots 
\label{eq_VA_exp}
\ee
the vector current leads to the pion-hadron interaction of the so-called Weinberg-Tomozawa interaction, while the axial vector current to the Yukawa coupling.
The former strength is determined by the pion decay constant while the latter contains one 
unfixed parameter, the axial coupling constant $g_A$.
In the present scheme it corresponds to the one of the constituent quark as discussed in section \ref{sec:Quark_model_estimate}.
By inserting the expansion (\ref{eq_VA_exp}), we find the $PP^*\pi$ and $P^*P^*\pi$ interaction Lagrangians
\be
{{{\cal L}}}_{P^*P\pi} 
&=&
- \frac{g_A}{f_\pi}
\left(  
P_\mu^{*a} \del^\mu ({\btau \cdot \bpi})_{ab} (P^\dagger)^b 
+ P^a \del^\mu ({\btau \cdot \bpi})_{ab}  (P^{*\dagger}_\mu)^b
\right) \ , 
\nonumber \\
{{{\cal L}}}_{P^*P^*\pi} 
&=&
-i \frac{g_A}{f_\pi} \epsilon^{\alpha \beta \gamma \delta } v_\alpha P_\beta^{*a} \del_\gamma (\btau \cdot \bpi)_{ab}
(P^{* \dagger})^b_\delta \ . 
\label{L_PP*pi}
\ee
As anticipated, the  strengths of these interactions are given by one coupling constant $g_A$, which is a consequence 
of heavy quark spin symmetry.  

\subsection{\label{sec:MesonDecaysI} Meson decays I, $D^* \to D \pi$}

To see the use of (\ref{L_PP*pi}) together with the heavy quark normalization, 
let us consider the simplest and important example of meson decays, 
$D^{*+}(\lambda, p) \to D^+(p^\prime) \pi^0(q)$, where 
$\lambda$ labels the polarization of $D^{*+}$.
The relevant matrix element for these charged states is ($P \sim D$)
\be
\fl
\bra P(p^\prime) \pi (q) | {{\cal L}}_{P^*P\pi}  | P^*(\lambda, p)\ket
=
- m_H \frac{i g_A}{f_\pi} q_\nu \epsilon^\nu(\lambda) e^{i(- p+ q + p^\prime)x}
\equiv  V e^{i(- p+ q + p^\prime)x} \ , 
\label{ME_L_PPpi}
\ee
where $\epsilon^\nu(\lambda)$ is the polarization vector of the $D^*$ meson, and 
we have used the relation for the matrix elements;
\be
\bra P(p^\prime)| P(x) | 0\ket &=& \sqrt{m_H} e^{+ip^\prime x}, 
\nonumber \\
\bra 0 | P^*_\mu(x) | P^*(\lambda, p)\ket &=& \sqrt{m_H} \epsilon_\mu(\lambda) e^{-ip x}, 
\nonumber \\
\bra \pi(q) | \pi(x) | 0\ket &=& e^{+iqx} \ .
\ee
Here the masses of $D$ and $D^*$ mesons are regarded sufficiently heavy and set equal to $m_H$.  
For later convenience, we summarize the other matrix elements 
which are needed for the computations of the transition amplitudes 
$PP^* \to P^*P$ and $\bar P \bar P^* \to \bar P^* \bar P$
in table~\ref{table_PP*pi}.
By using the relations of (\ref{eq_cc_state}), one can verify these relative signs.

\begin{table}[htp]
\caption{Relative signs of various $P^{(*)}P^{(*)} \pi$ couplings including antiparticles. }
\begin{center}
\begin{tabular}{c c c c }
\hline
$\bra P \pi | {{{\cal L}}} | P^*\ket $  
   & $\bra P^* \pi | {{{\cal L}}} | P \ket$ 
       & $\bra \bar P^* \pi | {{{\cal L}}} | \bar P \ket$ 
           & $ \bra \bar P \pi | {{{\cal L}}} | \bar P^*\ket$ \\ \hline
 + & $-$ & + & $-$ \\          
\hline
\end{tabular}
\end{center}
\label{table_PP*pi}
\end{table}%

Now the decay width is computed by 
\be
\Gamma = \frac{1}{2m_{D^*}}
\int \frac{\rmd^3p^\prime}{2E_D(\bi p^\prime) (2\pi)^3} \frac{\rmd^3q}{2E_\pi(\bi q) (2\pi)^3} 
(2 \pi)^4 \delta^{4}(p - p^\prime - q) |V|^2 \ .
\label{Gamma_general}
\ee
Note that in the heavy mass limit, 
$
m_{D^*} \sim E_D \sim m_H
$,
heavy meson mass $m_H$ dependence in $E_D(\bi p^\prime)$ in the denominator and that in $|V|^2$ 
in the numerator cancel.  
The heavy mass independence is reasonable because
the decay $D^{*+} \to D^+ \pi^0$ occurs through the spin flip 
of the light quark which should not depend on the heavy mass of the spectator heavy quark.  
Fixing the polarization $\epsilon(\lambda)^\nu$ of initial $D^*$ meson and 
performing the phase space integral 
\be
\fl
\int \frac{\rmd^3p^\prime}{2E_P(\bi p^\prime) (2\pi)^3} \frac{\rmd^3q}{2E_\pi(\bi q) (2\pi)^3} 
(2 \pi)^4 \delta^4(P_f - P_i)
= \frac{q}{16 \pi^2(E_D + E_\pi)} \int d \Omega 
\ee
together with the angle average of 
$|q_\nu \epsilon^\nu|^2 \to |\bi q \cdot  \bi \epsilon|^2 \to q^2/3 $, 
we find 
\be
\Gamma_{D^{*+} \to D^+ \pi^0} = 
\frac{q^3}{24\pi} \left( \frac{m_H}{m_D^*} \right)^2 
\left( \frac{g_A}{f_\pi} \right)^2
\to
\frac{q^3}{24 \pi } \left( \frac{g_A}{f_\pi} \right)^2 \ .
\label{Gamma1}
\ee
Using the experimental data
\be
\Gamma_{D^{*+} \to D^+ \pi^0} &=& 83.4\; {\rm keV} \times 30.7\; \% \ , 
\nonumber\\
m_{D^{*+}} &=& 2010\;  {\rm MeV},\; \;  
m_{D^+} = 1870\; {\rm MeV}, \; \;  
m_{\pi^0} = 135\; {\rm MeV}, 
\nonumber \\
q &=& 38\; {\rm MeV}, \; \; f_{\pi} = 93 \; {\rm MeV},
\ee  
we find
\be 
g_A \sim 0.55 \ .
\ee
This value is obtained in the limit  $m_{D}, m_{D^*} \to \infty$ using the formula (\ref{Gamma1}).  
If we take into account their finite values, we find $g_A \sim 0.53$.  
This estimates uncertainties of few percent at minimum in the discussions 
based on the leading terms of the heavy quark symmetry.  

\subsection{Meson decays II, $K^* \to K \pi$}

Now it is instructive to demonstrate another textbook like calculation for the decay 
$K^* \to K \pi$, which is the analog of $D^* \to D \pi$ by replacing 
the charm quark by the (anti)strange quark.  
A commonly used Lagrangian in a flavor SU(3) symmetric form is written as 
\be
{{\cal L}}_{K^* K \pi} &=& - ig K^{\dagger} \del^\mu (\btau \cdot \bpi)  K^*_\mu + (h.c.)
\nonumber \\
&=& - ig (K^- \del^\mu \pi^0 + \sqrt{2} \bar K^0 \del_\mu \pi^-) K^{*+}_\mu + \cdots
\ . 
\label{eq_L_K*Kpi}
\ee
where $g$ is the coupling constant of a vector meson with two pseudoscalar mesons.
In the SU(3) limit it is the $\rho \to \pi \pi$ coupling constant and is given to be 
$g \sim 6$~\cite{Hosaka:2001ux,Bando:1984ej,Bando:1987br}.   
In this convention, the normalization is
\be
\bra 0 | K^{(*)}(x) | K^{(*)}(p)\ket =  e^{-ipx} \ , 
\ee
such that there are $2E$ particles in a unit volume.  
The matrix element of the above Lagrangian is then
(again for the neutral pion decay) 
\be
\bra K^+(\bi q) \pi^0(-\bi q) | {\cal L}_{K^* K \pi} | K^{*+}({\rm at} \; {\rm rest})\ket
= -i g q_\mu \epsilon^\mu \ .
\ee
Therefore, we find the formula
\be
\Gamma_{K^{*+} \to K^+ \pi^0} = \frac{q^3}{24\pi m_{K^*} (E_K + E_\pi)} g^2 \ . 
\ee
Here if we break $SU(3)$ symmetry and take heavy mass limit for the strange quark, 
$m_{K^*} \sim E_K \to m_H$, we find the total decay width
\be
\Gamma_{tot} = \frac{q^3}{8\pi } \left( \frac{g}{m_H} \right)^2 \ . 
\label{Gamma_tot2}
\ee
By using the experimental data
\be
\Gamma_{K^{*+} \to K^+ \pi^0} = 50 \times \frac{1}{3} \; {\rm MeV}, \; \; \; 
m_{K^*} = 890\; {\rm MeV}, \; \; \; 
q = 290  \; {\rm MeV} \ , 
\ee
we find 
\be
g \sim 6.4
\label{g_at_K} \ , 
\ee
which is close to the coupling constant of the $\rho$ meson decay, $g(\rho \to \pi \pi) \sim 6$.

Comparing equations  (\ref{Gamma1}) and (\ref{Gamma_tot2}) we find 
\be
\frac{g_A}{f_\pi} = \frac{g}{m_H}\, .
\label{eq_general_GT}
\ee
This is nothing but the generalized Goldberger-Treiman relation, 
implying that the coupling constant $g$ scales as the meson mass $m_H$, 
when $g_A$ is independent of $m_H$ as we shall discuss in the next subsection.  
In other words, flavor symmetry breaking for the coupling constant $g$ defined by the Lagrangian
(\ref{eq_L_K*Kpi}) scales as that of the corresponding meson masses.  
As a matter of fact, the coupling constant $g$ for the decay of $D^*$ is estimated to be $g \sim 12$, 
which is different from $g \sim 6$ estimated from the decay of $K^*$ by about factor two.  
This difference is explained by the difference in the masses of $D^*$ and $K^*$ mesons, 
$m_{D^*} \sim 2000$ and  $m_{K^*} \sim 890$ MeV within about 10 \% accuracy.  

\subsection{\label{sec:Quark_model_estimate} Quark model estimate}

\begin{figure}[h]
\begin{center}
\includegraphics[width=0.4\linewidth, clip]{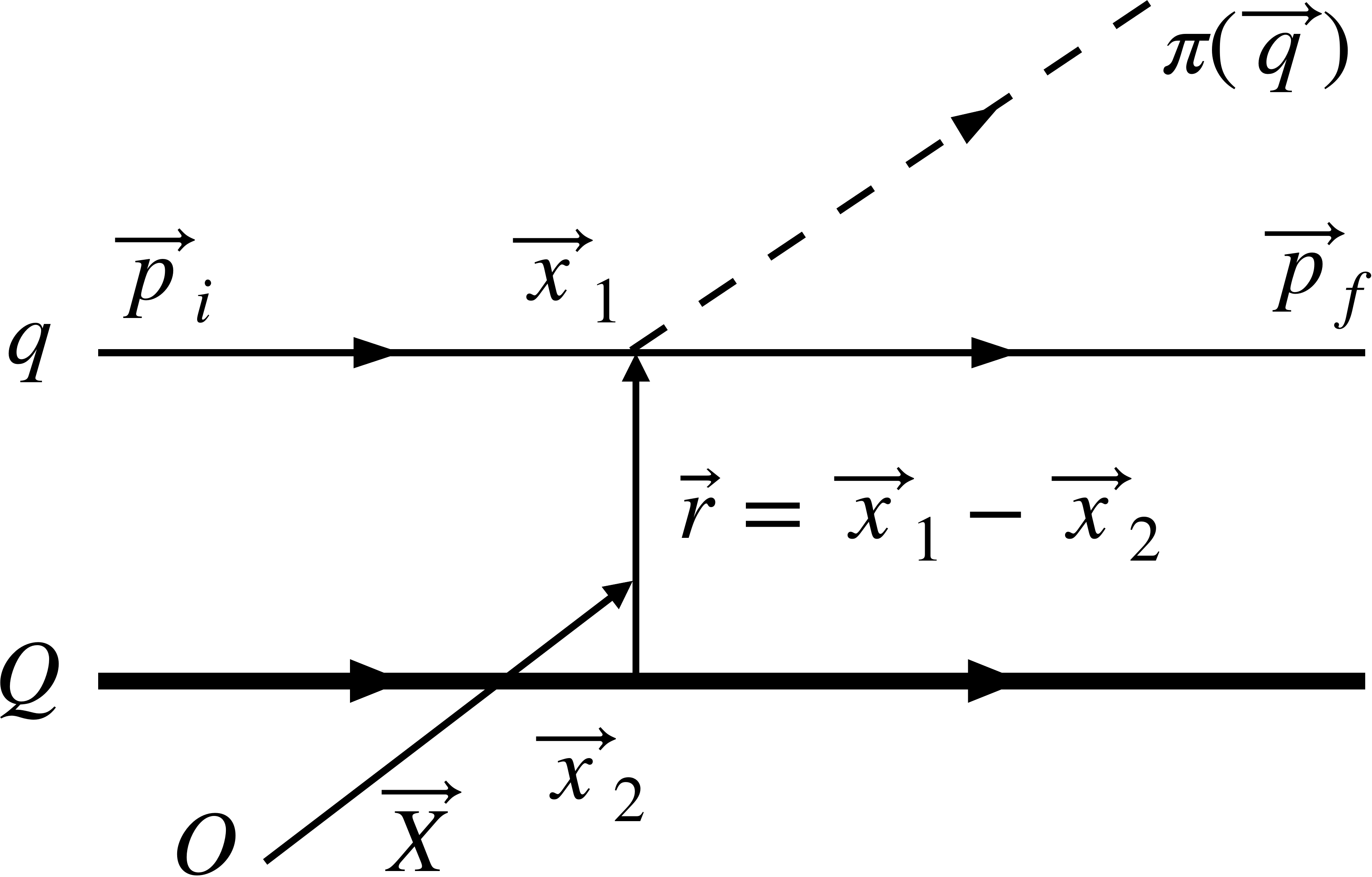}
\end{center}
\vspace{-5mm}
\caption{Quark model diagram for the decay of $D^* \to D \pi$.  }
\label{fig_D*decayquark}
\end{figure}

In this subsection we show that the coupling constant $g_A$  
in the Lagrangian (\ref{L_PP*pi}) is nothing but the quark axial-vector coupling constant 
in the non-relativistic quark model.  
Let us start with the $\pi qq$ Lagrangian in the axial vector type
\be
{\cal L}_{\pi qq} = \frac{g_A^q}{2 f_\pi} \bar q \gamma_\mu \gamma_5  \del^\mu \pi q 
\ , 
\label{eq_Lpiqq1}
\ee
where we have denoted the quark axial-vector coupling by $g_A^q$, and 
ignored isospin structure for simplicity.
This Lagrangian operates to the light quark-pion vertex at position $\bi x_1$ 
as shown in figure~\ref{fig_D*decayquark}, 
where assignments of various variables are also shown.  
For example, the 
the center of mass  and relative coordinates are defined by 
\be
\bi X = \frac{m \bi x_1 + M \bi x_2}{m+M}\, , \; \; \; 
\bi r = \bi x_1 - \bi x_2 \ , 
\ee
with $m, M$ being the masses of the light and heavy quarks, 

In the non-relativistic limit, the matrix element of (\ref{eq_Lpiqq1}) 
for $\bar D^*$(at rest) $\to \bar D(-\bi q) \pi (+\bi q)$ is given as~\cite{Nagahiro:2016nsx}
\be
{\cal L}_{\pi qq}  \to  
i \frac{g_A^q}{2 f_\pi}
\chi_f^\dagger \left[  \frac{\omega_\pi}{m} \bsigma \cdot \bi p_i 
- 
\left( 1 + \frac{\omega_\pi}{2m}\right) \bsigma \cdot \bi q \right]
\chi_i e^{+i \omega_\pi t - i \bi q \cdot \bi x_1} \ , 
\label{eq_Lpiqq2}
\ee
where $\chi_{i, f}$ are the quark wave functions of the initial $D^*$ and final $D$ meson, 
$(\omega_\pi, \bi q)$ the energy and momentum of the pion, 
$\bi p_i$ the momentum of the quark in the initial $\bar D^*$ meson.  
For a notational reason in the definition of the quark model wave function as explained below, here we 
 consider the decay of $\bar D^*$  rather than $D^*$.  

The wave functions $\chi_{i, f}$ are  written as 
a product of the plain wave for the center of mass and internal part including spin, $\phi_{i,f}(\bi r)$
\be
\chi_{i,f}(t,  \bi X, \bi r) = \exp \left( -i \omega_{i, f}t + i \bi P_{i,f} \cdot \bi X \right) \phi_{i,f} (\bi r)
 \ , 
\ee
where $\omega_{i, f}$ are the energies of the initial $\bar D^*$ and final $\bar D$ mesons.  
Expressing $\bi p_i$ by the relative momentum $\bi p_{r}$ as 
\be 
\bi p_{i} = \frac{m}{m+M} \bi P_{i} + \bi p_r \ , 
\ee
we can perform the $t$ and $\bi X$-integral leading to the total energy-momentum conservation, 
leaving the $\bi r$ integral as
\be
\fl
\frac{ig_A^q}{2f_\pi} \int \rmd^3 r \ \phi_f^\dagger (\bi r) 
\left[ \frac{\omega_\pi}{m} 
\left( \frac{m}{m+M} \bsigma \cdot \bi P_i - i \bsigma \cdot {\bi \nabla}_r \right)
 \left( 1 + \frac{\omega_\pi}{2m}\right) \bsigma \cdot \bi q \right] \phi_i(\bi r) 
 e^{ - i \tilde {\bi q} \cdot \bi r} \ , 
\ee
where effective momentum transfer is defined by
\be
\tilde {\bi q} = \frac{M}{m+M} \bi q \ , 
\ee
and the relative momentum $\bi p_r$ is replaced by ${- i \bi \nabla}_r$.
Using the harmonic oscillator wave function for both $\phi_i$ and $\phi_f$, 
\be
\phi(\bi r) = \frac{1}{\sqrt{4 \pi}} \frac{2\alpha^{3/2}}{\pi^{1/4}}
\exp \left( -\frac{\alpha^2}{2} r^2 \right) \ , 
\ee
with the size parameter $\alpha$, after some computation, we find 
\be
\bra D \pi | {\cal L}_{\pi qq} | D^*\ket 
&=&
\frac{i g_A^q}{2f_\pi}
\left(
1 + \frac{\omega_\pi}{2m} \left(-1 + \frac{M}{m+M} \right) 
\right)
|\bi q| \bra \bsigma \cdot \hat \bi q \ket F(\bi q^2)  \ , 
\label{eq_LpiDDquark}
\\
F(\bi q^2) &=& \int \rmd^3 r\
\phi_f^\dagger (\bi r) \phi_i (\bi r) e^{i \bi q \cdot \bi r} \ .
\ee
For small $|\bi q|$, we set the form factor $F(\bi q^2) \to 1$.  
For the spin matrix element $\bra \bsigma \cdot \hat \bi q \ket$, we evaluate the transition 
\be 
D^*(S=1, S_z=1) \to D(S=S_z=0), \; \; \hat \bi q \sim \hat \bi z \ . I 
\ee
Having the spin wave functions 
\be
|S =1, S_z = 0\ket = \frac{1}{\sqrt{2}}(\uparrow \downarrow + \downarrow \uparrow )\, ,  \; \; \; 
|S =S_z = 0\ket = \frac{1}{\sqrt{2}}(\uparrow \downarrow - \downarrow \uparrow ) \ , 
\ee
and with the understanding that the spin operator acts on the first (left) spin state for the light quark, 
we find
$
\bra S=0 | \bsigma \cdot \hat \bi q | S = 1 \ket =1 \ .
$

These results are compared with the matrix element (\ref{ME_L_PPpi}), 
where we may set 
$q^\mu = (0,0,0,|\bi q|)$ and $q^\mu \epsilon_\mu = - |\bi q|$.
Suppressing the second term in the heavy quark limit $M \to \infty$,
we find the relation 
\be
g_A = g_A^q
\ee
in the limit $|\bi q| \to 0$.

Usually in the quark model $g_A$ is assumed to be unity, $g_A^q = 1$.
However, it is known that this overestimates the axial couplings of various hadrons.
For the nucleon it is known that the quark model predicts $g_A^N = 5/3$~\cite{Hosaka:2001ux}.
In the quark model, the nucleon $g_A$ is defined to be the matrix element of spin and isospin operator
\be
\sum_{n=1,2,3} \sigma_i(n) \tau^a(n) \ .
\ee
Therefore, effectively the reduction of $g_A^q \sim 0.7$ is needed to reproduce the data.  
Similarly heavy baryon transitions such as $\Sigma_c^{(*)} \to \Lambda_c$ 
consistently implies small $g_A^q$~\cite{Nagahiro:2016nsx}.
How baryon $g_A$ is computed is found in Refs.~\cite{Hosaka:2001ux,Nagahiro:2016nsx}.
\section{Meson exchange potential}
\label{sec:Meson_exchange_potential}

In this section, we derive meson exchange potentials for the study of 
hadronic molecules.  
Starting from the classic method for the derivation, 
we revisit the OPEP for the nucleon ($N$). 
We find it is useful to recognize important and universal features of meson exchange potentials.  

\subsection{Simple exercise}
\label{sec:Simple_exercise}

Let us illustrate a simple example for a scalar field $\phi$ interacting with 
the exchange of a scalar $\pi$ meson of mass $m$.  
The extension to the case of physical pion will be done later in a rather straightforward manner.  
The model Lagrangian is 
\be
{\cal L} = \frac{1}{2} (\del_\mu \pi)^2 - \frac{m^2}{2}  \pi^2 - \frac{g}{2} \pi \phi^2 
+ ({\rm kinetic\ terms\ of\ } \phi) \ ,
\label{eq_L_pi-phi-phi}
\ee
from which the equation of motion and a special solution for $\pi$ are obtained as
\be
(\del^2 + m^2) \pi(x) &=& - \frac{g}{2}  \phi(x)^2 \ ,
\nonumber \\
\hspace*{18mm} \pi(x) &=& - \frac{g}{2}  \int \rmd^3 y\ \bra x | \frac{1}{\del^2 + m^2} | y \ket \phi(y)^2
\ .
\label{eq_EOQ_pi}
\ee
We note that in this example, the coupling constant $g$ carries dimension of unit mass.  

The potential energy for $\phi$ is given by the energy shift $\Delta E$ due to the interaction; 
\be
\Delta E &=& + \frac{g}{2} \int \rmd^3 x\  \pi(x) \phi^2(x)
\nonumber \\
&=& 
\left( \frac{g}{2}\right)^2 \int \rmd^3 x\rmd^3 y\  \phi^2(x)  \bra x | \frac{- 1}{\del^2 + m^2} | y \ket \phi(y)^2
 \ ,
\ee 
where in the second line we have used the solution of (\ref{eq_EOQ_pi}).
Inserting the complete set and representing the propagator in the momentum representation, 
$\int \rmd^3p/(2\pi)^3 |\bi{p}\ket \bra \bi{p}|$, 
we find the expression
\be
\Delta E 
&=&
  \int \rmd^3 x\rmd^3 y\  \phi^2(\bi{x})  
\left[
\int \frac{\rmd^3 q}{(2\pi)^3} \frac{- g^2}{-p^2 + m^2}  e^{i \bi{q}\cdot (\bi{x} - \bi{y})}
\right]
\phi(\bi{y)}^2 \ .
\label{eq_DeltaE}
\ee

\begin{figure}[h]
\begin{center}
\includegraphics[width=0.5\linewidth, clip]{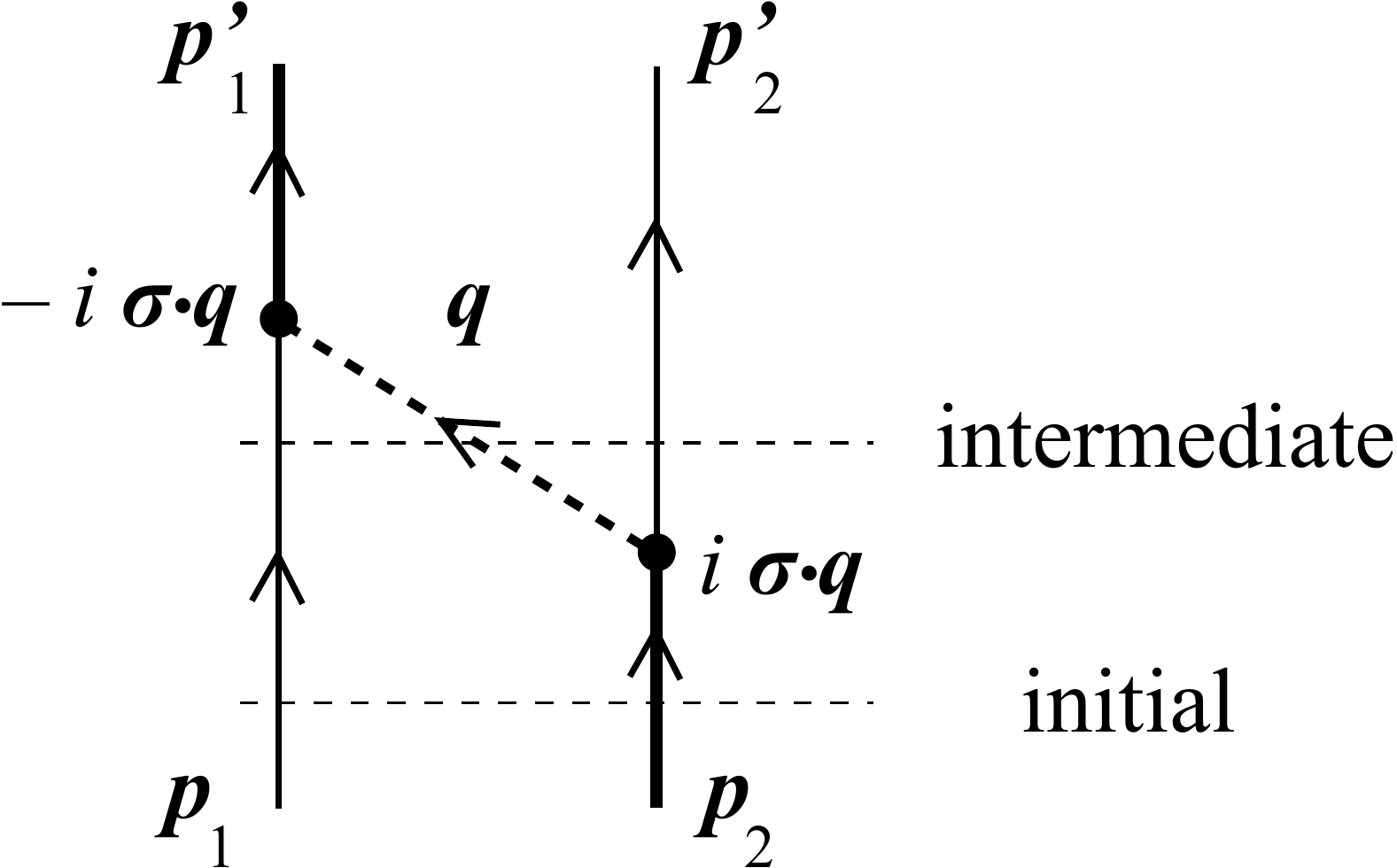}
\end{center}
\vspace{-5mm}
\caption{One meson exchange potential.  
The vertex structure of $\bsigma \cdot \bi q$ is needed for the one-pion exchange potential (OPEP) of the nucleon. 
For $\phi\phi$ or $NN$ the line widths of bold or normal are irrelevant.
It will become relevant when discussing the potential for $P^{(*)}P^{(*)}$.}
\label{fig_OPEP}
\end{figure}

Now we regard $\phi$ as a field operator and expand in momentum space.
Then consider a scattering process of 
$\bi p_1, \bi p_2 \to \bi p_1^\prime, \bi p_2^\prime$ 
as shown in figure~\ref{fig_OPEP}.
Taking the matrix element
$\bra \bi p_1^\prime, \bi p_2^\prime| \Delta E | \bi p_1, \bi p_2\ket$ and 
performing $x, y$, and $q$ integrals, we find 
\be
\Delta E 
=
(2\pi)^3\delta(\bi p_1 + \bi p_2 - \bi p_1^\prime - \bi p_2^\prime)
\frac{-g^2}{-q^2 + m^2}  \ .
\ee

Remarks are in order.
\begin{itemize}
\item 
The energy shift $\Delta E$ (\ref{eq_DeltaE}) is for the entire volume 
$V \sim (2\pi)^3\delta(\bi p_1 + \bi p_2 - \bi p_1^\prime - \bi p_2^\prime) \to \delta^3(0)$, 
and also for the normalization of $2E$ $\phi$-particles per unit volume.  
In the center-of-mass frame, energies of the two particles are the same and 
conserved, $E = \sqrt{\bi p^2 + M^2}$, where $M$ is the mass of $\phi$.  

Therefore, 
the energy shift per unit volume and per particle is given by 
\be
\Delta E
=
\frac{1}{(2E)^2}\frac{- g^2}{-q^2 + m^2}  \ .
\label{eq_DeltaE2}
\ee

\item
In the non-relativistic limit for the $\phi$ particle, 
we can take the static limit where the energy transfer $q^0$ is neglected such that 
$- q^2 \to + \bi{q}^2$, and $E \sim M$.
This defines the potential in momentum space
\be
V(\bi{q}) = - \frac{1}{(2M)^2}\frac{g^2 }{\bi{q}^2 + m^2}  \ ,
\label{eq_OPEP2}
\ee
and in turn
\be
V(\bi{r}) = \int \frac{\rmd^3 q}{(2\pi)^3} \frac{1}{(2M)^2}\frac{- g^2}{\bi{q}^2 + m^2}  e^{i \bi{q}\cdot \bi{r}}
= 
- 4 \pi \left( \frac{g}{2M} \right)^2 \frac{e^{-mr}}{r} 
\label{eq_OPEP1}
\ee
in coordinate space.

\item
Though obvious but not very often emphasized, 
the potential appears always 
attractive when the coupling square $g^2$ is positive.
If the coupling structure has spin dependence this is no longer the case, otherwise always so.
This is understood by the formula of second order perturbation theory where the intermediate 
state of $\phi \phi \pi$ in the pion-exchange process is in higher energy state than the initial 
$\phi \phi$ (see \fref{fig_OPEP}) .  
This fact is in contrast with what we know for the Coulomb force. 
The reason is that the latter is given by the unphysical component of the photon, which 
is manifest in the sign of the metric.  
\end{itemize}

\subsection{OPEP for the nucleon-nucleon $NN$}
\label{sec:OPEP_for_NN}

Now the most familiar and important example is the 
OPEP.
In this section we will discuss OPEP for the nucleon, because the nucleon system is 
the best established, and can share common features with heavy hadrons.  
Since the pion is the pseudoscalar particle, the pion nucleon coupling is given 
either by the pseudoscalar or axial vector (pseudovector) form, 
\be
{\cal L}_{\pi NN; PS} &=& - i g_{\pi NN} \bar N \gamma _5  \btau \cdot \bpi N  \ ,
\nonumber \\
{\cal L}_{\pi NN; PV} &=& - \frac{g_A^N}{2f_\pi}  
\bar N \gamma_\mu \gamma _5  \btau \cdot \del^\mu \bpi N \ .
 \label{eq_L_piNN}
 \ee
When the nucleons are on mass-shell, it is shown that the matrix elements 
for $N(\bi{p}) \to N(\bi{p}^\prime) \pi(\bi{q})$ in the two schemes 
are equivalent by using the equation of motion 
for the nucleon.  
The equivalence of the two expressions leads to the familiar Goldberger-Treiman 
relation
\be
\frac{g_{\pi NN}}{m_N} = \frac{g_A^N}{f_\pi} \ .
\label{eq_GT}
\ee

In the non-relativistic limit, the equivalent matrix elements reduce to 
\be
\bra N(\bi{p}_1^\prime) \pi^a (\bi{q}) | {\cal L}_{\pi NN} | N(\bi{p}_1) \ket
&\to&
- i \frac{g_A^N}{2f_\pi} \bsigma_1 \cdot \bi{q}\ \tau^a_1 \ 2m_N  \ ,
\nonumber \\
\bra N(\bi{p}_2^\prime)| {\cal L}_{\pi NN} | N(\bi{p}_2)  \pi^a (\bi{q}) \ket
&\to&
+ i \frac{g_A^N}{2f_\pi}  \bsigma_2 \cdot \bi{q}\ \tau^a_2 \ 2m_N  \ ,
\label{eq_ME_LpiNN}
 \ee
where the two-component nucleon spinors are implicit and $a$ on $\pi$ and 
$\tau$-matrix is an isospin index.  
The extra factor 
$2m_N$ on the right-hand side appears due to the normalization of the nucleon (fermion) field
when the state is normalized such that there are $2E \sim 2m_N$ nucleons in a unit 
volume.  
The positivity of the coupling square as discussed in the previous subsection 
is ensured by $\pm i$ in (\ref{eq_ME_LpiNN}).  
Inserting these coupling structures into the general form of (\ref{eq_OPEP2}), we find the 
OPEP for the nucleon in the momentum space
\be
 V^{NN}_\pi(\bi{q})=-\left(\frac{g_A^N}{2f_\pi} \right)^2
 \frac{(\bsigma_1\cdot\bi{q}\,)(\bsigma_2\cdot\bi{q}\,)}{\bi{q}\,^2+m^2_{\pi}}\btau_1\cdot\btau_2 
  \ .
 \label{eq_OPEP_NN}
\ee

Sometimes, this form is called the bare potential because the Lagrangian (\ref{eq_L_piNN}) 
does not consider the finite size structure of the nucleons and pions.  
The OPEP depends on $\bi{q}$, a feature consistent with 
the low energy theorems of chiral symmetry; interactions of the Nambu-Goldstone bosons 
contain their momenta.  
At low energies the $NN$ interaction (\ref{eq_OPEP_NN}) is of order ${\cal O}(\bi{q}^2)$.
In particular at zero momentum the interaction  vanishes.  
In contrast, when $\bi{q} \to \infty$, the interaction approaches a constant.  
This requires a careful treatment for the large momentum or short range behavior of the interaction.  

To see this point in more detail, 
let us decompose the spin factor $\bsigma_1\cdot\bi{q}  \bsigma_2\cdot\bi{q}$ into the central and tensor parts
\be
\fl
 V^{NN}_\pi(\bi{q}) 
 = \left(\frac{g_A^N}{2f_\pi} \right)^2\frac{1}{3}
 \left[
 \left( -1+\frac{m^2_\pi}{\bi{q}\,^2+m^2_{\pi}} \right)  \bsigma_1\cdot\bsigma_2
 +S_{12}(\hat{ \bi q})\frac{-\bi{q}\,^2}{\bi{q}\,^2+m^2_{\pi}}
 \right]\btau_1\cdot\btau_2  \ ,
 \label{eq_OPEP_NN2}
\ee
where the tensor operator is defined by 
\be
S_{12}(\hat{\bi q})=3(\bsigma_1\cdot\hat{\bi q})(\bsigma_2\cdot\hat{\bi q})-\bsigma_1\cdot\bsigma_2
 \ .
\label{eq_def_S12}
\ee
The first term of (\ref{eq_OPEP_NN2}) is the spin and isospin dependent central force, which has been 
further decomposed into the constant and the Yukawa terms.  
The constant term takes on the form of the $\delta$-function 
in the coordinate space.
This singularity appears because the nucleon is treated as a point-like particle.  
In reality, nucleons have finite structure and the delta function is smeared out. 

In the chiral perturbation scheme starting from the bare interaction of
(\ref{eq_OPEP_NN}) the constant ($\delta$-function) term is kept and 
higher order terms are systematically computed by perturbation.  
In this case, low energy constants are introduced order by order together with 
a form factor with a cutoff to limit the work region of the perturbation series~\cite{Cohen:1998bv,Epelbaum:2009sd}.
To determine the parameters experimental data are needed.
This is possible for the $NN$ force but not for hadrons in general.  
Alternatively, in nuclear physics the constant term is often subtracted.  
One of reasons is that the hard-core in the nuclear force suppresses the wave function 
at short distances and the $\delta$-function term is practically ineffective.  
Then form factors are introduced to incorporate the structure of the nucleon, and the cutoff parameters there 
are determined by experimental data.  

In this paper, we employ the latter prescription, namely subtract the constant term and multiply the form factor.  
As in (\ref{eq_OPEP_NN2}), the constant and Yukawa terms in the central force have opposite signs, and hence 
part of their strengths are canceled.  
The inclusion of the form factor in the Yukawa term is to weaken its strength, which is 
partially consistent with the role of the constant term.  
In practice, the central interaction of the OPEP is not very important for low energy properties.
Rather the dominant role is played by the tensor force.  
We will see the important role of the tensor force in the subsequent sections.  

So far we have discussed only the OPEP. 
In the so-called realistic nuclear force, to reproduce experimental data such as phase shifts 
and deuteron properties, 
more boson exchanges are included such as $\sigma$, $\rho$ and $\omega$ 
mesons~\cite{Nagels:1976xq,Rijken:1998yy,Machleidt:1987hj}.  
Their masses are fixed at experimental data except for less established $\sigma$.  
Coupling constants and cutoff masses in the form factors are 
determined by experimental data of $NN$ phase shifts and deuteron properties.  
The resulting potentials work well for $NN$ scatterings up to several hundred MeV, 
and several angular momentum (higher partial waves).  
However, if we restrict discussions to low energy properties, which is the case 
for the present aim for exotic hadronic molecules,  meson exchanges 
other than the pion exchange are effectively taken into account by the form factors.  
As discussed in the next section, we will see this for the deuteron.  

Having said so much, let us summarize various formulae for the OPEP for $NN$. 
Subtracting the constant term with the form factor included we find
\be
\fl
 V^{NN}_\pi(\bi{q}) 
\to \left(   \frac{g_A^N}{2f_\pi}   \right)^2\frac{1}{3}
 \left[
\frac{m^2_\pi}{\bi{q}\,^2+m^2_{\pi}} \bsigma_1\cdot\bsigma_2
 +S_{12}(\hat{q})\frac{-\bi{q}\,^2}{\bi{q}\,^2+m^2_{\pi}}
 \right] F(\bi{q})\
 \btau_1\cdot\btau_2,
\ee
For the form factor we employ the one of dipole type
\be
F(\bi{q})=\left(
\frac{\Lambda^2 - m_\pi^2}{\Lambda^2 - q^2}
\right)^2 
\to 
\left(
\frac{\Lambda^2 - m_\pi^2}{\Lambda^2 + \bi{q}^2} 
\right)^2 \ .
\label{eq_def_FF}
\ee
The potential in the $r$-space is obtained by performing the Fourier transformation as
\begin{eqnarray}
  V_{NN}^\pi(\bi r)
&=& \int \frac{\rmd^3q}{(2\pi)^3}  V(\bi{q}) e^{i\bi{q}\cdot\bi{r}}
\nonumber \\  
  &=&\left(\frac{g_A^N}{2f_\pi} \right)^2\frac{1}{3}
  \left[
  \bsigma_1\cdot\bsigma_2
C(r;m_\pi,\Lambda) 
+S_{12}(\hat{r})T(r;m_\pi,\Lambda)
  \right]\btau_1\cdot\btau_2  \ .
 \label{eq:NNpiV}
\end{eqnarray}
where
$C(r;m,\Lambda)$ and $T(r;m,\Lambda)$ are 
given by by
\be
  C(r;m,\Lambda) &=& \frac{m^2}{4\pi}\left[\frac{e^{-mr}}{r}-\frac{e^{-\Lambda r}}{r} 
 -\frac{(\Lambda^2-m^2)}{2\Lambda}
e^{-\Lambda r} \right], \label{eq:poteC}\\
 T(r;m,\Lambda)
 &=&
\frac{1}{4\pi} \left( \left(3+3mr+m^2r^2\right)\frac{e^{-mr}}{r^3}
 -\left(3+3\Lambda r+\Lambda^2r^2\right)
 \frac{e^{-\Lambda r}}{r^3} \right.
  \nonumber \\
 &&
 + \left. \frac{m^2-\Lambda^2}{2}(1+\Lambda r)\frac{e^{-\Lambda r}}{r}
 \right)  \ .
 \label{eq:poteT}
 \ee

\subsection{Deuteron}
\label{sec:Deuteron}

It is instructive to discuss  how the OPEP alone  explains basic properties 
of the deuteron by adjusting the cutoff parameter in the form factor.  
It implies the importance of OPEP especially for low energy hadron dynamics.  
Furthermore, we will see the characteristic role of the tensor force which couples 
partial waves of different orbital angular momenta by two units.  
Inclusion of more coupled channels gains more attraction, and hence more chances 
to generate hadronic molecules.  
The importance of the OPEP for the $NN$ interaction is discussed nicely 
in the classic textbook~\cite{Ericson:1988gk}.  

The deuteron is the simplest composite system of the proton and neutron.  
It has  spin 1 and isospin 0.  
The main partial wave in the orbital wave function is $S$-wave 
with a small $D$-wave admixture of about 4 \%.  
It has  binding energy of 2.22 MeV and  size of 
about 4 fm  (diameter or relative distance of the proton and neutron).  
Because the interaction range of OPEP is $\sim 1/m_\pi \sim 1.4$ fm while the deuteron size  
is sufficiently larger than that, the nucleons in the deuteron spend most of their time 
without feeling the interaction.  
This defines loosely bound systems and is the defining condition for a hadronic molecule.  

The main S-wave component of the wave function $\psi(r)$ 
can be written as those of the free space 
\be
\psi(r) = A \frac{e^{-r\sqrt{2\mu B}}}{r} \ ,
\ee
where $\mu, B$ and $A$ are the reduced mass of the nucleon, binding energy and normalization constant.  
By using this the root mean square distance can be computed as 
\be
\bra r^2 \ket^{1/2} = \frac{1}{2\sqrt{\mu B}} \ .
\label{eq:rms}
\ee
The binding energy and the mass of the nucleon give $\bra r^2 \ket^{1/2} \sim 4 $ fm, consistent with 
the data.

Now it is interesting to show that these properties are reproduced 
by solving the coupled channel Schr\"odinger equation with only the OPEP included.  
Explicit form of the coupled channel equations are found in many references, 
and so we show here only essential results.
Employing the axial vector coupling constant for the nucleon $g_A^N \sim 1.25$ 
and choosing
 the cutoff parameter at $\Lambda$ = 837 MeV the binding energy is reproduced.
At the same time 
experimental data for the scattering length\footnote{Throughout this article, we define that 
  the positive (negative) scattering length stands for the attraction (repulsion) 
  at the threshold.}
and effective range are well reproduced
as shown in the third raw of table~\ref{table:cutoffNN}.  
Note that since $g_A \sim 1.25$ is fixed, the cutoff $\Lambda$ is the only parameter here.  

The cutoff value $\Lambda = 837$ MeV is consistent with the intrinsic hadron (nucleon) size.
By interpreting the form factor related to the finite structure of the nucleon,  we 
may find the relation 
\be
\bra r^2 \ket^{1/2} = \frac{\sqrt{6}}{\Lambda} \sim  0.6\; {\rm fm} \ .
\ee
The size 0.6 fm corresponds to the core size of the nucleon 
with the pion cloud removed.  
In table~\ref{table:cutoffNN}, results are shown also for those 
when other meson exchanges are included~\cite{Yamaguchi:2011xb}.  
By tuning the cutoff parameter $\Lambda$ around the suitable range 
as consistent with the nucleon size, low energy properties are reproduced.   

 \begin{table}[h]
  \caption{\label{table:cutoffNN} The cutoff parameter $\Lambda_N$
   determined to reproduce the deuteron binding energy, $B=2.22$ MeV, in
  the various meson exchange models.
  The scattering length $a$ and the effective range $r_e$ of the $^3S_1$
  channel are also shown.
  The experimental values of $a$ and $r_e$ are 
  $a=-5.42$ fm and
  $r_e=1.70$ fm, respectively.}
  
    \begin{center}   
   \begin{tabular}{cccc}
    \hline\hline
    Meson ex.& $\Lambda_N$ [MeV]& $a$ [fm] & $r_e$ [fm] \\ 
    &  & -5.42 (Exp)  & 1.70 (Exp)    \\   
    \hline
    $\pi$ & 
    837 & -5.25& 1.49\\
    $\pi\rho\omega$ &
    839&-5.25 &1.49 \\    
    $\pi\sigma$ &
    681&-6.51 &1.51 \\
    $\pi\rho\omega\sigma$ &
    710&-5.27 &1.53 \\
    \hline\hline
   \end{tabular}
  \end{center}   
 \end{table}

\subsection{OPEP for $P^{(\ast)}\bar{P}^{(\ast)}$}
\label{sec:OPEP_for_PPbar}

For the interaction of heavy $P$ and $P^*$ mesons, we use the Lagrangian and matrix elements 
of (\ref{L_PP*pi}) and (\ref{ME_L_PPpi}).  
In deriving the potential, we need to be a bit careful about the normalization of the state; 
there are $2E$ particles in unit volume.
This requires to divide amplitudes by 
$\sqrt{2E}$ per one external leg as was done for $NN$~\footnote{\label{note:sqrt2} In the previous publications 
by some of the present authors and others the factor $\sqrt{2}$ was missing~\cite{Yasui:2009bz,Yamaguchi:2011xb,Yamaguchi:2011qw,Ohkoda:2011vj,Ohkoda:2012hv,Yamaguchi:2013ty,Yamaguchi:2013hsa}.  
It is verified also by the former collaborator (S. Yasui, private communications).  
In this article this problem has been corrected.  
Accordingly, 
it turns out that the OPEP plays an important role for e.g. $X(3872)$, while not so 
for $Z_c$ and $Z_b$ as discussed in sections~\ref{sec:X3872} and \ref{sec:charged}. 
The baryon systems such as $\bar D N$ will be discussed elsewhere. } .  
The OPEP for the $P^{(\ast)}\bar{P}^{(\ast)}$ is given by
\begin{eqnarray}
 \fl & &V^{\pi}_{P\bar{P}{\mathchar`-}P^\ast \bar{P}^\ast}(r)
  =\left(\frac{g_A}{2f_\pi}\right)^2\frac{1}{3}\left[
 \bvarepsilon_1\,^\ast\cdot\bvarepsilon_2\,^\ast
 C(r;\mu,\Lambda) 
 +S_{\varepsilon^\ast\varepsilon^\ast}(\hat{r})
 T(r;\mu,\Lambda) \right]\btau_1\cdot\btau_2 \ ,
  \label{eq_VPPtoP*P}
  \\
 \fl & &V^{\pi}_{P\bar{P}^\ast{\mathchar`-}P^\ast \bar{P}}(r) 
= 
\left(\frac{g_A}{2f_\pi}\right)^2\frac{1}{3}\left[
 \bvarepsilon_1\,^\ast\cdot\bvarepsilon_2\,
 C(r;\mu,\Lambda)
 +S_{\varepsilon^\ast\varepsilon}(\hat{r})
 T(r;\mu,\Lambda) 
 \right]\btau_1\cdot\btau_2 \ ,
 \label{eq_VPP*toP*P}\\
 \fl & &V^{\pi}_{P\bar{P}^\ast{\mathchar`-}P^\ast \bar{P}^\ast}
  (r) 
  = 
  \left(\frac{g_A}{2f_\pi}\right)^2\frac{1}{3}
 \left[
 \bvarepsilon_1\,^\ast\cdot\bi{S}_2\,C(r;m_\pi,\Lambda)
 +S_{\varepsilon^\ast S}(\hat{r})T(r;m_\pi,\Lambda)
 \right]\btau_1\cdot\btau_2 \ ,
  \label{eq_VPP*toP*P*}\\
 \fl & &V^{\pi}_{P^\ast\bar{P}^\ast{\mathchar`-}P^\ast\bar{P}^\ast}  
  (r)= -\left(\frac{g_A}{2f_\pi}\right)^2\frac{1}{3}
 \left[
 \bi{S}_1\cdot\bi{S}_2\,C(r;m_\pi,\Lambda)
 +S_{S S}(\hat{r})T(r;m_\pi,\Lambda)
 \right]\btau_1\cdot\btau_2  \ .
  \label{eq_VP*P*toP*P*}
\end{eqnarray}
Here the axial coupling $g_A$ is for $P^{(*)}$ (or for the light quark), and 
$\bvarepsilon$ and $\bi{S}$ are the spin transition operator between $P \leftrightarrow  P^*$, and spin one operator for $P^*$, respectively.
The polarization vector plays the role of spin transition of $P \leftrightarrow P^*$ and $\bar P \leftrightarrow \bar P^*$.  
The tensor operator is 
$S_{{\cal O}_1{\cal O}_2}(\hat{r})=3({\cal O}_1\cdot \hat{r}) ({\cal O}_2\cdot \hat{r})-{\cal O}_1\cdot{\cal O}_2$
with ${\cal O}=\bvarepsilon^{(\ast)}$ or $\bi{S}$.
In actual studies for $X$ and $Z$ states, the total isospin of a $P^{(*)}\bar P^{(*)}$ system must be specified.  
The isospinors of these particles are
\be
\bar P^{(*)} = 
\left(
\begin{array}{c}
u \bar Q \\
d \bar Q 
\end{array}
\right), \; \; \; 
P^{(*)} = 
\left(
\begin{array}{c}
- Q \bar d \\
Q \bar u 
\end{array}
\right)  \ ,
\label{eq:PQqbar}
\ee
and the $\btau$ matrices in (\ref{eq_VPPtoP*P})-(\ref{eq_VP*P*toP*P*}) are understood to operate 
these isospin states.  
When $\bar P^{(*)}$ is replaced by $P^{(*)}$, an extra minus sign appears at each vertex reflecting 
the charge-conjugation or G-parity of  the pion as shown in table~\ref{table_PP*pi}.  
Finally we note that 
in (\ref{eq_VPPtoP*P}) and (\ref{eq_VPP*toP*P}), the mass is replaced by  an effective one $\mu$
by taking into account the energy transfer as discussed in the next subsection.  

\subsection{Effective long-range interaction}
\label{sec:Effective_long-range_interaction}

In the derivation of  $P^{(\ast)}\bar{P}^{(\ast)}$ potential, (\ref{eq_VPP*toP*P*}) and (\ref{eq_VP*P*toP*P*}) we have assumed
the static approximation, where 
the energy transfer $p^0$ is neglected for the exchanged pion
\be
\frac{1}{q^2 - m_\pi^2} \to - \frac{1}{\bi q^2 + m_\pi^2} \ .
\ee
However, when the masses of the interacting particles changes  such as in 
(\ref{eq_VPPtoP*P}) and (\ref{eq_VPP*toP*P}), 
the effective mass of the exchanged pion may change from that in the free space 
due to finite energy transfer.  

To see how this occurs let us start with the expression 
\be
V(\bi r) = - \left(\frac{g_A}{2 f_\pi} \right)^2
\int \frac{\rmd^3q}{(2\pi)^3} 
\frac{\bi \bvarepsilon^* \cdot \bi q \bvarepsilon \cdot \bi q}{-(E_{P^*}(\bi p)-E_P(\bi p^\prime))^2 + \bi q^2 + m_\pi^2} 
e^{i\bi q \cdot \bi r} \ ,
\label{eq_V_nonstatic}
\ee
where we have included the energy transfer with
\be
E_{P^*}(\bi p) = \sqrt{m_{P^*}^2 + \bi p^2}, \; \; \; 
E_{P}(\bi p^\prime) = \sqrt{m_{P}^2 + \bi p^{\prime 2}}, \; \; \; 
\bi q = \bi p - \bi p^\prime \ .
\ee
For heavy particles, we may approximate
\be
E_{P^*}(\bi p) \sim m_{P^*}, \; \; \; E_{P}(\bi p) \sim m_{P} \ .
\ee
The ignored higher order terms are of order 
\be
\frac{\bra {\bi p}^2 \ket}{2m_{P^{(*)}}} &\sim& 20\; {\rm MeV}, \; \; {\rm for\; charm} \ ,
\nonumber \\
\frac{\bra {\bi p}^2 \ket}{2m_{P^{(*)}}} &\sim& 8\; {\rm MeV}, \; \; {\rm for\; bottom} \ ,
\ee
or less when the molecule size is of order 1 fm or larger.
These values can be neglected as compared with the mass differences of 
$m_{D^*} - m_D \sim 140$ MeV and 
$m_{B^*} - m_B \sim 45$ MeV.  
In the integration over $\bi q$, the pion energy is $E_\pi(\bi q) > m_\pi$.  
Therefore, 
\begin{itemize}

\item
If $m_{P^*}-m_P < m_\pi$ which is the case of $B^*, B$ mesons, the integrand of (\ref{eq_V_nonstatic}) is regular 
but the exchanged pion mass is effectively reduced as 
\be
\mu^2 \equiv m_\pi^2 - (m_{P^*}-m_P)^2 < m_\pi^2 \ ,
\ee
where $\mu$ is regarded as an effective mass.  
Therefore the interaction range is extended.  
Some consequences of this effective long range interactions are discussed in~\cite{Geng:2017jzr}.  

\item 
If $m_{P^*}-m_P > m_\pi$ which is marginally the case of $D^*, D$ mesons, 
the integrand of (\ref{eq_V_nonstatic})  hits the singularity and generates an imaginary part.  
The integral is still performed, and the resulting $\bi r$-space potential is given by 
\be
-\int \frac{\rmd^3q}{(2\pi)^3} 
\frac{1}{\bi q^2 - \vert \mu\vert ^2 - i \epsilon} 
e^{i\bi q \cdot \bi r}
=
-\frac{1}{4\pi} \frac{e^{i\vert \mu\vert r}}{r}  \ .
\label{eq_complex_OPEP}
\ee
The function ${e^{i\vert\mu \vert r}}/{r}$ represents the outgoing wave for the decaying pion with momentum $|\bi k| = \vert \mu \vert$.  
The plus sign is determined by the boundary condition implemented by $+ i \epsilon$.  

\end{itemize}

\subsection{Physical meaning of the imaginary part}
\label{sec:Physical_meaning_of_the_imaginary_part}

The presence of imaginary part implies an instability of a system.  
For $D\bar D^*$ systems, it corresponds to the decay of $D\bar D^* \to D\bar D\pi$ if this process is allowed kinematically.  
To show this explicitly, let us first consider the matrix element of the complex OPEP (\ref{eq_complex_OPEP}) 
by a bound state $\bra \bi r | B \ket = \varphi(\bi r)$, 
\be
\bra B | V | B \ket
&=&
\int \rmd^3 r\ \varphi^*(\bi r) V(\bi r) \varphi(\bi r)
\nonumber \\
&=&
\left(\frac{g_A}{2 f_\pi} \right)^2
\int 
\frac{\rmd^3p}{(2\pi)^3}
\frac{\rmd^3q}{(2\pi)^3}
\tilde \varphi^*(\bi p+\bi q) \frac{\bvarepsilon^* \cdot \bi q \bvarepsilon \cdot \bi q}{\bi q^2 - \mu^2 - i \epsilon} 
 \tilde \varphi(\bi p)  \ ,
\ee
where the momentum wave function is defined by
\be
\tilde \varphi(\bi p)
=
\int \rmd^3 r\ e^{- i\bi p \cdot \bi r} \varphi(\bi r)
\ee
and is normalized as 
\be
\int \frac{\rmd^3 p}{(2\pi)^3} \tilde \varphi(\bi p)^* \tilde \varphi(\bi p) = \int \rmd^3 x \ \varphi^*(\bi x)  \varphi(\bi x) = 1 \ .
\ee

Decomposing the denominator of the interaction by using $\mu^2 = m_\pi^2 - (m_{D^*}^2 - m_{D}^2)$
and 
$\Delta = m_{D^*} - m_D + E_\pi(\bi q^2)$
\be
\frac{1}{\bi q^2 - \mu^2 - i \epsilon} 
&=&
- \frac{1}{(m_{D^*} - m_D - E_\pi(\bi q^2) + i \epsilon)\Delta}  \ ,
\ee
we find
\be
\fl
\bra B | V | B \ket
&=&
\left(\frac{g_A}{2 f_\pi} \right)^2
\int 
\frac{\rmd^3p}{(2\pi)^3}
\frac{\rmd^3q}{(2\pi)^3}
\tilde \varphi^*(\bi p+\bi q) 
\frac{\bvarepsilon^* \cdot \bi q \bvarepsilon \cdot \bi q}{m_{D^*} - m_D - E_\pi(\bi q^2) + i \epsilon} \tilde \varphi(\bi p) 
\frac{1}{\Delta}
\nonumber \\
\fl
&=&
- \left(\frac{g_A}{2 f_\pi} \right)^2
\int 
\frac{\rmd^3p}{(2\pi)^3}
\frac{\rmd^3q}{(2\pi)^3}
\frac{\rmd^3p^\prime}{(2\pi)^3}
(2\pi)^3\delta^3(\bi p^\prime + \bi p + \bi q)
\nonumber \\
\fl
&\times&
\tilde \varphi^*(\bi p^\prime) 
\frac{\bvarepsilon^* \cdot \bi q \bvarepsilon \cdot \bi q}{m_{D^*} - m_D - E_\pi(\bi q^2) + i \epsilon}
 \tilde \varphi(\bi p) 
\frac{1}{\Delta} \ .
\label{eq_Vmatrixelement}
\ee
By using the identity
\be
\frac{1}{x+ i \epsilon} = \frac{\rm P}{x} - i \pi \delta(x) \ ,
\label{eq_Satoformula}
\ee
where P stands for the principal value integral, the imaginary part of (\ref{eq_Vmatrixelement}) is written as 
\be
\fl
{\rm Im}\ \bra B | V | B \ket
&=&
+\frac{1}{2}
\left(\frac{g_A}{2 f_\pi} \right)^2
\int 
\frac{\rmd^3p}{(2\pi)^3}
\frac{\rmd^3q}{2E_\pi(\bi q) (2\pi)^3}
\frac{\rmd^3p^\prime}{(2\pi)^3}
(2\pi)^4\delta^4(p^\prime + p + q - P)
\nonumber \\
\fl
&\times&
\tilde \varphi^*(\bi p^\prime)
\bvarepsilon^* \cdot \bi q  \bvarepsilon \cdot \bi q \tilde \varphi(\bi p)  \ .
\label{eq_ImVmatrixelement}
\ee

\begin{figure}[h]
\begin{center}
\includegraphics[width=0.8 \linewidth]{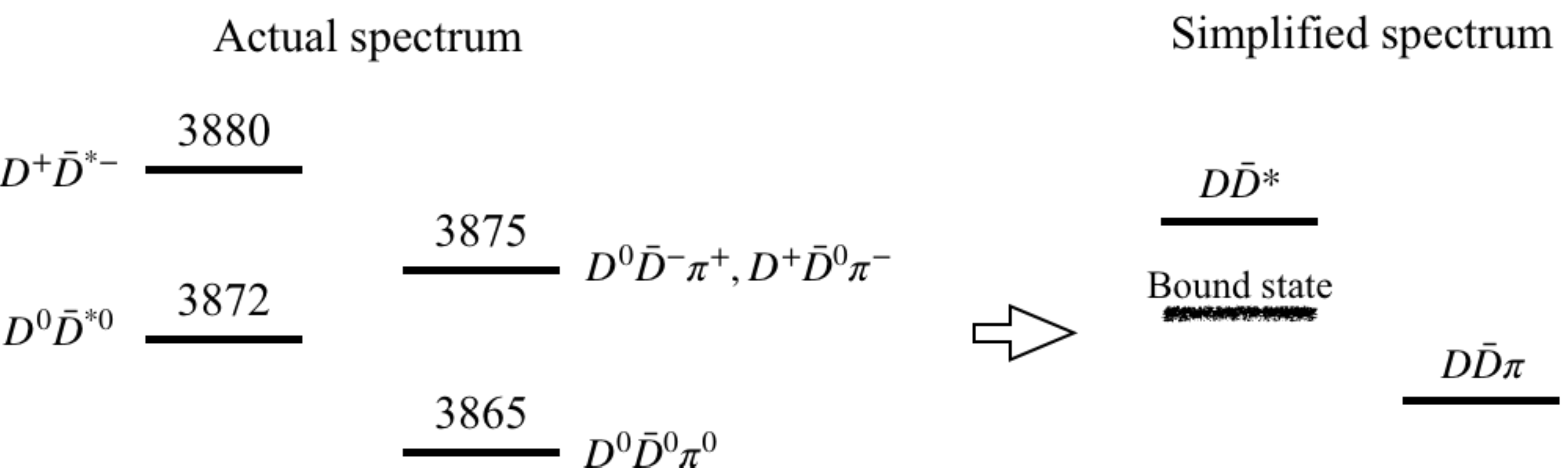}
\end{center}
\vspace{-5mm}
\caption{The spectrum of $D\bar D^*$ and $D\bar D\pi$. Actual case with isospin breaking (left) and a simplified one (right).}
\label{fig_DDpi_spectrum}
\end{figure}

We can now show explicitly that the imaginary part (\ref{eq_ImVmatrixelement}) is related 
to a part of the decay processes 
of the quasi-bound state $\varphi$.  
For illustrative purpose, we consider 
the three-body decay of $D\bar D^* \to D \bar D \pi$ of isospin symmetric case
as shown in \fref{fig_DDpi_spectrum}.
There actual small mass differences in the charged and neutral particles 
are ignored in the right panel.  

\begin{figure}[h]
\begin{center}
\includegraphics[width=0.8 \linewidth]{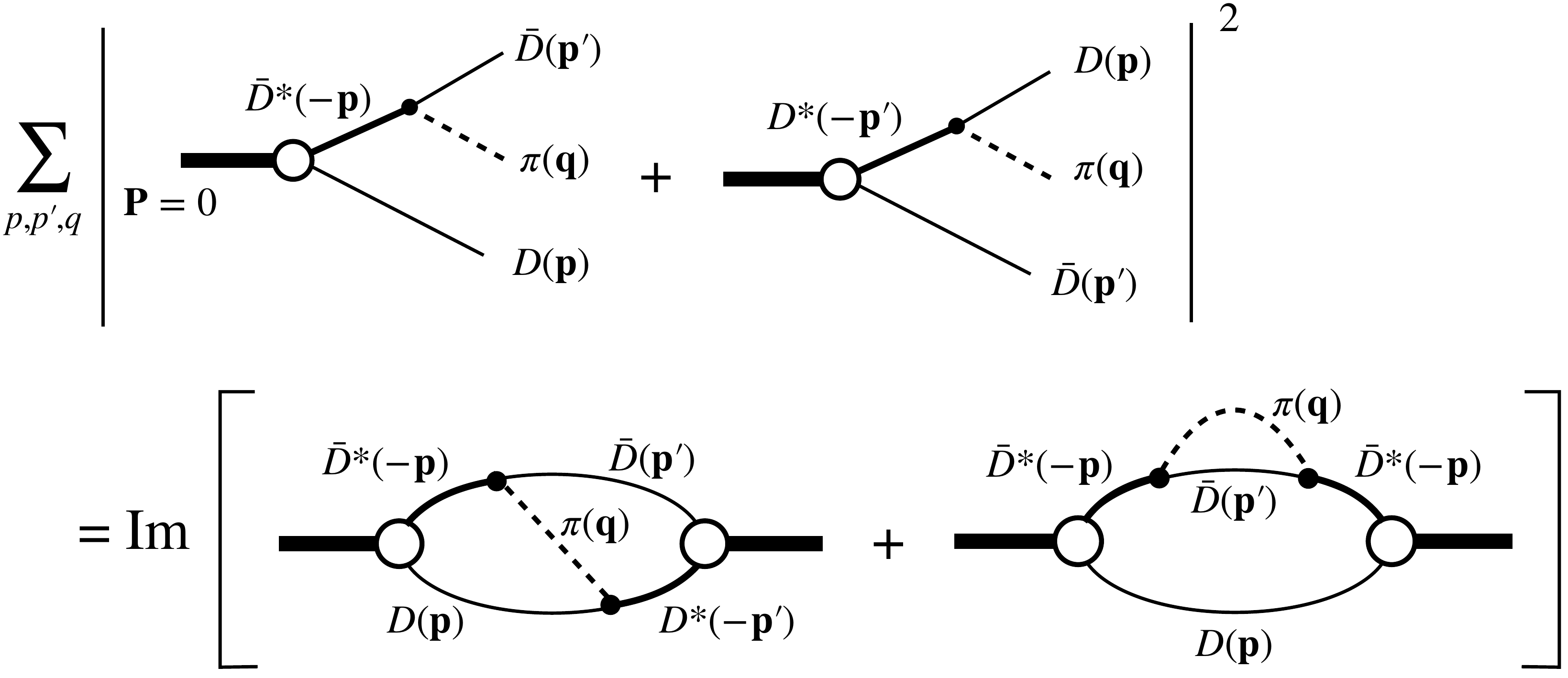}
\end{center}
\vspace{-5mm}
\caption{The optical theorem for a $D\bar D^*$ bound state decaying into $D\bar D\pi$.}
\label{fig_cutkosky}
\end{figure}

The three-body decay  is computed by the diagrams in the first (upper) line 
of \fref{fig_cutkosky}, which are for the decay
of the quasi-bound state at rest ($\bi P = 0$) into $D(\bi p), \bar D(\bi p^\prime), \pi(\bi q)$.  
Note that there are two possible processes for a given set of the final state momenta 
$\bi p, \bi p^\prime, \bi q$, whose amplitudes are added coherently.  
Denoting the interaction vertex of $\bar D^* \to \bar D \pi$ as $h \bvarepsilon \cdot \bi q$, 
where $h = g_A/2f_\pi$
the amplitude is written as  
\be
\sqrt{2M_B} (\tilde \varphi(\bi p) + \tilde \varphi(\bi p^\prime)) h \bvarepsilon \cdot \bi q
 \ .
\ee
Here the factor $\sqrt{2M_B}$ is  for the normalization of the initial state; there are $2M_B$ particles in a unit volume. 
Squaring this and multiplying the three-body phase space the decay rate is computed by
\be
\fl
\Gamma &=& 
\frac{1}{2M_B} 
\int 
\frac{\rmd^3p}{2E_D(\bi p)(2\pi)^3} 
\frac{\rmd^3p^\prime}{2E_D(\bi p^\prime )(2\pi)^3} 
\frac{\rmd^3q}{2E_\pi(\bi q)(2\pi)^3} 
(2\pi)^4 \delta^4(p+p^\prime+q- P) 
\nonumber \\
\fl
&\times&
2M_B h^2
(\varphi^*(\bi p) + \varphi^*(\bi p^\prime)) 
\bvarepsilon^* \cdot \bi q \bvarepsilon \cdot \bi q
(\varphi(\bi p) + \varphi(\bi p^\prime))  \ .
\ee

Now let us consider the diagrams of the second line of \fref{fig_cutkosky}.
The left diagram is computed by
\be
\fl
A(M_B)  &=& 
\int 
\frac{\rmd^3p}{2E_D(\bi p)(2\pi)^3} 
\frac{\rmd^3p^\prime}{2E_D(\bi p^\prime )(2\pi)^3} 
\frac{\rmd^3q}{2E_\pi(\bi q)(2\pi)^3} 
(2\pi)^3 \delta^3(\bi p + \bi p^\prime + \bi q)
\nonumber \\
\fl
&\times&
\sqrt{2M_B} \tilde \varphi^*(\bi p) {h \bvarepsilon \cdot \bi q}
\frac{1}{M_B - E_D(\bi p) - E_D(\bi p^\prime) - E_\pi(\bi q) + i \epsilon}
{h \bvarepsilon \cdot \bi q} \sqrt{2M_B} \tilde \varphi(\bi p^\prime)
\nonumber \\
\fl
&=&
2M_B h^2 \int 
\frac{\rmd^3p}{2E_D(\bi p)(2\pi)^3} 
\frac{\rmd^3p^\prime}{2E_D(\bi p^\prime )(2\pi)^3} 
\frac{\rmd^3q}{2E_\pi(\bi q)(2\pi)^3} 
(2\pi)^3 \delta^3(\bi p + \bi p^\prime + \bi q)
\nonumber \\
\fl
& & 
\hspace*{1cm}\times\; 
\tilde \varphi^*(\bi p)  
\frac{\bvarepsilon^* \cdot \bi q \bvarepsilon \cdot \bi q }
{M_B - E_D(\bi p) - E_D(\bi p^\prime) - E_\pi(\bi q) + i \epsilon}
\tilde \varphi(\bi p^\prime)  \ .
\ee
Here we have used the time ordered perturbation theory and taken into account only the 
terms that contribute to the decay.  
Similarly we obtain the amplitude for the right diagram by the replacement 
$\varphi(\bi p^\prime) \to \varphi(\bi p)$ in the numerator.
Therefore the sum of the diagrams is 
\be
\fl
A(M_B)  &=& 
2M_B h^2 \int 
\frac{\rmd^3p}{2E_D(\bi p)(2\pi)^3} 
\frac{\rmd^3p^\prime}{2E_D(\bi p^\prime )(2\pi)^3} 
\frac{\rmd^3q}{2E_\pi(\bi q)(2\pi)^3} 
(2\pi)^3 \delta^3(\bi p + \bi p^\prime + \bi q)
\nonumber \\
\fl
& & 
\hspace*{1cm}\times\; 
\frac{(\varphi^* (\bi p) + \varphi^*(\bi p^\prime)) 
\bvarepsilon \cdot \bi q \bvarepsilon \cdot \bi q (\varphi(\bi p) + \varphi(\bi p^\prime))}
{M_B - E_D(\bi p) - E_D(\bi p^\prime) - E_\pi(\bi q) + i \epsilon} \frac{1}{2}
 \ .
\ee
Picking up the imaginary part, we find 
\be
\fl
{\rm Im}\ A(M_B)
&=&
- \frac{M_B}{2}h^2 \int 
\frac{\rmd^3p}{2E_D(\bi p)(2\pi)^3} 
\frac{\rmd^3p^\prime}{2E_D(\bi p^\prime )(2\pi)^3} 
\frac{\rmd^3q}{2E_\pi(\bi q)(2\pi)^3} 
(2\pi)^4 \delta^4(p + p^\prime + q - P)
\nonumber \\
\fl
& & 
\hspace*{1cm}\times\; 
(\varphi^* (\bi p) + \varphi^*(\bi p^\prime)) 
\bvarepsilon \cdot \bi q \bvarepsilon \cdot \bi q (\varphi(\bi p) + \varphi(\bi p^\prime)) \ .
\label{eq_ImAmatrixelement}
\ee
The optical theorem says that by writing the $S$-matrix as $S = 1 - iT$,
\be
S^\dagger S = 1 \to 2 {\rm Im} T = |T|^2 \ .
\ee
Therefore, considering the normalization of this particle ($2M_B$ particles in a unit volume), 
we find that the imaginary part agrees with the decay width.
This is nothing but an explicit check of the optical theorem.  

We see that the off diagonal integral in (\ref{eq_ImAmatrixelement}), 
$
\varphi^* (\bi p) 
\cdots
\varphi(\bi p^\prime),
$
agrees with the the potential matrix element (\ref{eq_ImVmatrixelement})
modulo a kinematical factor.
The difference appears due to different normalization factors in the state 
decaying into two particles and that of bound states.  
The diagonal part 
$
\varphi^* (\bi p) 
\cdots 
\varphi(\bi p)
$
corresponds to the imaginary part of the self energy of $\bar D^*$ and 
is not included in the potential matrix element.

\subsection{The quark model and the hadronic model}
\label{sec:quark-hadronic-models}

Here we define meson and exotic states using a quark model.
By doing so, we can combine the quark model and the hadron model smoothly into a quark-hadron hybrid model.
Such a model enables us
to handle the physics of resonances, long range interactions like OPEP, and rather complicated systems, by a hadron model but
with the quark degrees of freedom effectively included.
Also, by constructing hadrons from the quark degrees of freedom,
the charge conjugation of the hadron systems 
can be defined in a more consistent way as shown below.

To obtain observables from the model Lagrangian, one has to choose the initial and/or the final state.
A $q_\alpha \qbar_\beta$ meson with a certain spin structure 
 can be defined by using the fermion bilinear as \cite{Peskin:1995ev}
\begin{eqnarray}
|q_\alpha \bar q_\beta; \tilde n \Gamma\ket &=& 
\sqrt{2M\over 2m_\alpha 2m_\beta}~\phi ~ 
\tilde n^*\cdot \Big(\bar \psi_\alpha\;\Gamma \;\psi_\beta\Big)|0\ket ~,
\end{eqnarray}%
where $\psi_\alpha$ is the field operator, 
$\Gamma$ stands for the sixteen 4$\times$4 matrices,
$\tilde n$ stands for the spin orientation of the state,
and the mark ${\tilde n}^*$ corresponds to the complex conjugate of $\tilde n$.
For a vector meson, $\tilde n^* \cdot(\bar\psi\Gamma\psi)$ corresponds to ${1\over \sqrt{2}}{\bi \epsilon}^*_\mu(\bar\psi\gamma^\mu\psi)$,
and for a pseudoscalar meson, it corresponds  to $({-i\over \sqrt{2}}) (\bar\psi i\gamma^5\psi)$.
The suffix $\alpha$ or $\beta$ stands for other quantum numbers, such as color and flavor.
 $\phi$ is a relative motion wave function of the quark and the antiquark. 
$M$ is the meson mass while $m_\alpha$ and $m_\beta$ are the quark and the antiquark masses.
The normalization of this state is taken as 
$\bra q_{\alpha'} \bar q_{\beta'}|q_\alpha \bar q_\beta\ket = 2M\delta_{\alpha'\alpha}\delta_{\beta'\beta}(2\pi)^3\delta^3({\bi K}'-{\bi K})$, where
$\bi K$ and $\bi K'$ are the center of mass momenta of the initial and final $q\bar q$ mesons, which we set to be zero in the following.
The above expression can be reduced to
\begin{eqnarray}
|q_\alpha \bar q_\beta; \tilde n \Gamma {\bi r}\ket 
&=
\sqrt{2M\over 2m_\alpha 2m_\beta} 
&\int {{\rm d}^3{\bi k}\over (2\pi)^3}e^{-i{\bi k}{\bi r}}  \phi({\bi k})
\nonumber\\&&\times
 \tilde n^*\cdot \sum_{s,t}  \Big({\bar u}^s_\alpha(k) \Gamma  v^t_\beta(-k)\Big) {a}^s_{\alpha,{\bi k}}{}^\dag b^t_{\beta, -{\bi k}}{}^\dag|0\ket 
 ~,
 \label{eq:qqbarmeson}
\end{eqnarray}%
where  ${\bi r}={\bi r}_{q}-{\bi r}_{\bar q}$.
Note that in this definition 
the ${a}^s{}^\dag$ operator stays always on the left side of the $b^t{}^\dag$ operator,
not vice versa, for the $q\bar q$ meson.

Charge conjugation, $C$, changes the creation operators of the quarks to those of the antiquarks:
\begin{equation}
C {a}^s_{\alpha,{\bi k}}{}^\dag C =  {b}^s_{\alpha,{\bi k}}{}^\dag ~,
~~~ C{b}^s_{\alpha,{\bi k}}{}^\dag C= {a}^s_{\alpha,{\bi k}}{}^\dag ~.
\end{equation}%
So, the state $|q_\alpha \bar q_\beta; \tilde n \Gamma\ket$ is transformed into
\begin{eqnarray}
C|q_\alpha \bar q_\beta; \tilde n \Gamma{\bi r}\ket 
&= &
\sqrt{2M\over 2m_\alpha 2m_\beta} 
 \int {{\rm d}^3{\bi k}\over (2\pi)^3}\rme^{+\rmi{\bi k}{\bi r}} \phi({\bi k}) 
\nonumber\\&&\times
\tilde n^*\cdot \sum_{s,t}  \Big({\bar u}_\alpha^s(k) \Gamma  v^t_\beta(-k)\Big) 
{b}^s_{\alpha, {\bi k}}{}^\dag a^t_{\beta,-{\bi k}}{}^\dag|0\ket 
\nonumber\\
&=&
\sqrt{2M\over 2m_\alpha 2m_\beta} 
\int {{\rm d}^3{\bi k}\over (2\pi)^3}e^{-i{\bi k}{\bi r}} \phi(-{\bi k}) 
\nonumber\\&&\times
\tilde n^*\cdot \sum_{s,t}  \Big({\bar u}_\alpha^s(-k) \Gamma  v^t_\beta(k)\Big) 
{b}^s_{\alpha,-{\bi k}}{}^\dag a^t_{\beta,{\bi k}}{}^\dag|0\ket 
\nonumber\\
&=&
-\sqrt{2M\over 2m_\alpha 2m_\beta} 
\int {{\rm d}^3{\bi k}\over (2\pi)^3}e^{-i{\bi k}{\bi r}} (-1)^\ell\phi({\bi k}) 
\nonumber\\&&\times
\tilde n^*\cdot \sum_{s,t}  \Big({\bar u}_\beta^t(k) \gamma^0\gamma^2\Gamma^T \gamma^2\gamma^0 v^s_\alpha(-k)\Big) 
a^t_{\beta,{\bi k}}{}^\dag{b}^s_{\alpha,-{\bi k}}{}^\dag |0\ket  ~.
\end{eqnarray}%
Here the coordinate ${\bi r}$ is changed to $-{\bi r}$ after the charge conjugation because
it is defined as ${\bi r}_{q}-{\bi r}_{\bar q}$. 
The minus sign in the last equation 
comes from anticommutation of the fermion operators $ a^\dag$ and ${b}^\dag$. 
There we also use
\begin{eqnarray}
{\bar u}^s(k)&=&i(\gamma^2\gamma^0v^s(k))^T,{\rm ~~~and~~~}
v^s(k)=-i({\bar u}^s(k)\gamma^0\gamma^2)^T ~.
\end{eqnarray}%

For simplicity, let us omit the orbital part of the wave function, $\phi$, 
and assume $\ell=0$.
In a nonrelativistic quark model, 
the higher order term of $O(p/m)$ in the spinors is usually taken care of in operators
as relativistic effects. 
For further computations here, we follow the convention of \cite{Peskin:1995ev}.
The nonrelativistic spinors are taken as 
\begin{eqnarray}
&&u^s(k)=\sqrt{m}\left(\begin{array}{c}\xi^s\\ \xi^s\end{array}\right) {\rm ~and~}
v^s(k)=\sqrt{m}\left(\begin{array}{c}\xi^{-s}\\ -\xi^{-s}\end{array}\right),\\
&&\xi^{-s}=-i\sigma_y (\xi^s)^*,  \\
&&\xi^1
=\left(\begin{array}{c}1\\0\end{array}\right)
,
\xi^2
=\left(\begin{array}{c}0\\1\end{array}\right)
,
\\
&&\xi^{-1}
=\left(\begin{array}{c}0\\1\end{array}\right)
{\rm ~and~}  
\xi^{-2}
=\left(\begin{array}{c}-1\\0\end{array}\right),
\end{eqnarray}%
where spin up and down  correspond to $s$ = 1 and 2, respectively, for both of quarks and antiquarks.

First let us consider the pseudoscalar $Q\bar q$ meson,
$|Q\bar q;SS_z\ket=|Q\bar q;00\ket$, and its behavior under the charge conjugation. 
For a pseudoscalar meson we take $\Gamma=i\gamma^5$ and $\tilde n=i/\sqrt{2}$
in \eref{eq:qqbarmeson},
\begin{eqnarray}
{1\over \sqrt{2M}}|Q\bar q;00\ket
&=& {-i\over \sqrt{2}} {1\over \sqrt{2m_Q2m_q}}
\sum_{s,t}\Big({\bar u}^s_Q i\gamma^5  v^t_q\Big) {a}^s_{Q}{}^\dag b^t_{q}{}^\dag|0\ket  ~.
\end{eqnarray}%
Then we have
\begin{eqnarray}
{1\over \sqrt{2M}}|Q\bar q;00\ket
&=&{-i\over \sqrt{2}}(-i) \sum_{s,t}\xi^s {}^\dag \xi^{-t} {a}^s_{Q}{}^\dag b^t_{q}{}^\dag|0\ket 
\nonumber\\
&=& 
  -{1\over \sqrt{2}} \Big(
  \xi^1 {}^\dag \xi^{-2} {a}^1_{Q}{}^\dag b^2_{q}{}^\dag
+  \xi^2 {}^\dag \xi^{-1} {a}^2_{Q}{}^\dag b^1_{q}{}^\dag
  \Big)|0\ket 
\nonumber\\
&=& 
  -{1\over \sqrt{2}}\Big(
  -{a}^1_{Q}{}^\dag b^2_{q}{}^\dag
+  {a}^2_{Q}{}^\dag b^1_{q}{}^\dag
  \Big)|0\ket 
\nonumber\\
&=& 
{1\over \sqrt{2}}\Big(
  |Q^\up \bar q^\dw\ket
 -|Q^\dw \bar q^\up\ket
  \Big).
\end{eqnarray}%
This corresponds to the $D$ meson when $Q$ and $q$ are taken as the charm and the light quarks, respectively.
In the last equation, we define $|q\ket=a^\dag|0\ket$ so that 
its normalization becomes 1 instead of $2E$ as in the nonrelativistic quark model.

When charge conjugation is applied to this state, we have
\begin{eqnarray}
C{1\over \sqrt{2M}}|Q\bar q;00\ket
&=&-{-i\over \sqrt{2}}{1\over \sqrt{2m_Q2m_q}}\sum_{s,t}\Big({\bar u}^t_q 
\gamma^0\gamma^2 i(\gamma^5)^T  \gamma^2\gamma^0v^s_Q\Big) {a}^t_{q}{}^\dag b^s_{Q}{}^\dag|0\ket 
\nonumber\\
&=& {-i\over \sqrt{2}}{1\over \sqrt{2m_Q2m_q}}\Big({\bar u}^t_q i\gamma^5  v^s_Q\Big) {a}^t_{q}{}^\dag b^s_{Q}{}^\dag|0\ket 
\nonumber\\
&=& 
{1\over \sqrt{2}}\Big(
  |q^\up \bar Q^\dw\ket
 -|q^\dw \bar Q^\up\ket
  \Big)
\nonumber\\
&=&+{1\over \sqrt{2M}}|q\bar Q;00\ket ~.
\label{eq:CCps}
\end{eqnarray}%
This state can be regarded as the $\bar D$ meson, meaning that the charge conjugate of the $D$ meson 
is ($+1$) times the $\bar D$ meson.
Or, when both of the $Q$ and $q$ quarks are taken to be the charm quarks, this state corresponds to the $\eta_c$ meson,
 whose $C$-parity is ($+1$).

Next we consider the  vector meson, 
$|Q\bar q;SS_z\ket=|Q\bar q;11\ket$, and its behavior under charge conjugation. 
Now we take 
$\tilde n$ to be ${{\bi \epsilon}^+/ \sqrt{2}}={1\over 2}(0,1,-i,0)^\mu$
and $\Gamma=\gamma^\mu$.
\begin{eqnarray}
{1\over \sqrt{2M}}|Q\bar q;11\ket
&=& {1\over 2}(0,1,i,0)_\mu{1\over \sqrt{2m_Q2m_q}}\sum_{s,t}\Big({\bar u}^s_Q \gamma^\mu  v^t_q\Big) {a}^s_{Q}{}^\dag b^t_{q}{}^\dag|0\ket 
 ~.
\end{eqnarray}%
Then we have
\begin{eqnarray}
{1\over \sqrt{2M}}|Q\bar q;11\ket
&=&\sum_{s,t}\xi^s {}^\dag {1\over 2}(\sigma_x + i\sigma_y)\xi^{-t} {a}^s_{Q}{}^\dag b^t_{q}{}^\dag|0\ket 
\nonumber\\
&=& 
 \xi^1 {}^\dag \xi^{-1} {a}^1_{Q}{}^\dag b^1_{q}{}^\dag|0\ket 
 ={a}^1_{Q}{}^\dag b^1_{q}{}^\dag|0\ket,
\nonumber\\
&=& 
|Q^\up \bar q^\up\ket ~,
\end{eqnarray}%
which corresponds to the $D^*$ meson, when $Q$ and $q$ are the charm and nonflavor quarks, respectively.
Under the charge conjugation, it becomes
\begin{eqnarray}
C{1\over \sqrt{2M}}|Q\bar q;11\ket
&=&-{1\over 2}(0,1,i,0)_\mu 
\nonumber\\
&&\times  {1\over \sqrt{2m_Q2m_q}}\sum_{s,t}\Big({\bar u}^t_q \gamma^0\gamma^2(\gamma^\mu)^T  \gamma^2\gamma^0v^s_Q\Big) {a}^t_{q}{}^\dag b^s_{Q}{}^\dag|0\ket 
\nonumber\\
&=& -{1\over 2}(0,1,i,0)_\mu  
{1\over \sqrt{2m_Q2m_q}}\sum_{s,t}\Big({\bar u}^t_q \gamma^\mu  v^s_Q\Big) {a}^t_{q}{}^\dag b^s_{Q}{}^\dag|0\ket 
\nonumber\\
&=&-{a}^1_{q}{}^\dag b^1_{Q}{}^\dag|0\ket
\nonumber\\
&=& 
-|q^\up \bar Q^\up\ket ~.
\end{eqnarray}%
This corresponds to ($-1$) times the ${\bar D}{}^*$ meson.
Or, when both of the $Q$ and $q$ quarks are taken to be the charm quarks, this state corresponds to the $J/\psi$ meson,
 whose $C$-parity is ($-1$).
This result and \eref{eq:CCps} are in accordance with what has been anticipated in 
\eref{eq_cconjugate}.

Finally, we consider a four-quark system such as $D^{(*)}\bar D{}^{(*)}$ 
and its charge conjugation.
The $C$-parity can be defined for the neutral systems.
The charge conjugation changes $D$ to $\bar D$, and $D^*$ to $-{\bar D}{}^*$.
Thus the $C$-parity changes $D^{(*)}\bar D{}^{(*)}$
associated with the orbital relative wave function, $\psi_L({\bi r})$, with ${\bi r}={\bi r}_D-{\bi r}_{\bar D}$, as
\begin{eqnarray}
C:~D{\bar D}\psi_L({\bi r}) &\to & {\bar D}D\psi_L(-{\bi r})=(-1)^L D{\bar D}\psi_L({\bi r}), \\
C:~D{\bar D}{}^* \psi_L({\bi r}) &\to & {-{\bar D}}D^*\psi_L(-{\bi r})=-(-1)^L D^*{\bar D}\psi_L({\bi r}), \\
C:~D{}^*{\bar D} \psi_L({\bi r}) &\to & {-{\bar D}{}^*}D\psi_L(-{\bi r})=-(-1)^L D{\bar D}{}^*\psi_L({\bi r}), \\
C:~D^*{\bar D}{}^* \psi_L({\bi r})&\to & {\bar D}{}^*D^*\psi_L(-{\bi r})=(-1)^{L+S}D^*{\bar D}{}^*\psi_L({\bi r}),
\end{eqnarray}%
where $L$ is the orbital angular momentum of the $D^{(*)}$ and $\Dbar^{(*)}$ relative motion,
and $S$ is the total spin $(=0,1,2)$.
Thus, the $C$-parity eigenstates of the $D{\bar D}$ systems are also eigenstates of the parity,
\begin{eqnarray}
J^{PC}&=&L^{++}: [D\bar D]_+ \psi_{L=even}
\label{eq:JPCDD}\\
&&L^{--}: [D\bar D]_- \psi_{L=odd}
\end{eqnarray}%
with
\begin{eqnarray}
{}[D\bar D]_\pm&=&{1\over \sqrt{2}}\Big(D{\bar D} \pm {\bar D}D\Big) ~.
\label{eq:JPC-DD-}
\end{eqnarray}%
Similarly,
those with $D{\bar D}{}^*$ or $D^*{\bar D}$ are
\begin{eqnarray}
J^{PC}
&=&J^{++}:{1\over \sqrt{2}}\Big( [D{\bar D}{}^*]_- -[D{}^*{\bar D}]_-\Big)\psi_{L=even}, \\
&&J^{+-}: {1\over \sqrt{2}}\Big( [D{\bar D}{}^*]_+ +[D{}^*{\bar D}]_+\Big)\psi_{L=even}, \\
&&J^{-+}: {1\over \sqrt{2}}\Big( [D{\bar D}{}^*]_- +[D{}^*{\bar D}]_-\Big)\psi_{L=odd}, \\
&&J^{--}: {1\over \sqrt{2}}\Big( [D{\bar D}{}^*]_+ -[D{}^*{\bar D}]_+\Big)\psi_{L=odd}
\label{eq:JPCDDstar}
\end{eqnarray}%
with
\begin{eqnarray}
{}[D{\bar D}{}^*]_\pm&=&{1\over \sqrt{2}} (D{\bar D}{}^* \pm {\bar D}D^*),~~~
{}[D{}^*{\bar D}]_\pm = {1\over \sqrt{2}} (D^*{\bar D} \pm {\bar D}{}^*D) ~.
\label{eq:JPC-DDstar-}
\end{eqnarray}%
For $D^*{\bar D}{}^*$, the 
simultaneous eigenstates of the parity and the $C$-parity
relate to the angular momentum $L$ and the total spin $S$ as
\begin{eqnarray}
J^{PC}
&=&J^{++}:[D^*{\bar D}{}^*]_{S=even,+} \psi_{L=even} , \\
&& J^{+-}:[D^*{\bar D}{}^*]_{S=odd ,-} \psi_{L=even}  , \\
&& J^{-+}:[D^*{\bar D}{}^*]_{S=odd,+}  \psi_{L=odd}  , \\
&& J^{--}:[D^*{\bar D}{}^*]_{S=even,-} \psi_{L=odd} 
\label{eq:JPCDstarDstar}
\end{eqnarray}%
with
\begin{eqnarray}
{}[D{}^*\bar D{}^*]_{S,\pm}
&=&{1\over \sqrt{2}}\Big(D{}^*{\bar D}{}^* \pm {\bar D}{}^*D{}^*\Big)\Big|_S ~.
\label{eq:JPC-DstarDstar-}
\end{eqnarray}%

In the quark model, the relations concerning the rearrangement 
between  $c\bar c$-$q\bar q$ and $c\bar q$-$q\bar c$ 
are derived in a systematic manner.
For this, first we note that there are two color configurations for the $q\qbar c\cbar$ systems:
the one where the quark-antiquark pairs $q\qbar$ and  $c\cbar$ are color singlet
and the other where the pairs $q\cbar$ and $c\qbar$ are color singlet.
These two configurations are related to hidden charm,   
 $\omega\Jpsi$ etc., and open charm  $D^{(*)}\Dbar{}^{(*)}$ configurations, respectively.
They are two independent bases although 
they are not orthogonal to each other from the quark model point of view.  
In fact, they are related by rearrangement factors.
The color rearrangement factor is 1/3.
In the following we demonstrate the rearrangement for the spin and isospin parts 
and omit the color factor for simplicity.

The spin rearrangement factor of the $q\qbar c\cbar$ $S$-wave system 
can be obtained as
\begin{eqnarray}
|(s_{12}s_{34})SM\ket 
&=&
\sum_{s_{14},s_{32}}
(-1)^{-s_{12}-s_{32}} 
\sqrt{(2s_{12}+1)(2s_{34}+1)(2s_{14}+1)(2s_{32}+1)}
\nonumber\\
&&\times 
\left\{
\begin{array}{ccc}
{1\over 2}&{1\over 2}&s_{12}\\
{1\over 2}&{1\over 2}&s_{34}\\
s_{32}&s_{14}&S\\
\end{array}
\right\}
|(s_{32}s_{14})SM\ket,
\end{eqnarray}%
where $s_{ij}=s_i+s_j$ with the spin of $i$th quark $s_i$, 
the total spin $S$ corresponds to $S=\sum_i s_i=s_{12}+s_{34}=s_{14}+s_{32}$, and
the array with the braces is the 9-$J$ symbol.
The factor $(-1)^{-s_{12}-s_{32}}$ appears
because we define the mesons by the $q\qbar$ (not $\qbar q$) states.
The numerical factors are listed in \tref{tbl:rearrangement} together with
the two meson states for the isospin 0 systems.
The table shows, for example, that the rearrangement of $\eta\eta_c$ ($J^{PC}=0^{++}$) 
consists of $[D\Dbar]_+$ and  the total spin-0 $[D^*\Dbar{}^*]_{0,+}$ states in the spin space as
\begin{equation}
\eta\eta_c(0^{++}) = {1\over 2}[D\Dbar]_+ -{\sqrt{3}\over 2} [D^*\Dbar{}^*]_{0,+}.
\end{equation}
One can see in \tref{tbl:rearrangement} that the above relation between $J^{PC}$ and the phase
in \eref{eq:JPCDD}--\eref{eq:JPC-DstarDstar-}
appears for each $C$-parity.

\begin{table}[tbp]
\caption{Rearrangement of the $q\qbar$-$c\cbar$-type mesons and $D\Dbar $ mesons.
The definition of $[D{}^{(*)} \Dbar{}^{(*)}]$ are shown in 
\eref{eq:JPC-DD-},
\eref{eq:JPC-DDstar-} and
\eref{eq:JPC-DstarDstar-}.
$[D\Dbar{}^*]_\pm $ means $[D\Dbar{}^*]_+$ for $J^{PC}=1^{+-}$,
and $[D\Dbar{}^*]_-$ for $1^{++}$.
The quark configuration of the $\eta$ meson is assumed to be $(u\ubar+d\dbar)/\sqrt{2}$.}
\begin{center}
\renewcommand\arraystretch{1.2}
\begin{tabular}{clcccccccccc}\hline
$J^{PC}$ &             
& $[D\Dbar]_+ $ 
& $[D^*\Dbar{}^* ]_{S=0,+}$ 
& $[D\Dbar{}^*]_\pm $ 
& $[D^*\Dbar]_\pm $ 
& $[D^*\Dbar{}^*]_{S=1,-}$ 
& $[D^*\Dbar{}^*]_{S=2,+}$ 
\\ \hline
$0^{++}$ & $\eta\eta_c$  & $      {   1\over  2}$&$-{\sqrt{  3}\over 2}$&$                    $&$                    $&$                    $&\\
$0^{++}$ & $\omega\Jpsi$ & $-{\sqrt{  3}\over 2}$&$     -{   1\over  2}$&$                    $&$                    $&$                    $&\\
$1^{+-}$ & $\eta\Jpsi$   & $                    $&$                    $&$      {   1\over  2}$&$      {   1\over  2}$&$-\sqrt{   1\over  2}$&\\
$1^{+-}$ & $\omega\eta_c$& $                    $&$                    $&$      {   1\over  2}$&$      {   1\over  2}$&$ \sqrt{   1\over  2}$&\\
$1^{++}$ & $\omega\Jpsi$ & $                    $&$                    $&$-\sqrt{   1\over  2}$&$ \sqrt{   1\over  2}$&$                   0$&\\
$2^{++}$ & $\omega\Jpsi$ & $                    $&$                    $&$                    $&$                    $&$                    $&1\\ \hline
\end{tabular}
\end{center}
\label{tbl:rearrangement}
\end{table}%

Up to now, the $D$ meson corresponds to $c\qbar$ and the $\Dbar$ meson to $q\cbar$.
The neutral $D\Dbar$ states consist of the isospin 0 and 1 states,  
$((c\ubar u\cbar)+(c\dbar d\cbar))/\sqrt{2}$ and 
$((c\ubar u\cbar)-(c\dbar d\cbar))/\sqrt{2}$, respectively.
Using relation \eref{eq:PQqbar}, we replace the $c\qbar$-$q\cbar$ expression,
$[D^{(*)}\Dbar{}^{(*)}]_\pm$, by
\begin{eqnarray}
-\sqrt{   1\over  2}\Big([D^{(*)+}D{}^{(*)-}]_\pm - [D^{(*)0}\Dbar{}^{(*)0}]_\pm\Big) ~~~(I=0)\\
+\sqrt{   1\over  2}\Big([D^{(*)+}D{}^{(*)-}]_\pm + [D^{(*)0}\Dbar{}^{(*)0}]_\pm\Big) ~~~(I=1).
\end{eqnarray}
In the next section we discuss the $X$(3872), whose $J^{PC}=1^{++}$. 
The rearrangement becomes
\begin{eqnarray}
\omega\Jpsi(1^{++}) &=& -\sqrt{   1\over  2}[D\Dbar{}^*]_- +\sqrt{   1\over  2}[D^*\Dbar]_-
,
\end{eqnarray}
which, if the state has isospin 0, becomes
\begin{eqnarray}
{1\over 2}
\Big(([D^{+}D{}^{*-}]_- - [D^{0}\Dbar{}^{*0}]_-) -([D^{*+}D{}^{-}]_- - [D^{*0}\Dbar{}^{0}]_- )\Big).
\end{eqnarray}
We will denote above state 
simply by $(D\bar D{}^*-D^*{\Dbar})/\sqrt{2}$, or just by $D\bar D{}^*$ 
in the isospin basis if there is no room for confusion. 
Or, in the following section on $X(3872)$, when we write $D^{+}D{}^{*-}$ and $D^{0}\Dbar{}^{*0}$ in the particle basis, they mean
\begin{eqnarray}
{1\over \sqrt{2}}([D^{+}D{}^{*-}]_- - [D^{*+}D{}^{-}]_-) ~~{\rm and}~~
{1\over \sqrt{2}}([D^{0}\Dbar{}^{*0}]_- - [D^{*0}\Dbar{}^{0}]_-),
\end{eqnarray}
respectively.
So, by this notation, the isospin eigenstates of $I(J^{PC})=0(1^{++})$ are
\begin{eqnarray}
\sqrt{   1\over  2}\Big(D^{+}D{}^{*-} - D^{0}\Dbar{}^{*0}\Big) 
\end{eqnarray}
as usual.

In much of the literature on hadron models, such as \cite{Swanson:2003tb,Thomas:2008ja},
the charge conjugate of $D^*$ is defined as ${\bar D}{}^*$, not $-{\bar D}{}^*$.
This can be realized when the $\Dbar{}^{(*)}$ meson 
is taken to be ${\bar c} q$,  not $q\bar c$.
In such a hadron model, 
the $C$-parity $\pm 1$ eigenstates are given by $(D{\bar D}{}^*\pm D^*{\bar D})/\sqrt{2}$, respectively.
This difference in the phase is due only to the definition.
An extra factor appears when the observables are calculated,
which compensates for the above difference.


\section{$X$(3872)}
 \label{sec:X3872}

\subsection{The observed features of $X$(3872)}

 \label{sec:X3872_review}

The $X$(3872) state (also $\chi_{c1}(3872)$) was first observed in 2003 by Belle in the weak decay of the $B$ meson,
$B^{\pm} \to J/\psi \, \pi^+ \, \pi^- \, K^{\pm}$ \cite{Choi:2003ue} and was confirmed by
CDF~\cite{Acosta:2003zx}, D0~\cite{Abazov:2004kp}, $BABAR$~\cite{Aubert:2004ns}, LHCb~\cite{Aaij:2011sn}, CMS~\cite{Chatrchyan:2013cld}
ATLAS~\cite{Aaboud:2016vzw} and BESIII~\cite{Ablikim:2013dyn} collaborations. 
Since then a considerable amount of $X$(3872)-related data has been accumulated. 
We summarize major experimental data samples in Table \ref{table:X3872exp}.

The observed mass of $X$(3872) extracted from the $B \to J/\psi \, \pi^+ \pi^- K$ mode
in the recent measurement is $3871.85 \pm 0.27 \pm 0.19$ MeV \cite{Choi:2011fc}. 
Those from the $p \bar p$ and the $p p$ collisions in the final state $J/\psi \pi^+ \pi^- +$ anything 
are $3871.61 \pm 0.16 \pm 0.19$ MeV \cite{Aaltonen:2009vj} and 
$3871.95 \pm 0.48 \pm 0.12$ MeV \cite{Aaij:2011sn}, respectively.
The average mass given by the particle data group in 2018 \cite{Tanabashi:2018oca} 
in the $J/\psi X$ modes is 
$3871.69 \pm 0.17$ MeV, which is $0.01 \pm 0.19$ MeV above the $D^0 \dsz$ threshold 3871.68 MeV.

The observed masses of the $X$(3872) in the $B \to \dsz D^0 K$ decay mode are
$3872.9 ^{+0.6} _{-0.4}  {}^{+0.4} _{-0.5}$ MeV by Belle \cite{Adachi:2008sua} and
$3875.1 ^{+0.7} _{-0.5}  \pm 0.5$ MeV by $BABAR$ \cite{Aubert:2007rva}.
The observed mass of the $X$(3872) in the $B \to D^0 \bar D^0 \pi^0 K$ decay mode is
$3875.2 \pm 0.7 ^{+0.9} _{-1. 8}$ MeV  \cite{Gokhroo:2006bt}.
The masses observed in the $B \to \dsz D^0 K$ and $B \to D^0 \bar D^0 \pi^0 K$ decay mode are heavier than those in the $J/\psi X$ modes.

The full width is less than 1.2 MeV \cite{Choi:2011fc}.  The observed full widths of the charmonia in the same energy region
as the $X(3872)$ are $27.2 \pm 1.0$ MeV for $\psi(3770)$, $11.3 ^{+3.2}_{-2.9}$ MeV for $\eta_c(2S)$, $201 ^{+154}_{-67} {}^{+88}_{-82}$ MeV for $\chi_{c0}(3860)$ and
$24 \pm 6$ MeV for $\chi_{c2}(3930)$ \cite{Tanabashi:2018oca}.
Therefore the width of $X(3872)$ is unusually small as compared with the other charmonia, which is 
one of the striking features of the $X(3872)$.

As for the spin-parity quantum numbers of the $X$(3872), the angular distributions and correlations
of the $\pi^+ \pi^- J/\psi$ final state have been studied by CDF \cite{Abulencia:2006ma}.
They concluded that the pion pairs originate from $\rho^0$ mesons and that
the favored quantum numbers of the  $X$(3872) are $J^{PC} = 1^{++}$ and $2^{-+}$.
The radiative decays of $X(3872) \to \gamma J/\psi$ have been observed 
 \cite{Aubert:2006aj,Aubert:2008ae,Bhardwaj:2011dj,Aaij:2014ala}, 
which implies that the  $C$-parity of $X$(3872) is positive.
Finally, LHCb performed an analysis of the angular correlations in $B^+ \to X(3872) K^+$, $X(3872) \to \pi^+ \pi^- J/\psi$, $J/\psi \to \mu^+ \mu^-$ decays 
and confirmed the eigenvalues of 
total spin angular momentum, parity and charge conjugation of the $X(3872)$ state to be $1^{++}$ \cite{Aaij:2013zoa,Aaij:2015eva}.
%
%
  \begin{table}[htbp]
    \caption{\label{table:X3872exp} Experimental status of $X(3872)$. In the fourth and fifth columns, short notations are employed. 
    S($\sigma$) means the significance in unit of $\sigma$ and OQ means Observed quantities.
    M, W, BF and CS mean Mass, Width, Branching Fraction and Cross Section, respectively.}
   \begin{center}
     \begin{tabular}{llllll}
      \hline\hline
      Exp. & Mode & Yield & S($\sigma$) & OQ & Ref. \\ \hline \hline
      Belle & $B^\pm \to X(3872) (\to J/\psi \pi^+ \pi^-) K^\pm$ & $35.7 \pm 6.8$       & 10.3  & M, W &  \cite{Choi:2003ue} \\ 
            & $B \to X(3872) (\to D^{\ast 0} \bar D^0) K$      & $50.1^{+14.8}_{-11.1}$ & 6.4   & M, W, BF & \cite{Adachi:2008sua} \\
            & $B^+ \to X(3872) (\to J/\psi \gamma) K^+$        & $30.0^{+8.2}_{-7.4}$   & 4.9   & BF & \cite{Bhardwaj:2011dj} \\
            & $B^0 \to X(3872) (\to J/\psi \gamma) K^0_S$      & $5.7^{+3.5}_{-2.8}$    & 2.4   & BF & \cite{Bhardwaj:2011dj} \\ 
            & $B^+ \to X(3872) (\to J/\psi \pi^+ \pi^-) K^+$   & $152 \pm 15$           &       & M, W, BF & \cite{Choi:2011fc} \\
            & $B^0 \to X(3872) (\to J/\psi \pi^+ \pi^-) K^0_S$ & $21.0 \pm 5.7 $        & 6.1   & M, W, BF & \cite{Choi:2011fc}\\
     CDF II & $J/\psi \pi^+ \pi^-$ in $p \bar p$ at $\sqrt{S} = 1.96$ TeV & $730 \pm 90$  & 11.6  & M, W & \cite{Acosta:2003zx} \\
     CDF    & $J/\psi \pi^+ \pi^-$ in $p \bar p$ at $\sqrt{S} = 1.96$ TeV &            &        & $J^{PC}$ & \cite{Abulencia:2006ma} \\
     D0     & $J/\psi \pi^+ \pi^-$ in $p \bar p$ at $\sqrt{S} = 1.96$ TeV & $522 \pm 100$  & 5.2  & M & \cite{Abazov:2004kp} \\
     \babar\  & $B^- \to X(3872) (\to J/\psi \pi^+ \pi^-) K^-$ & $25.4 \pm 8.7$         &        &  M, BF & \cite{Aubert:2004ns} \\
              & $B^+ \to X(3872) (\to J/\psi \gamma) K^+$      & $19.2 \pm 5.7$         & 3.4    &  BF & \cite{Aubert:2006aj} \\
              & $B^+ \to X(3872) (\to D^{\ast 0} \bar D^0) K^+$ & $27.4 \pm 5.9$        & 4.6    &  BF & \cite{Aubert:2007rva} \\
              & $B^0 \to X(3872) (\to D^{\ast 0} \bar D^0) K^0_S$ & $5.8 \pm 2.7$       & 1.3    &  BF & \cite{Aubert:2007rva} \\
              & $B^+ \to X(3872) (\to J/\psi \pi^+ \pi^-) K^+$   & $93.4 \pm 17.2$      & 8.6    & M, W, BF & \cite{Aubert:2008gu} \\
              & $B^0 \to X(3872) (\to J/\psi \pi^+ \pi^-) K^0_S$   & $9.4 \pm 5.2$      & 2.3    & M, W, BF & \cite{Aubert:2008gu} \\
              & $B^+ \to X(3872) (\to J/\psi \gamma) K^+$        & $23.0\pm 6.4 \pm 0.6 $   & 3.6   & BF & \cite{Aubert:2008ae} \\
              & $B^+ \to X(3872) (\to \psi(2S) \gamma) K^+$      & $25.4\pm 7.3 \pm 0.7 $   & 3.5   & BF & \cite{Aubert:2008ae} \\
     LHCb     & $J/\psi \pi^+ \pi^-$ in $pp$ at $\sqrt{S} = 7$ TeV & $565 \pm 62$  &  & M, W, CS & \cite{Aaij:2011sn}\\
              & $B^+ \to X(3872) (\to J/\psi \gamma) K^+$        & $591 \pm 48$   &   & BF & \cite{Aaij:2014ala} \\
              & $B^+ \to X(3872) (\to \psi(2S) \gamma) K^+$      & $36.4 \pm 9.0$   & 4.4  & BF & \cite{Aaij:2014ala} \\
              & $B^+ \to X(3872) (\to J/\psi \rho^0) K^+$        & $1011 \pm 38$   &  16 & $J^{PC}$ & \cite{Aaij:2013zoa,Aaij:2015eva} \\
     CMS    & $J/\psi \pi^+ \pi^-$ in $pp$ at $\sqrt{S} = 7$ TeV &  $11910 \pm 490$ &  & CS & \cite{Chatrchyan:2013cld}\\         
     ATLAS    & $J/\psi \pi^+ \pi^-$ in $pp$ at $\sqrt{S} = 8$ TeV &   &  & CS & \cite{Aaboud:2016vzw}\\
     BESIII   & $e^+ e^- \to \gamma X(3872) (\to J/\psi \pi^+ \pi^-)$ & $20.0 \pm 4.6 $  & 6.3 & M, CS, BF & \cite{Ablikim:2013dyn}\\
          	  & $e^+ e^- \to \gamma X(3872) (\to J/\psi \pi^+ \pi^- \pi^0)$ & $45 \pm 9 \pm 3 $  & 5.7 & M, W, BF & \cite{Ablikim:2019zio}\\
              \hline\hline
     \end{tabular}
   \end{center}
  \end{table}
%
%
\par
Since the first observation of the $X$(3872), 
it has received much attention because its features are difficult to explain
if a simple $c \bar c$ bound state of the quark potential model is assumed \cite{Barnes:2003vb}.
The interaction between heavy quark and heavy antiquark is well understood and known that it can be approximately expressed by a Coulomb-plus-linear potential,
a feature that is confirmed  by lattice QCD studies~\cite{Donoghue:1992dd,Bali:2000gf}. 
The charmonium states with $J^{PC} = 1^{++}$ are the $\chi_{c1}$ states. 
The observed mass of the ground state of the $\chi_{c1}(1P)$ is ($3510.67 \pm  0.05$) MeV and the quark potential model gives similar mass~\cite{Barnes:2005pb}.
The first excited state of the $\chi_{c1}$ is the $\chi_{c1}(2P)$ and that state has, so far, not been observed. The predicted mass of the $\chi_{c1}(2P)$ in quark potential models is in
between $3925$ MeV and $3953$ MeV~\cite{Barnes:2005pb}.  The observed mass of 
$X(3872)$ is 53-81 MeV smaller than these predictions.  This is one of the strong grounds for the identification of the $X$(3872) as a non-ccbar structure.
A variety of structures have been suggested for the $X$(3872) from the theoretical side, 
such as a tetraquark structure 
\cite{Maiani:2004vq,Ebert:2005nc,Terasaki:2007uv,Dubnicka:2010kz,Kim:2016tys}, 
 $D^0 \dsz$ molecule \cite{Close:2003sg,Voloshin:2003nt,Swanson:2003tb,Tornqvist:2004qy,AlFiky:2005jd,Ding:2009vj,Lee:2009hy,Gamermann:2009uq,Li:2012cs,Guo:2013sya,Wang:2013kva,Baru:2013rta,Garzon:2013uwa,Jansen:2013cba,Wang:2013daa,Tomaradze:2015cza,Baru:2015nea,Braaten:2015tga,Wang:2017dcq,Schmidt:2018vvl} 
and a charmonium-molecule hybrid \cite{Matheus:2009vq,Danilkin:2010cc,Coito:2010if,Coito:2012vf,Takizawa:2012hy,Ferretti:2013faa,Chen:2013pya,Takeuchi:2014rsa,Ferretti:2014xqa,Zhou:2017dwj}.
The structure of the $X(3872)$ has also been studied with the lattice QCD approach \cite{Yang:2012mya,Prelovsek:2013cra,Lee:2014bea,Padmanath:2015era}.

One of the important properties of the $X$(3872) is its isospin structure.
The branching fractions measured by Belle \cite{Abe:2005ix} is
\begin{equation}
\label{TT_eq:1_1}
  \frac{Br(X \rightarrow \pi^+ \pi^- \pi^0 J/\psi)}{Br(X \rightarrow \pi^+ \pi^- J/\psi)} 
  = 1.0 \pm 0.4 \pm 0.3 \, ,
\end{equation}
and $0.8 \pm 0.3$ by $BABAR$ \cite{delAmoSanchez:2010jr}. 
Here the two-pion mode originates from the isovector $\rho$ meson 
while the three-pion mode comes from the isoscalar $\omega$ meson. 
So, the equation (\ref{TT_eq:1_1}) indicates strong isospin violation. 
Recently, BESIII observed $X(3872) \to \omega J/\psi$ decay with a significance of more than 5$\sigma$ and 
the relative decay ratio of $X(3872) \to \omega J/\psi$ and $\pi^+ \pi^- J/\psi$ was measured to be ${\cal R} = 1.6^{+0.4}_{-0.3} \pm 0.2$ \cite{Ablikim:2019zio}.
The kinematical suppression factor 
including the difference of the vector meson decay width 
were studied \cite{Braaten:2005ai,Suzuki:2005ha}, and 
the production amplitude ratio \cite{Suzuki:2005ha} 
was obtained by using Belle's value \cite{Abe:2005ix}
\begin{equation}
\label{TT_eq:1_2}
  \left| \frac{A(\rho J/\psi)}{A(\omega J/\psi)} \right|
  = 0.27 \pm 0.02 \, .
\end{equation}
Typical size of the isospin symmetry breaking ratios is at most a few \%.
It is interesting to know what the origin of this strong isospin symmetry breaking is.
In \cite{Gamermann:2009uq}, this problem was studied using the chiral unitary model, and
the effect of the $\rho$-$\omega$ mixing has been discussed in \cite{Terasaki:2009in}.
It was reported that both of these approaches can explain the observed ratio given in \eref{TT_eq:1_1}.
In the charmonium-molecule hybrid approach, the difference of the $D^0\dsz$ and the $D^+D^{*-}$ 
thresholds produces sufficient isospin violation to naturally explain the experimental results \cite{Takizawa:2012hy,Takeuchi:2014rsa}. 
The Friedrichs-model-like scheme can also explain the isospin symmetry breaking \cite{Zhou:2017txt}. 
Recently, a new Isospin=1 decay channel, $X(3872) \to \pi^0 \chi_{c1}$ has been observed \cite{Ablikim:2019soz}.

$X$(3872) production at high energy hadron colliders has been
studied in \cite{Braaten:2004fk,Braaten:2004ai,Bignamini:2009sk,Zanetti:2011ju,Chatrchyan:2013cld,Cho:2013rpa,Denig:2014fha,Torres:2014fxa,Larionov:2015nea,Wang:2015rcz,Abreu:2016dfe,Meng:2013gga,Albaladejo:2017blx,Esposito:2017qef},
where unexpectedly large production rates have been observed at large transverse momentum transfers 
$p_{\perp} > 10$ GeV~\cite{Esposito:2016noz}.  
These rates are much larger than those for production of light nuclei such as the deuteron and $^3$He, and are about 5 \% of that for the $\psi(2S)$.  
This property is naively explained if $X(3872)$ has a small ``core'' component that is a compact structure such as the $\chi_{c1}(2P)$.  In a later subsection we will see that this will be realized in a model of $D \bar D^*$ molecular coupled with a $c \bar c$ core.  

The hadronic decays of the $X$(3872) are investigated in \cite{Swanson:2004pp,Braaten:2005ai,Navarra:2006nd,Braaten:2007dw,Dubynskiy:2007tj,Braaten:2007ft,Fleming:2008yn,Artoisenet:2010va,Fleming:2011xa,Braaten:2013poa,Cho:2013rpa,Guo:2014hqa,Meng:2014ota,Mehen:2015efa,Kang:2016jxw,Ferretti:2018tco}.
As for radiative decays,
as seen in \cite{Colangelo:2007ph,Dong:2008gb,Nielsen:2010ij,Wang:2010ej,Harada:2010bs,Mehen:2011ds,Dubnicka:2011mm,Ke:2011jf,Badalian:2012jz,Margaryan:2013tta,Guo:2013nza,Guo:2014taa,Cardoso:2014xda,Badalian:2015dha,Takeuchi:2016hat},
the existence of a core seems to be required, but the results depend on details of the wave function. 

Another important issue is whether the charged partner of 
$X(3872)$ exists 
as a measurable peak or not.
$BABAR$ has searched such a state in the $X(3872) \to \pi^- \pi^0 J/\psi$ 
channel and found no signal \cite{Aubert:2004zr}. Belle has also studied such a state using 
much
accumulated data but found no signal \cite{Choi:2011fc}.
The hybrid picture, where the coupling to the $c\cbar$ core is
essential to bind the neutral $X(3872)$, is consistent with the
absence of a charged $X(3872)$.

Since the $X(3872)$ has many interesting properties, and many studies of it have been done, several review papers have been written from various viewpoints \cite{Swanson:2006st,Godfrey:2008nc,Brambilla:2010cs,Olsen:2014qna,Hosaka:2016pey,Chen:2016qju,Esposito:2016noz,Lebed:2016hpi,Ali:2017jda,Guo:2017jvc}.

 \label{sec:X-intro} 
\subsection{$D^{(\ast)}\bar{D}^{(\ast)}$ molecule with OPEP}
\label{sec:DDbar_molecule_with_OPEP}

 In this subsection, 
 we demonstrate 
 the analysis
 of the $X(3872)$ as a $D^{(\ast)}\bar{D}^{(\ast)}$ molecule with
 $I^G(J^{PC})=0^+(1^{++})$.
 As for the interaction between $D^{(\ast)}$ and $\bar{D}^{(\ast)}$ mesons, 
 we employ only the OPEP in~\eref{eq_VPPtoP*P}-\eref{eq_VP*P*toP*P*}.
 In the $D^{(\ast)}\bar{D}^{(\ast)}$ coupled channel system, 
 the possible $D^{(\ast)}\bar{D}^{(\ast)}$ components with positive charge conjugation 
 are 
 \begin{eqnarray}
  \frac{1}{\sqrt{2}}(D\bar{D}^\ast-D^\ast\bar{D})(^3S_1, {^3D_1}), \quad  D^\ast\bar{D}^\ast(^5D_1),
   \label{eq:X3872-components}
 \end{eqnarray}
 where $(^{2S+1}L_J)$ denotes the total spin $S$, the orbital angular momentum $L$ 
 and the total angular momentum $J$~\cite{Tornqvist:1993ng,Ohkoda:2011vj}.
 We note that the phase convention in~\eref{eq:X3872-components}
 is different from the one in the literature~\cite{Tornqvist:1993ng,Ohkoda:2011vj} as discussed 
 in~\sref{sec:quark-hadronic-models}.
 In this basis, 
 the matrix elements of the OPEP
 are given by~\cite{Tornqvist:1993ng,Ohkoda:2011vj}
 \begin{eqnarray}
  V_\pi(r)&=&\left(\frac{g_A}{2f_\pi}\right)^2\left(
	\begin{array}{ccc}
	 {\cal C}_\pi&-\sqrt{2}{\cal T}_\pi &-\sqrt{6}T_\pi \\
	 -\sqrt{2}{\cal T}_\pi& {\cal C}_\pi+{\cal T}_\pi& -\sqrt{3}T_\pi\\
	 -\sqrt{6}T_\pi& -\sqrt{3}T_\pi& C_\pi-T_\pi\\			 
	\end{array}
   \right),
  \label{eq:OPEP-X3872}  
 \end{eqnarray}
  where ${\cal C}_\pi=C(r;\mu,\Lambda)$,
  ${\cal T}_\pi=T(r;\mu,\Lambda)$,
  $C_\pi=C(r;m_\pi,\Lambda)$, and $T_\pi=T(r;m_\pi,\Lambda)$,
  respectively.
  The functions $C(r;m,\Lambda)$ and $T(r;m,\Lambda)$ are defined
  in~\eref{eq:poteC} and \eref{eq:poteT}.  
  The functions ${\cal C}_\pi$ and ${\cal T}_\pi$ with $\mu$ emerge because the nonzero energy transfer 
  in the $D$-$D^\ast$ transition is taken into account in the potential $V^\pi_{P\bar{P}^\ast{\mathchar`-}P^\ast\bar{P}}(r)$
  as explained in~\sref{sec:OPEP_for_PPbar}.  
  In the charm sector, the mass $\mu$ becomes imaginary,
  i.e. 
  $\mu^2=m^2_\pi-(m_{D^\ast}-m_{D})^2 < 0$.  
  In this subsection, the hadron masses summarized in~\tref{table:Hadronmass} are used, which are 
  the isospin averaged masses.
  Then, $\mu^2=(37.3i)^2$ [MeV$^2$] is obtained, and the $\cal{C}_\pi$ and $\cal{T}_\pi$ 
  become complex
  as seen in \eref{eq_complex_OPEP}.
  The explicit expression of the imaginary central and tensor potentials is given in \cite{Suzuki:2005ha}.
  In this analysis, 
  we consider only the real part of the potential,
  because 
  the imaginary part is small for small $\mu$.

  \begin{table}[htbp]
    \caption{\label{table:Hadronmass} Hadron masses used in the numerical calculation.
    The masses of the pion, pseudoscalar meson $P=D,B$ and vector $P^\ast=D^\ast,B^\ast$ meson 
    are shown in the $ud$, charm and bottom sectors. 
    The pion mass is given as the averaged mass of $\pi^+$, $\pi^0$ and $\pi^-$~\cite{Tanabashi:2018oca}.
    The $P^{(\ast)}$ mass is the averaged mass of $P^{(\ast)0}$ and $P^{(\ast)\pm}$~\cite{Tanabashi:2018oca}.
    The mass difference $\Delta M_{PP^\ast}$ between the masses of $P$ and $P^\ast$ is also shown.
    The values are in units of MeV.}
   \begin{center}
     \begin{tabular}{cccc}
      \hline\hline
      & $\pi$ [MeV]& & \\ \hline
      $ud$&137 & & \\ \hline\hline
      & $P$ [MeV]& $P^\ast$ [MeV]& $\Delta M_{PP^\ast}$ [MeV]\\ \hline
      charm & 1867 &2009 & 145\\   
      bottom &5279 &5325 & 46\\
      \hline\hline
     \end{tabular}
   \end{center}
  \end{table}

  The Hamiltonian of the $D^{(\ast)}\bar{D}^{(\ast)}$ coupled channel
  system is 
  \begin{eqnarray}
   H=K+V_{\pi},
    \label{eq:Ham1++}
  \end{eqnarray}
  where the kinetic term 
  $K$ is given by
  \begin{eqnarray}
   K= {\rm diag}\left(
			-\frac{1}{2\mu_{D\bar{D}^\ast}}\Delta_0,-\frac{1}{2\mu_{D\bar{D}^\ast}}\Delta_2,
			-\frac{1}{2\mu_{D^\ast\bar{D}^\ast}}\Delta_2+\Delta m_{DD^\ast}
		       \right).
  \end{eqnarray}
  Here we define 
  \begin{eqnarray}
   \Delta_{\ell}=\frac{\partial^2}{\partial r^2}+\frac{2}{r}\frac{\partial}{\partial r} + \frac{\ell(\ell+1)}{r^2} ,    
    \label{eq:Delta-ell}
  \end{eqnarray}
  for the state of orbital angular momentum $\ell$,
  the reduced mass 
  \begin{eqnarray}
   \mu_{D^{(\ast)}\bar{D}^{(\ast)}}=\frac{m_{D^{(\ast)}}m_{\bar{D}^{(\ast)}}}{m_{D^{(\ast)}}+m_{\bar{D}^{(\ast)}}},
    \label{eq:mu-P*P*}
  \end{eqnarray}  
  and 
  \begin{eqnarray}
   \Delta m_{DD^\ast}=m_{D^\ast}-m_{D} .
    \label{eq:delta-m}
  \end{eqnarray}

  The $D^{(\ast)}\bar{D}^{(\ast)}$ 
  systems are studied by
  solving the Schr\"odinger equation with the Hamiltonian $H$~\eref{eq:Ham1++}. 
  In the potential 
  $V_\pi$, there are two parameters, 
  the coupling constant $g_A$ and the cutoff parameter $\Lambda$.
  The coupling $g_A$ is determined by the $D^\ast\rightarrow D\pi$ decay
  as shown in~\sref{sec:MesonDecaysI}.  
  The cutoff $\Lambda$ is a free parameter, while
  it can be evaluated by the ratio of the size of hadrons.
  In~\cite{Yasui:2009bz,Yamaguchi:2011xb}, the cutoff $\Lambda$ for the heavy meson
  is determined by the relation $\Lambda/\Lambda_N=r_N/r_D$, with
  the nucleon cutoff $\Lambda_N$, and the sizes of the nucleon and $D$ meson $r_N$, and $r_D$,
  respectively.
  The nucleon cutoff is determined to reproduce the deuteron properties as discussed 
  in~\sref{sec:Deuteron}, 
  and we use $\Lambda_N=837$ MeV.
  The ratio of the hadron sizes $r_N/r_D=1.35$ is obtained by the quark model 
  in~\cite{Yasui:2009bz}.
  Thus, $\Lambda=1.13$ GeV is obtained.

  To start with, the $D^{(\ast)}\bar{D}^{(\ast)}$ system is solved for the standard parameters 
  $(g_A, \Lambda)$ = (0.55, 1.13 GeV).  
  We have found that the OPEP provides an attraction but is not strong enough to generate a bound or resonant state.  
  The resulting scattering length is $a = 0.64$ fm for the $S$-wave $D \bar{D}^\ast$ channel. 
  By changing the parameter set $(g_A, \Lambda)$ 
  by a small amount of value 
  toward more attraction, a bound state is accommodated.  

\begin{figure}[t]  
   \begin{center}
    \includegraphics[width=11cm,clip]{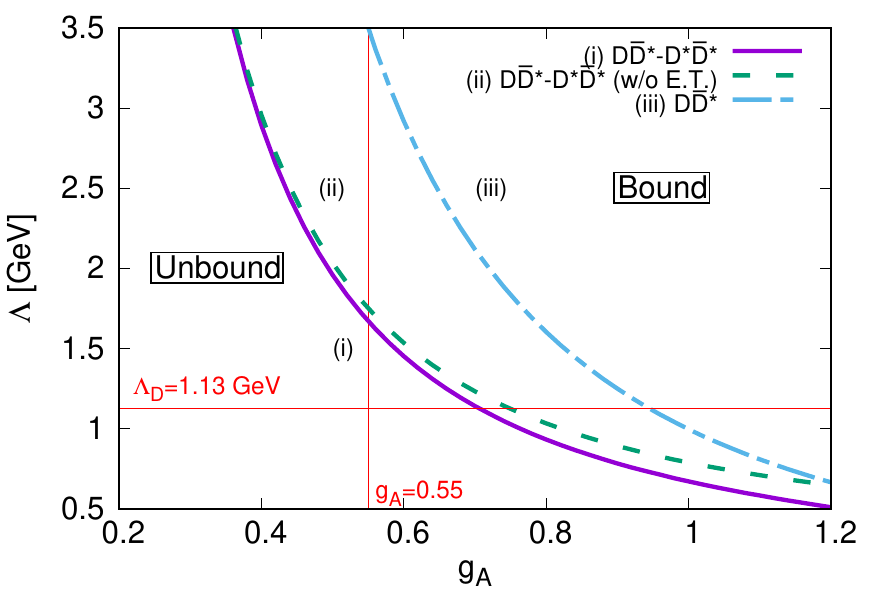}    
   \end{center}
   \caption{\label{fig:X3872-ulim-1}
   Boundary lines separating the regions where a bound state exists or not.
   The solid (i) and long-dashed (ii) lines are the results 
   with and without the energy transfer, respectively.
   The dot-dashed line (iii) is the result without the $D^{(\ast)}\bar{D}^{(\ast)}$ channel
   (see the text for details).
   The right sides of 
   these lines (i), (ii) and (iii) are the region where the system is bound 
   in the cases of (i), (ii) and (iii), respectively, 
   while the left side is the region where the the system is unbound.
   The vertical and horizontal solid lines shows the values at $g_A=0.55$  and
   at $\Lambda=\Lambda_D=1.13$ GeV, respectively.   
   }
  \end{figure}

  To see better the properties of the interaction, we show parameter regions on the $(g_A, \Lambda)$ plane 
  which allow bound states or not.  
  In \fref{fig:X3872-ulim-1}, boundaries of the two regions are plotted for three cases  
  depending on how the system is solved; 
  (i) the full calculations with all coupled-channels of $D^{(\ast)} \bar{D}^{(\ast)}$ states included 
  and with energy transfer properly taken into account
  in the potential~\eref{eq:OPEP-X3872}, 
  (ii) calculations in the full coupled channels but with the energy transfer ignored (static approximation), 
  and (iii) calculations with a truncated coupled channels removing the $D^\ast \bar{D}^\ast$ states. 
  Those lines indicate the correlation between $g_A$ and $\Lambda$. 
  If the coupling $g_A$ is small, the cutoff $\Lambda$ 
  should be large to produce the bound state, and vice versa.

  The lines for (i) and (ii) are similar, 
  which is a consequence of the fact that the energy transfer is not very important here.  
  Nevertheless, the dashed line (ii) is slightly on the right side (or upper side) of the solid line (i). 
  When $g_A$ = 0.55, $\Lambda  = 1.6$ GeV on the line (i), while 
  $\Lambda  = 1.7$ GeV on the line (ii). 
  Hence, introducing the energy transfer produces more attraction due to smaller effective mass 
  or equivalently to longer force range. 
  Even for $\mu=0$, the result is almost the same as that in the case (i). 

\begin{figure}[htbp]
      \begin{center}
       \begin{tabular}{cc}
	 Central&Tensor \\
	\includegraphics[width=7cm,clip]{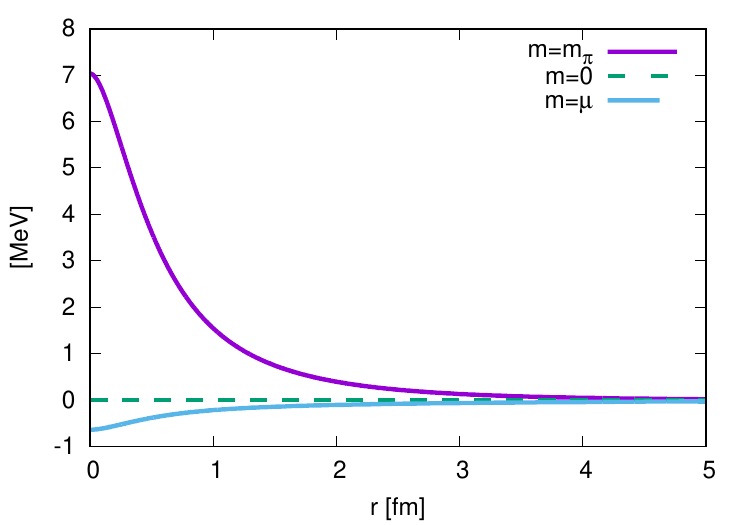}& 
	    \includegraphics[width=7cm,clip]{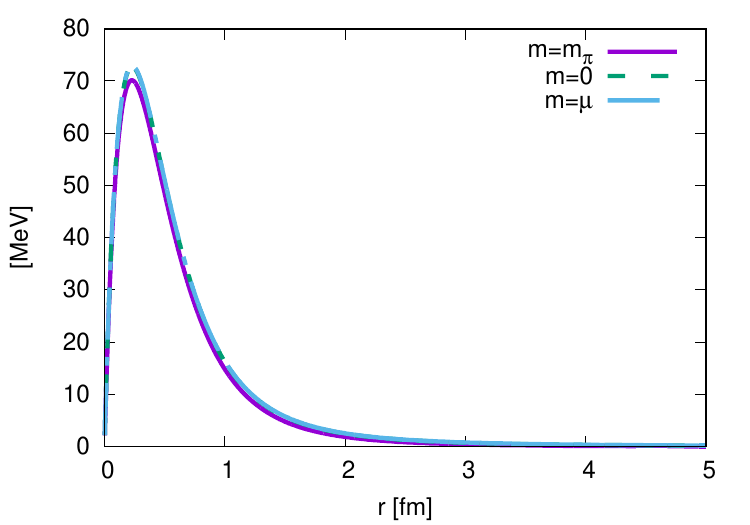}\\
       \end{tabular}
       \caption{\label{fig:potential_variousmu}
       The central and tensor
       components of the OPEP
       for the $X(3872)$ \eref{eq:OPEP-X3872}, $\left(\frac{g_A}{2f_\pi}\right)^2C(r;m,\Lambda)$ 
       and $\left(\frac{g_A}{2f_\pi}\right)^2T(r;m,\Lambda)$ respectively,        
       with various 
       effective pion masses and $(g_A,\Lambda)=(0.55,1.13\,{\rm GeV})$.
       The solid, dashed and dashed-dot lines correspond to the potentials with 
       $m=m_\pi(=137), 0, \mu(=37.3i)$ MeV, 
       respectively.
       }
      \end{center}
     \end{figure}

  The central and tensor potentials $C(r;m,\Lambda)$ and 
  $T(r;m,\Lambda)$ 
  for the $X(3872)$ in \eref{eq:OPEP-X3872} 
  are shown in~\fref{fig:potential_variousmu},
  where the potentials with various effective pion masses 
  are compared, $m=m_\pi, 0, \mu$.
  The potential with $m=m_\pi$ corresponds to
  the $D^\ast\bar{D}^\ast$ potential and
  the $D\bar{D}^\ast$ one in the static approximation,
  where the energy transfer is ignored.
  The potential with $m=\mu$ is the $D\bar{D}^\ast$ potential 
  ${\cal C}_\pi$ taking into account the energy transfer. 
  In~\fref{fig:potential_variousmu}
  we plot only the real part of the potential.
  We also show the potential with $m=0$, which is in the 
  limit of the small mass of the transfer pion.
Since the central potential is proportional to the effective mass of the transfer pion, $m^2$,
  for $m^2=m^2_\pi > 0$ in the static approximation, 
  the overall sign of the potential $C(r;m_\pi,\Lambda)$ is positive, 
  while for $m^2=\mu^2 <0$ with the energy transfer, 
  the sign of the potential $C(r;\mu,\Lambda)$ is negative.
  The central potential vanishes for $m=0$.  
  On the other hand,
  the tensor force does not depend on the effective pion mass strongly as shown in~\fref{fig:potential_variousmu}.

  Naively, one would expect that a longer range potential yields more interaction strength, 
  which we do not see 
  here.  
  One reason is that the central force has the factor $m^2$ 
  as discussed.  
  Another reason is that the tensor force is mostly effective at shorter distances than $1/m$, due to the $S$-$D$ coupling.  
  In momentum space, it is due to the $\bi q^2$ dependence in the numerator
  \eref{eq_OPEP_NN} which 
  increases 
  the tensor force for 
  large $\bi q^2$.

  Turning to \fref{fig:X3872-ulim-1}, the line (iii) 
  shows the result without the $D^{\ast}\bar{D}^\ast$ channel.
  This line is far above 
  the lines (i) and (ii), indicating that the attraction is 
  significantly reduced.
  Since the coupling to $D^{\ast}\bar{D}^\ast$ component with the $D$-wave induces
  the tensor force as shown 
  in~\eref{eq:OPEP-X3872},
  ignoring this component 
  decreases the attraction due to the tensor force significantly.
  Hence, the full-coupled channel analysis of 
  $D\bar{D}^\ast$ and $D^\ast\bar{D}^{\ast}$ 
  is important when the tensor force of the OPEP is considered.  
  
  \begin{figure}[t]  
   \begin{center}
    \includegraphics[width=11cm,clip]{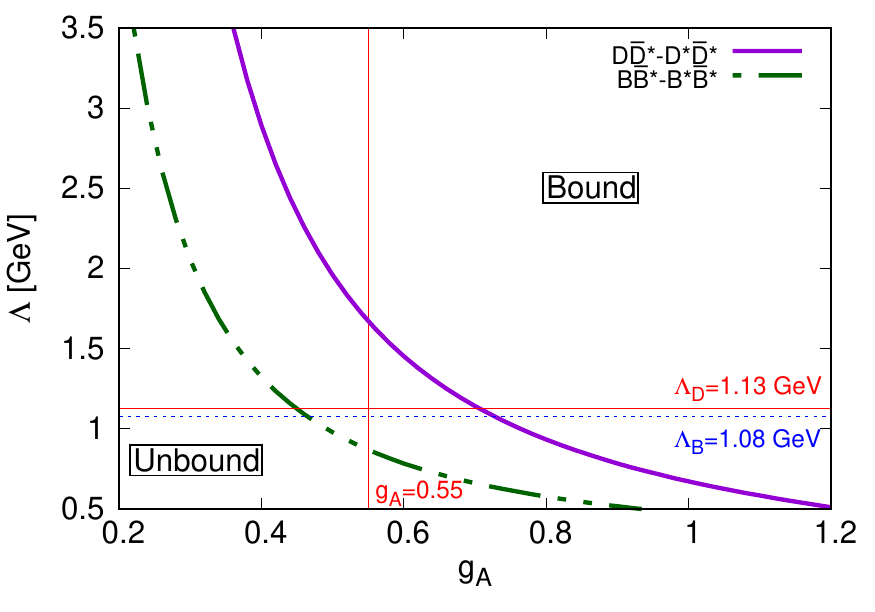}
   \end{center}
   \caption{\label{X3872_ulim_7-3}
   The boundaries of the 
   $D^{(\ast)}\bar{D}^{(\ast)}$ (solid line)
   and
   $B^{(\ast)}\bar{B}^{(\ast)}$ (dashed double-dotted line)
   bound states in the $(g_A,\Lambda)$ plane.
   The boundary of the $D^{(\ast)}\bar{D}^{(\ast)}$ state is the same as the boundary (i) in~\fref{fig:X3872-ulim-1}.
   The vertical solid line shows the value of 
   $g_A=0.55$, and
   the horizontal solid and dashed lines show the values of $\Lambda=\Lambda_D=1.13$ GeV
   and $\Lambda=\Lambda_B=1.08$ GeV, respectively.   
   }
  \end{figure} 

Finally, 
the $B^{(\ast)}\bar{B}^{(\ast)}$ bound state in the bottom sector is studied.
We employ the same potential as used in the $D^{(\ast)}\bar{D}^{(\ast)}$
system \eref{eq:OPEP-X3872} 
because the potential is given as the leading term of the $1/M_{P}$ expansion and thus 
the potential form is heavy flavor independent in the heavy quark limit.
The cutoff $\Lambda_B$ for the $B$ meson is also evaluated by the hadron size in the similar way to the cutoff $\Lambda_D$. 
In~\cite{Yasui:2009bz}, 
the ratio of the hadron size is obtained by $r_N/r_B=1.29$, and thus
the cutoff $\Lambda_B$ is obtained by $\Lambda_B=1.29\Lambda_N=1.08$ GeV.
This value can be the reference point here, while we also vary the cutoff to see the cutoff dependence.
The use of different $\Lambda$ for charm and bottom sectors 
is to take partly into account $1/$(heavy quark mass) corrections 
due to kinematics, because in the quark model meson size is a function of the reduced mass.

 In~\fref{X3872_ulim_7-3}, 
 the boundary line of the $B^{(\ast)}\bar{B}^{(\ast)}$ bound state is shown, 
 where it is compared with the boundary of the $D^{(\ast)}\bar{D}^{(\ast)}$ bound state, which
 is the same as shown in \fref{fig:X3872-ulim-1} (i).
 The bound region 
 for the $B^{(\ast)} \bar{B}^{(\ast)}$ system 
is larger than 
 that of the $D^{(\ast)}\bar{D}^{(\ast)}$. 
In the bottom sector, the kinetic term is suppressed by the large $B^{(\ast)}$ meson mass, about 5 GeV,
while the $D^{(\ast)}$ meson mass is about 2 GeV.
In addition, the small mass difference between $B$ and $B^\ast$, about 46 MeV,
magnifies the mixing rate of the $S$-$D$ coupled channel due to the tensor force, yielding more attraction.  
For the parameters 
$(g_A,\Lambda)=(0.55,1.08\,{\rm GeV})$, 
the bound state is found in the bottom sector,
where the binding energy is 6.3 MeV. 

Because of the attraction in the bottom sector, 
the bottom counter part 
of the $X(3872)$ is also expected to be 
formed as the $B^{(\ast)}\bar{B}^{(\ast)}$ bound state.
Verification in experiments is needed.

 \label{sec:X-OPEP} 
\subsection{Admixture of the $c\bar c$ core and the $D \bar D^*$ molecule}
\def\DDbarc{{D^{+}D^{*-}}}
\def\DDbarz{{D^0 \Dbar{}^{*0}}}

As discussed in the previous section,
the OPEP tensor term induces the $D^{(*)}\bar D{}^{(*)}$ $S$-$D$-wave channel mixing,
which gives an attraction to the $X$(3872) system.
This attraction is sizable, but seems not large enough to produce a bound state.
Another origin of the attraction is discussed in \cite{Takizawa:2012hy}, 
where $X$(3872) is assumed to be a shallow bound state of the 
coupled channels of $c\cbar$, $\DDbarz$ and the $\DDbarc$.
The coupling occurs between the bare $c\bar c$ pole 
and the isospin-0 $S$-wave $D\Dbar{}^*$ continuum.
A nearby $c\cbar(1^{++})$ state is $\chi_{c1}(2P)$, which has not been observed experimentally 
but was predicted by the quark model \cite{Godfrey:1985xj}.
The predicted mass of $\chi_{c1}(2P)$ is by about 80 MeV above the $D\Dbar{}^*$ threshold energy according to the quark model.
So, the coupling to the $c\cbar$ state pushes the low energy $D\Dbar{}^*$ continuum states downward,
 toward the threshold.
As a result, the coupling
provides an attraction for the isospin-0 $S$-wave $D\Dbar{}^*$.
This dynamically generates a pole, $X$(3872),
while the $c\bar c$ state gets a broad width, which makes the state difficult to observe.

The $c\cbar$-$D\bar D{}^*$ coupling occurs in the short range 
where the light quark pair in the $D \Dbar{}^*$ state can annihilate.
On the other hand, the size of $X$(3872)
is very large as shown later in \tref{tbl:rms}.
The volume of the interaction region is the order of $10^{-3}$ of that of the $X$(3872) \cite{Achasov:2015oia}.
Since most of the $X$(3872) wave function stays spatially outsize of the interaction region,
one may wonder whether such a short range coupling can be responsible to make the $X$(3872).
Actually, any potential of a finite range can make a bound state with an appropriate strength.
Suppose we employ a three-dimensional square-well potential of the range $a$ and the strength $V_0$.
Then one bound state at the threshold appears when 
$V_0\sim \pi^2/(8 \mu  a^2 )$. Since the reduced mass of the $D\bar D^*$ system is about 1 GeV,
the required strength $V_0$ to make a bound state is 50--200 MeV for $a\sim$ 0.5--1 fm.
This size of the strength is reasonable when considering that the typical mass difference of the 
hadrons, such as $D^*$-$D$ (140 MeV) or as $J/\psi$-$\eta_c$ (113 MeV), is the order of 100 MeV.
So, 
in this section,
we study $X$(3872) in a coupled channel model of 
$c\cbar$, $D^{(*)0} \Dbar{}^{(*)0}$ and $D^{(*)+} D^{(*)-}$.

To start with, we investigate a simple model of such, 
a model of coupled channels of $D^{} \bar D^{*}$ and $c\bar c$ 
where the interaction of $D \bar D^{*}$  takes place only 
through their coupling to $c\bar c$ channel.  
We call this $c \bar c$ model, where, 
in the absence of  OPEP, only the $S$-waves are relevant 
for $D \bar D^{*}$ channels.  
It is reported that 
by assuming a coupling between $c\bar c$ and $\DDbarz$ and $\DDbarc$, 
a shallow bound state appears below the $\DDbarz$ threshold;
but there is no peak structure found at the $\DDbarc$ threshold.
The  coupling structure is assumed as
\begin{equation}
\bra \DDbarc ({\bi q})|U|c\bar c\ket
=
 -\bra \DDbarz ({\bi q})|U|c\bar c\ket 
= {g_{c\bar c}\over \sqrt{\Lambda_q}}{ \Lambda_q^2\over q^2+ \Lambda_q^2}.
\label{eq:X-cc-DDbar-pot}
\end{equation}%
The coupling strength $g_{c\bar c}$ is taken so as to produce the observed mass of $X$(3872). 
The cutoff $\Lambda_q$ is roughly corresponds to the
inverse of the size of the region where the $q\bar q$ annihilation occurs, 
being $U\propto e^{-\Lambda_q r}/r$.
Here we show the results with $\Lambda_q$ = 0.5 GeV $\sim$ (0.4 fm)$^{-1}$ \cite{Takizawa:2012hy}.
When one uses smaller value for $\Lambda_q$, {\it e.g.} 0.3 GeV,
the model gives a sizable enhancement around the mass of $\chi_{c1}(2P)$, 3950 MeV, 
in the final $D\Dbar^*$ mass spectrum of the $B\to D\Dbar^*K$ decay.
Since such a structure is not observed, 
it 
can be a constraint to the interaction region from the $B$-decay experiments 
that $\Lambda_q$ is more than about 0.5 GeV \cite{Takizawa:2012hy}.

\begin{figure}[htbp]
   \begin{center}
    \includegraphics[width=9cm,clip]{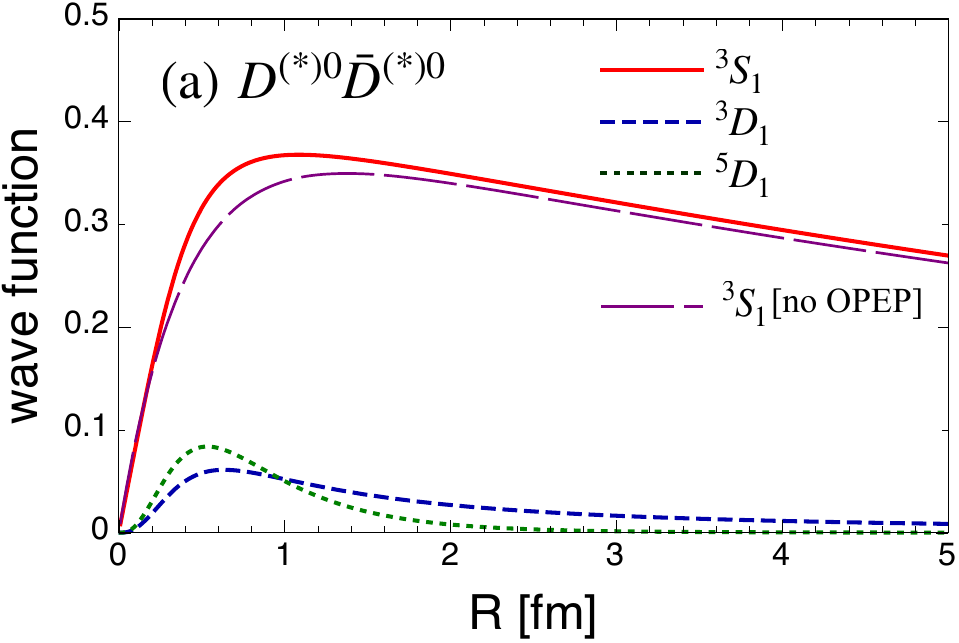}
    \includegraphics[width=9cm,clip]{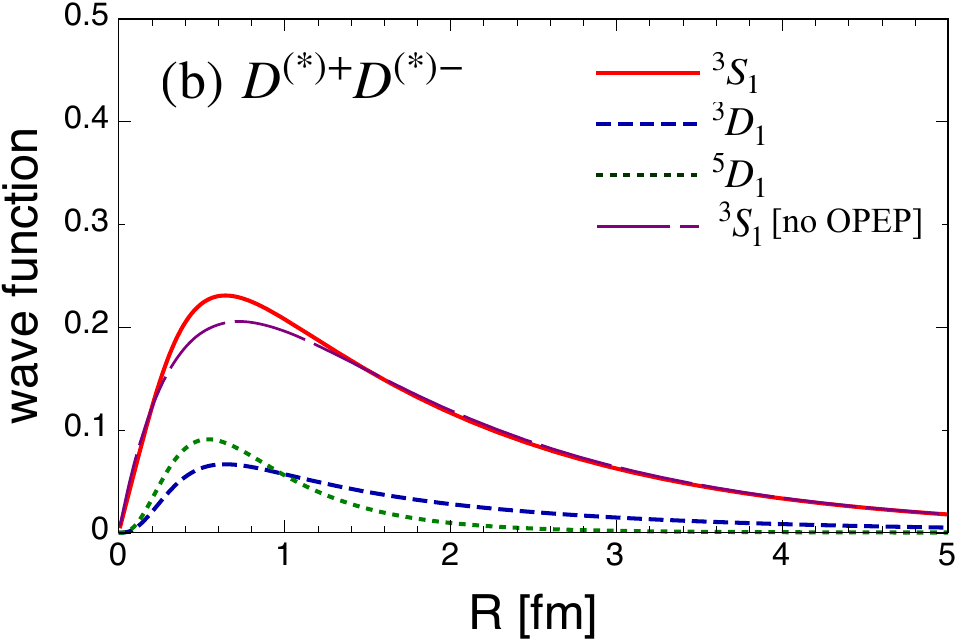}
   \end{center}
   \caption{
(a) The $\DDbarz$ and (b) the $\DDbarc$ wave function of $X$(3872) in the OPEP model; 
solid lines are for $D\Dbar{}^*({}^3S_1)$, dashed lines for $({}^3D_1)$ 
and dotted lines for $D^*\Dbar{}^*({}^5D_1)$. 
The long dashed lines are for those of \cite{Takizawa:2012hy}, 
which corresponds the $c\bar c$-model in tables \ref{tbl:rms} and \ref{tbl:Xcompo},
where the $X$(3872) consists of 
the $c\cbar$ and the $D\Dbar{}^*({}^3S_1)$ components, without OPEP. 
The sign of the wave functions is taken to be positive at small $r$.
 }
   \label{fig:X3872-DDwf1}
  \end{figure}

\begin{figure}[htbp]
   \begin{center}
    \includegraphics[width=9cm,clip]{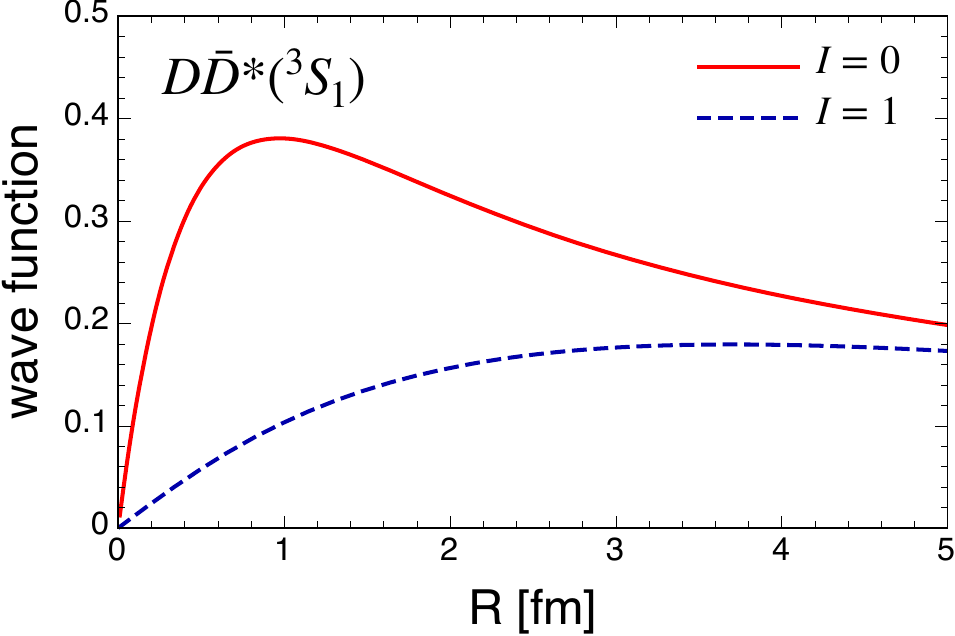}
   \end{center}
   \caption{The $D\Dbar{}^*$ wave function of $X$(3872) in the isospin basis.
   The solid (dashed) line is for the isospin 0 (1) wave function of the $c\bar c$ model.
The sign of the wave functions is taken to be positive at small $r$. }
   \label{fig:X3872-DDwf-isospin}
  \end{figure}

The mass of the charged $D$ meson is heavier than the neutral one by 
4.822$\pm$0.015 MeV and that of $D^*$ by 
3.41$\pm$0.07 MeV \cite{Tanabashi:2018oca}.
Therefore,
the threshold difference between $\DDbarz$ and $\DDbarc$ is about 8.2 MeV.
Since the $X(3872)$ mass is almost at the 
$\DDbarz$ threshold, 
the major component of the $X(3872)$  
is considered to be $\DDbarz$.
In such a situation,
it is convenient to look into $X$(3872) in the particle basis rather than in the isospin basis.
The wave functions of the $S$-wave $D\Dbar{}^*$ components 
of the $X$(3872) obtained by using the $c\bar c$ model are plotted 
by the long dashed curves in \fref{fig:X3872-DDwf1}.
In the $c \bar c$ model only the S-waves are relevant.
The $\DDbarz$ wave function is actually large in size and has a very long tail. 
Its root mean square distance (rms) is listed in \tref{tbl:rms}. 
Note that this number varies rapidly as the binding energy varies
because the rms becomes infinite as the binding energy goes zero
 as seen from \eref{eq:rms}.
The rms of the $\DDbarc$ component is 
much smaller than that of the
$\DDbarz$ because of the $\DDbarz$-$\DDbarc$ threshold difference. 
As seen from \fref{fig:X3872-DDwf1}, 
the amplitudes of the $\DDbarz$ and the $\DDbarc$ wave functions are similar in size
in the very short range region 
where the $D\Dbar{}^*$ state couples to the $c\cbar$;
the isospin-0 $D\Dbar{}^*$ state becomes a dominant component there
as shown in \fref{fig:X3872-DDwf-isospin}.
Probabilities of various components of the bound state are shown 
in the first line of  \tref{tbl:Xcompo}.
As was mentioned in \sref{sec:X-intro}, the production rate of $X$(3872) in the $p\bar p$ collision experiments
suggests that the amount of the $c\bar c$ component is expected to be several \%.
In the present $c\bar c$ model, the admixture is 8.6\%.
As we will show later, by introducing OPEP between the $D$ and $\Dbar$
mesons, this admixture reduces to 5.9 \%, 
which corresponds to the amount just required from the experiments.

\begin{table}[htbp]
\caption{Parameters and the root mean square distance (rms) of the two mesons in $X$(3872) 
in the $c\bar c$ model ($c\bar c$) \cite{Takizawa:2012hy}, in the OPEP model (OPEP) \cite{Yamaguchi:2019X}, 
and in the $c\bar c$-OPEP model ($c\bar c$-OPEP) \cite{Yamaguchi:2019X}.
The rms$_0$ (rms$_\pm$) means the rms between $D^{(*)0}$ and $\bar D{}^{(*)0}$
 ($D^{(*)+}$ and $D{}^{(*)-}$). BE is the binding energy in MeV.
\eref{eq:rms} stands for the values calculated by using \eref{eq:rms}.}
\begin{center}
\begin{tabular}{lccccccccc}\hline
&$g_{c\bar c}$ & $\Lambda_q$ (GeV) & $g_A$ & $\Lambda$ (GeV) & rms$_0$ (fm) &  rms$_\pm$ (fm) & BE(MeV)\\ \hline
$c\bar c$       & 0.05110 & 0.5& -    & -     & 8.39 & 1.56 & 0.16\\
OPEP            & -       &  - & 0.55 & 1.791 & 8.25 & 1.44 & 0.16\\
$c\bar c$-OPEP & 0.04445  & 0.5& 0.55 & 1.13  & 8.36 & 1.59 & 0.16\\ \hline
\eref{eq:rms}   & -       &  - & -    & -     & 7.93 & 1.11 & 0.16\\ \hline
\end{tabular}
\end{center}
\label{tbl:rms}
\end{table}%

\begin{table}[htbp]
\caption{Probabilities of each components of $X$(3872)
in the $c\bar c$ model ($c\bar c$) \cite{Takizawa:2012hy}, in the OPEP model (OPEP) \cite{Yamaguchi:2019X}, 
and in the $c\bar c$-OPEP model ($c\bar c$-OPEP) \cite{Yamaguchi:2019X}.
$|c_{c\bar c}|^2$ stands for the probability of the $c\cbar$ component, 
$|c_{0,\pm}({{}^3\!S})|^2$ stands for the $\DDbarz$ ($\DDbarc$) $({}^3\!S)$ component,
$|c_{0,\pm}({{}^{2S+1}\!D})|^2$ for the $\DDbarz$ ($\DDbarc$) $({}^{2S+1}\!D)$ component.
$D$ stands for the $D$-state probabilities, $\sum|c_{0,\pm}(^{2S+1}D)|^2$.
}
\begin{center}
\begin{tabular}{lcccccccccc}\hline
model & $|c_{c\bar c}|^2$ & 
$|c_0({{}^3\!S})|^2$& $|c_0({{}^3\!D})|^2$& $|c_0({{}^5\!D})|^2$ & 
$|c_\pm({{}^3\!S})|^2$& $|c_\pm({{}^3\!D})|^2$& $|c_\pm({{}^5\!D})|^2$ 
&$D$(\%)\\\hline
$c\bar c$ & 0.086 & 0.848 &-&-& 0.067 &-&-&-\\
OPEP  & - & 0.910 & 0.004 & 0.004 & 0.073 & 0.005 & 0.006 &2.0\\
$c\bar c$-OPEP  & 0.059 & 0.869 & 0.002 & 0.001& 0.065 &  0.002 & 0.001 & 0.6\\ \hline
\end{tabular}
\end{center}
\label{tbl:Xcompo}
\end{table}%

  \begin{figure}[htbp]
   \begin{center}
    \includegraphics[width=9cm,clip]{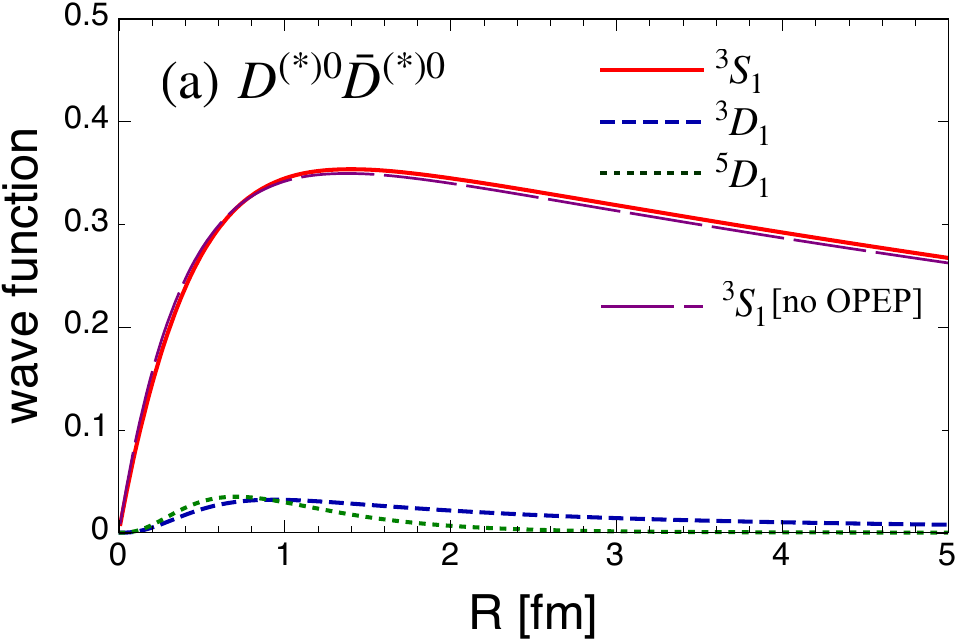}
    \includegraphics[width=9cm,clip]{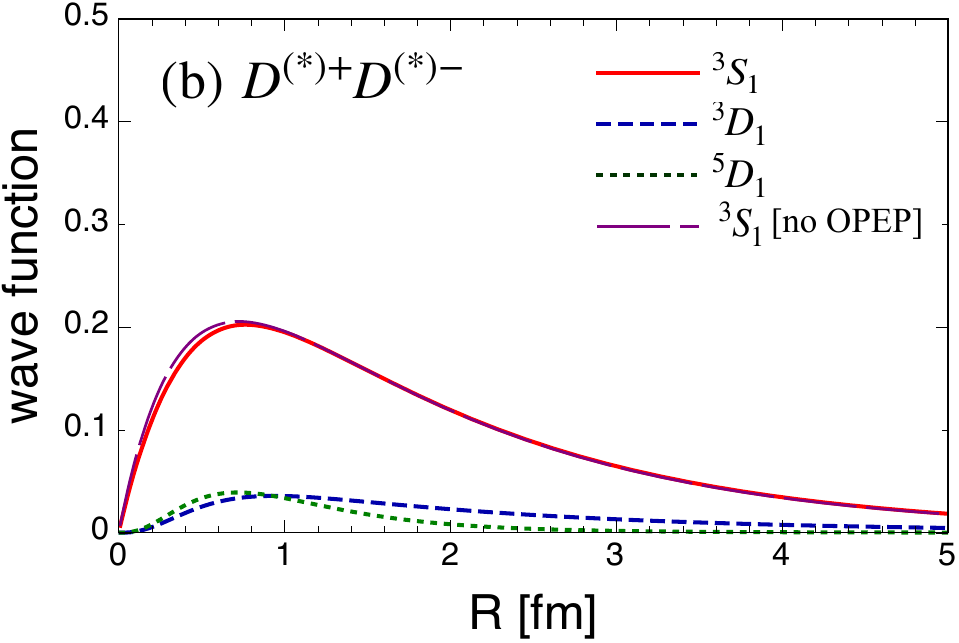}
   \end{center}
   \caption{
(a) The $\DDbarz$ and (b) the $\DDbarc$ wave function of $X$(3872) in the $c\bar c$-OPEP model with the same convention as in \fref{fig:X3872-DDwf1}.  
}
   \label{fig:X3872-DDwf2}
  \end{figure}

Now we consider models with OPEP included; 
the one denoted as OPEP in \tref{tbl:Xcompo} is the model with only 
the $D^{(*)} \bar D^{(*)}$ channels included as discussed in \sref{sec:X-OPEP}, 
and the other one denoted as $c \bar c$-OPEP  
is the $c\bar c$-$D^{(*)} \bar D^{(*)}$ coupled channel
model with the OPEP and their $S$-$D$ tensor couplings included~\cite{Yamaguchi:2019X}.  
The model space  is now taken to be ${c\bar c}$, $D^{(*)0} \Dbar{}^{(*)0}$ and $D^{(*)+} D^{(*)-}$ found in \eref{eq:X3872-components}:
\begin{eqnarray}
\Psi &=& c_{c\bar c}|c\cbar\ket+\psi_0+ \psi_\pm ,\\
\psi_{0,\pm}&=&c_{0,\pm }({}^3\!S)|D\Dbar{}^* ({}^3\!S_1)\ket+c_{0,\pm}({}^3\!D)|D\Dbar{}^* ({}^3\!D_1)\ket+c_{0,\pm}({}^5\!D)|D^*\Dbar{}^* ({}^5\!D_1)\ket , 
\nonumber\\
\end{eqnarray}
where $c_{c\bar c}$ is the amplitude of the $c\bar c$ component, 
$c_{0 }({}^3\!S)$ is that of the $\DDbarz ({}^3\!S_1)$ component, 
$c_{\pm }({}^3\!S)$ is that of the $\DDbarc ({}^3\!S_1)$ component, 
and so on.

The OPEP potential among the $I(J^{PC})=0(1^{++})$ $D^{(*)}\Dbar{}^{(*)}$ states are found  
 in 
 \eref{eq:OPEP-X3872}. 
In the particle base calculation, 
it is convenient to use the expression with the explicitly written isospin factor 
 \begin{eqnarray}
  V_\pi(r)&=&
  -
  \left(\frac{g_A}{2f_\pi}\right)^2\frac{1}{3}\left(
	\begin{array}{ccc}
	 {\cal C}_\pi&-\sqrt{2}{\cal T}_\pi &-\sqrt{6}T_\pi \\
	 -\sqrt{2}{\cal T}_\pi& {\cal C}_\pi+{\cal T}_\pi& -\sqrt{3}T_\pi\\
	 -\sqrt{6}T_\pi& -\sqrt{3}T_\pi& C_\pi-T_\pi\\			 
	\end{array}
   \right)\btau_1\cdot\btau_2,
  \label{eq:OPEP-X3872-particle_basis}  
 \end{eqnarray}
where $C_\pi$, $T_\pi$, ${\cal C}_\pi$, and ${\cal T}_\pi$ are the same as those defined for  \eref{eq:OPEP-X3872}. 
The $c\bar c$-$D\Dbar{}^*$ coupling is 
taken as \eref{eq:X-cc-DDbar-pot}.
The parameters are listed in \tref{tbl:rms}.
The OPEP cutoff $\Lambda$ in the OPEP-model 
is taken to be a free parameter to reproduce a bound state with the binding energy,
 0.16 MeV.
As for the $c\bar c$-OPEP model, the OPEP cutoff $\Lambda=\Lambda_D$ 
is the standard one obtained from the $D$-meson size as marked
in \fref{fig:X3872-ulim-1} in the previous subsection.
The $c\bar c$-$D\Dbar$ coupling strength, $g_{c\bar c}$, in the $c\bar c$-OPEP model
is taken to be a free parameter to fit the binding energy.
 
 In \tref{tbl:rms}, rms of the $\DDbarz$ and $\DDbarc$
 system are listed.
 The size of $X$(3872), governed mostly by the binding energy,
 does not depend much on details of the model.
The wave functions of each model are plotted in \fref{fig:X3872-DDwf1} and \fref{fig:X3872-DDwf2}.
The $D\Dbar{}^*$ ${}^3S_1$ wave functions are similar to each other, though 
they are slightly enhanced at the short distance in the OPEP model.
This is due to the tensor force; the $S$-$D$ coupling causes effectively 
an attraction in the $S$-wave channel which contains the square of the $D$-wave amplitude.  
In fact, the location of the maximum strength of the $D$-wave amplitude 
roughly coincides with where the ${}^3 S_1$ wave function is enhanced.

In the $c\bar c$ model, the attraction comes from the $c\bar c$-$D\Dbar^*$ coupling,
while the $S$-$D$-wave mixing by the OPEP tensor term provides the attraction in the OPEP model.
Their effects can be roughly estimated by the amounts of the ${c\bar c}$ components, $|c_{c\bar c}|^2$,
and the $D$-state probabilities, 
which are listed in \tref{tbl:Xcompo}.
In the ${c\bar c}$-OPEP model, where both of these attractions are introduced,
$|c_{c\bar c}|^2$ reduces from 8.6 \% to 5.9 \%, while the $D$-state probabilities
reduces from 2.0 \% to 0.6 \%.
The former reduces to 2/3, and the latter reduces to 1/3,
which are the rough share of the attraction in the $c\bar c$-$D\Dbar^*$ coupling
model with a reasonable cutoff for the OPEP.
The $D$-state probability and the $c\cbar$ probability depend much on the binding energy, or on slight change of $g_A$, 
whose value is determined in the heavy quark limit.
The size of the $c\bar c$ component can also vary as shown in the next subsection.
Therefore, it is difficult to estimate the relative importance of OPEP quantitatively.
Qualitatively, however, we can conclude that effects of the $c\bar c$-$D\Dbar{}^*$ coupling
and OPEP are comparable in $X$(3872).
One has to consider the coupled channel system of the $c\cbar$ pole,
 the $\DDbarz$ and the $\DDbarc$ scattering channels
with their mass difference,
and the $c\bar c$-$D\Dbar^*$ coupling and the OPEP, simultaneously,
to understand the feature of $X$(3872).

 \label{sec:X-core-vs-OPEP} 
\def\Gcc{{G^{(c\bar c)}}}
\def\Gmm{{G^{(mm)}}}
\def\Gmmz{{G^{(mm)}_0}}

\subsection{The decay spectrum of $X(3872)$}

The strong decay modes of $X(3872)$ 
observed up to now
are $\Jpsi\pi^+\pi^-$, $\Jpsi\pi^+\pi^-\pi^0$, $D^0\Dbar{}^{*0}$\cite{Tanabashi:2018oca}, and recently, $\chi_{c1}\pi^0$ \cite{Ablikim:2019soz}.
Here we discuss the strong decay of $X(3872)$,
especially the following two notable features to understand the $X$(3872) nature.
One is that a large isospin symmetry breaking is found in the final decay fractions: 
as seen in \eref{TT_eq:1_1},
the decay fractions of $X$(3872) going into 
$\Jpsi\pi^3$ and $\Jpsi\pi^2$ indicate that
amounts of the $\Jpsi\omega$ and $\Jpsi\rho^0$ components in $X$(3872) are comparable to each other
as shown in \eref{TT_eq:1_2}.
The other feature we would like to discuss here is that 
the decay width of the $X$(3872) is very small
for a resonance above the open charm threshold,
or for a resonance decaying through 
the $\rho$ and $\omega$ components, which themselves have a large decay width.

In the following we employ a model which consists of 
the $c\bar c$ core, $\DDbarz$, $\DDbarc$, $\Jpsi\omega$ and $\Jpsi\rho^0$.
The system here does not include the $\chi_{c1}\pi^0$ channel.
Since an amount of the observed fraction is about the same as that of the $\Jpsi\rho^0$,
this channel can probably be treated by a perturbational;
properties of the other channels will not change much if this $\chi_{c1}\pi^0$ channel is introduced.
Moreover, since its threshold is lower than the $X$(3872) by 230 MeV,
it will be necessary to consider the pion radiation from the $D^{(*)}\bar D{}^{(*)}$ components 
to obtain the $\chi_{c1}\pi^0$ fraction.
Here the model contains only the relative $S$-wave hadron systems which have thresholds close to each other. 

For the discussions of decay properties here, 
it is sufficient to consider 
the formation of a loosely bound $D \bar D^*$ states, which couple to the $c\bar c$
and to the $J\psi \rho$ and $J\psi \omega$ 
with finite decay widths for $\rho$ and $\omega$.
We assume effective couplings between $c\bar c$ and $D\Dbar{}^*$,
which gives the attraction as we discussed in the previous section,
and between $D\Dbar{}^*$ and $\Jpsi \omega(\rho^0)$, which expresses the rearrangements.
In this section we do not introduce OPEP;
the system is restricted only to the $S$-waves, and  
the attraction from the OPEP is effectively taken into account 
by introducing the central attraction between 
the $D^{(*)}$ and $\Dbar{}^{(*)}$. 
The widths of the $\omega$ and $\rho$ mesons are taken into account
as an imaginary part in the $J/\psi \omega$ and $J/\psi \rho$ propagators.
In this way, we consider that the model can simulate essential features of the decay properties of $X(3872)$.

From the quark model point of view,
the $D\Dbar{}^*$ states of total charge 0 are the $c\cbar u\ubar $ or $c\cbar d\dbar $ states,
which contain also the $\Jpsi\omega$ or $\Jpsi\rho^0$ state
with the appropriate color configuration.
The observed final  $\Jpsi \pi^3$ and $\Jpsi \pi^2$ decay modes are considered to come from these components. 
The rearrangement 
between  $D\Dbar{}^*$ and  $\omega\Jpsi$ or $\rho^0\Jpsi$ occurs 
at the short distance, where all four quarks exist in the hadron size region.
The coupling between the $c\bar c$ and $D\bar D{}^*$, however, is not
known. Therefore, it is treated as a phenomenological one, as shown below.
Note that there is no direct channel coupling between the
$c\cbar$ channel 
and the $\omega\Jpsi$ or $\rho^0\Jpsi$ channels in the present model setup.
They break the OZI rule,
and the latter breaks the isospin symmetry.

The model Hamiltonian for the $c\cbar$, $D^{0}\Dbar{}^{*0}$, $D^{+}D^{-*}$, $\Jpsi\omega$ and $\Jpsi\rho$ channels 
 is
taken as  \cite{Takeuchi:2014rsa}:
\begin{eqnarray}
H&=&
\left(\begin{array}{ccccccc}
m_{c\bar c} & U^{(mc)}{}^\dag \\
U^{(mc)} & H^{(mm)}\\
\end{array}
\right),
\\
H^{(mm)}&=&H^{(mm)}_0+V^{(mm)}, \\
H^{(mm)}_0&=&{\rm diag}( K_{D{\bar D}0}, K_{D{\bar D}\pm} ,K_{J/\psi\omega},K_{J/\psi\rho}), 
\label{eq:X-H0}
\\
K_{D{\bar D}0,\pm}&=& m_{D^{0,-}}+m_{D^{*0,+}}+{p^2\over 2 \mu_{D{\bar D}0,\pm}}
,
\\K_{\Jpsi\omega,\rho}&=& m_{\Jpsi}+m_{\omega,\rho}
+{p^2\over 2 \mu_{J/\psi\omega,\rho}} ,
\\
U^{(mc)}&=&
\left(\begin{array}{ccccccc}
-u   \\
u \\
0   \\
0    \\
\end{array}
\right),~~~
V^{(mm)}=
\left(\begin{array}{ccccccc}
 \tilde v & 0 & -v & -v \\
 0 &\tilde v& v & -v\\
 -v  & v &0 & 0\\
 -v & -v &0 & 0 \\
\end{array}
\right) ,
\label{eq:X-UV}
\end{eqnarray}

\begin{eqnarray}
u(q) &=& {g_{c\bar c}\over \sqrt{\Lambda_q}}L_{\Lambda_q}(q)
,~~~~L_{\Lambda_q}(q)={\Lambda_q^2\over \Lambda_q^2+q^2} ,
\label{eq:X-ccDD-coupling}
\\
v(q,q') &=& {v_0\over \Lambda_q^2}L_{\Lambda_q}(q)L_{\Lambda_q}(q'),~~~
\tilde v(q,q') = {\tilde v_0\over \Lambda_q^2}L_{\Lambda_q}(q)L_{\Lambda_q}(q') ,
\label{eq:X3872-W}
\end{eqnarray}%
where $m_{c\bar c}$ is the $c\cbar$ bare mass when the coupling to $D\Dbar{}^*$ is switched off.
The reduced masses, $\mu_{D{\bar D}0,\pm}$ and $\mu_{J/\psi\omega,\rho}$, are for the $D{}^0 \Dbar{}^{*0}$, 
$D^\pm D^{*\mp}$, $\Jpsi\omega$ and $\Jpsi\rho$ systems, respectively.
The coupling between the $c\cbar$ state and the $D\Dbar{}^*$ state
is expressed by the transfer potential, $U^{(mc)}$, 
which is chosen to be Lorentzian in the momentum space with the strength $g_{c\bar c}$.
The rearrangement between the $D\Dbar{}^*$ states and the $\Jpsi\omega$ and $\Jpsi\rho$ meson 
is expressed by a separable potential, $v$ in $V^{(mm)}$. 
The basis of the matrix expression in \eref{eq:X-H0} and in \eref{eq:X-UV} are 
($D^{(*)0}\Dbar{}^{(*)0}$, $D^{(*)\pm} D^{(*)\mp}$, $\Jpsi\omega$, $\Jpsi\rho^0$).
The strength of the interaction between the $D$ and $\Dbar$ mesons, $\tilde v_0$,
is taken to be the maximal value 
which does not create a bound state in the $B\bar B$ systems,
where no bound states has been observed yet. 
The strengths $g_{c\bar c}$ and $v_0$ are free parameters
under the condition that the mass of $X$(3872) can be reproduced.
The value of $\Lambda_q$ is the same as the one used in the previous section, 
$\Lambda_q$ = 0.5 GeV.
The parameters are summarized in \tref{tbl:X-param-width}.

\begin{table}[htbp]
\caption{Model parameters and the amounts of each component. 
Masses and widths are in MeV, and taken from \cite{Agashe:2014kda}
}
\begin{center}
\begin{tabular}{cccccccccc}\hline
$m_{J/\psi}$ &$m_\omega$ & $\Gamma_\omega$ & $m_\rho$ & $\Gamma_\rho$ & $g_{c\bar c}$ & $v_0$ & $\tilde v_0$\\ 
3096.916 & 782.65 & 8.49 & 775.26 & 147.8 & 0.04136 & 0.1929 & $-0.1886$\\ \hline
$|c_{c\bar c}|^2$ & $|c_0(^3S)|^2$& $|c_\pm(^3S)|^2$ & $|c_{J/\psi\omega}|^2$ & $|c_{J/\psi\rho}|^2$ \\
0.036 & 0.913 & 0.034 & 0.010 & 0.006\\\hline
\end{tabular}
\end{center}
\label{tbl:X-param-width}
\end{table}%

The amount of each component in the $X$(3872) bound state
is also listed in \tref{tbl:X-param-width}.
The bulk feature is similar to the models in the previous section:
the dominant component is $\DDbarz$ while the $\DDbarc$ component is considerably smaller
because of the threshold difference.
The amount of the $c\bar c$ component is somewhat smaller but still sizable.
The $J/\psi\omega$ and $J/\psi\rho$ components are small
comparing to the $D\Dbar{}^*$ components.
The fact that the $J/\psi\rho$ and  $J/\psi\omega$ components of $X$(3872)
are comparable in size is reproduced in the present model.

  \begin{figure}[tbp]
   \begin{center}
    \includegraphics[height=6cm]{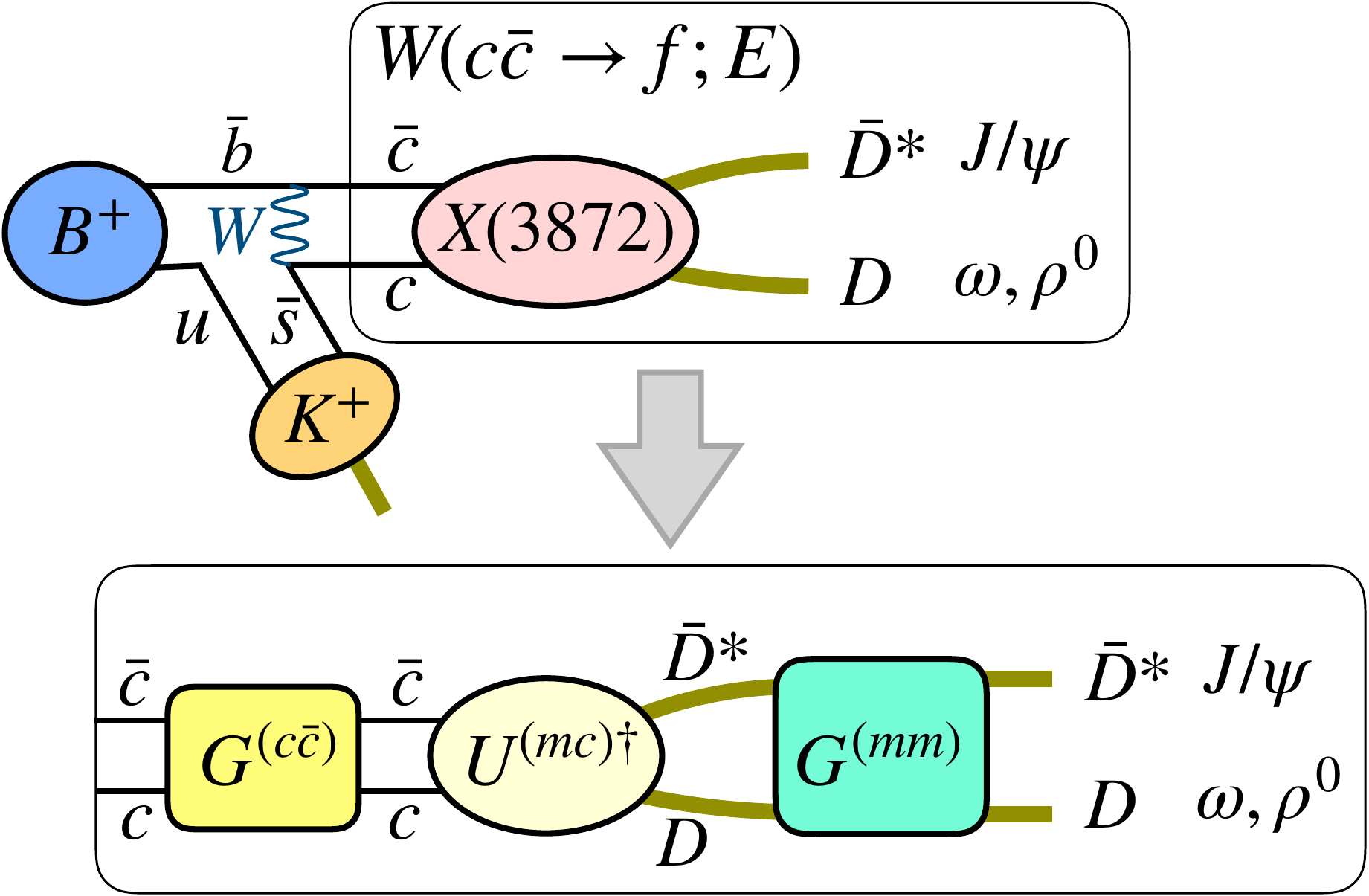}%
   \end{center}
   \caption{The $B$ meson weak decay. See text.}
   \label{fig:BtoXtof-2}
  \end{figure}

As listed in \tref{table:X3872exp},
the $X$(3872) is produced by various processes.
As a typical example, we discuss the $X$(3872) decay process in the $B$ meson weak decay in the following.
As illustrated in \fref{fig:BtoXtof-2}, the mass spectrum of $X$(3872) from the $B$ meson weak decay is proportional to 
the sum of the transfer strength from the $c\cbar$ to
the two-meson states, $f$, $W(c\cbar \rightarrow f;E)$, which can be expressed as
\begin{eqnarray}
\sum_f{\rmd W(c\cbar \rightarrow f;E)\over \rmd E} 
= -{1\over \pi}{\rm Im}\bra c\cbar|\Gcc(E)|c\cbar\ket .
\label{eq:X-W}
\end{eqnarray}
Here $\Gcc$ is the full propagator of the $c\cbar$ state, 
which can be written by using the self energy $\Sigma_{c\bar c}$ as:
\begin{eqnarray}
\Gcc(E)&=&(E-m_{c\bar c}-\Sigma_{c\bar c}(E))^{-1} ,
\label{eq:X-Gcc}
\\
\Sigma_{c\bar c}(E)&=&U^{(mc)}{}^\dag \Gmm(E) U^{(mc)} .
\label{eq:X-Sigcc}
\end{eqnarray}
Here we define the free and the full propagators within the two-meson space, 
$\Gmmz(E)$ and $\Gmm(E)$, 
respectively, with the decay widths as
\begin{eqnarray}
\Gmmz(E)&=&\Big(E-H_0^{(mm)}+\rmi{\Gamma\over 2}\Big)^{-1} ,\\
\Gmm(E)&=&\Big(E-H^{(mm)}+\rmi{\Gamma\over 2}\Big)^{-1} ,\\
\Gamma&=&{\rm diag}(0,0,\Gamma_\omega,\Gamma_\rho ),
\end{eqnarray}
where $\Gamma_{\omega,\rho}$ is the $\omega$ or $\rho$ decay width, respectively.
The $\rho$ meson width is taken to be energy dependent as discussed in \cite{Takeuchi:2014rsa}. 
The widths of $D^*$ mesons are neglected.
The width $\Gamma$ is ignored 
when the bound state energy or the component is calculated above.
It, however, is essential to include them when one investigates the decay spectrum.

In order to obtain the decay spectrum of each final two-meson channel separately,
we have rewritten the right-hand side of  \eref{eq:X-W} as follows.
Since the system has only one $c\bar c$ state in the present model, the above $\Gcc(E)$
and $\Sigma_{c\bar c}(E)$ are single channel functions of the energy $E$.
They become matrices when more than one $c\bar c$ states are introduced, but 
the following procedure can be extended in a straightforward way.
As seen from \eref{eq:X-Gcc}, 
the imaginary part of $\Gcc(E)$ comes only from the imaginary part of $\Sigma_{c\bar c}(E)$.
Therefore,
\begin{eqnarray}
{\rm Im}\, \Gcc&=& 
{\rm Im}\,(\Gcc^* \,(\Gcc^*)^{-1} \,\Gcc)
={\rm Im}\,(\Gcc^* \,\Sigma_{c\bar c} \,\Gcc)
\nonumber\\
&=&{\rm Im}\,(\Gcc^* \,U^{(mc)}{}^\dag \Gmm U^{(mc)} \,\Gcc) ,
\label{eq:ImG-1}
\end{eqnarray}%
where $^*$ stands for the complex conjugate. 
Using the following relation for a real potential
\begin{eqnarray}
{\rm Im}\, \Gmm^{-1} &=&  {\rm Im}\, \Gmmz^{-1}
\end{eqnarray}%
 and  Lippmann Schwinger equation for the propagator,
$G=(1+GV)G_0$, we have for Im$\Gmm$ on the right-hand side of \eref{eq:ImG-1} 
\begin{eqnarray}
{\rm Im}\, \Gmm
&=&
{\rm Im}\,  (\Gmm (\Gmm{}^*)^{-1}\Gmm{}^* )
\nonumber\\
&=&
{\rm Im}\,  (\Gmm (\Gmmz{}^*)^{-1}\Gmm{}^* )
\nonumber\\
&=&{\rm Im}\,\Big((1+\Gmm V^{(mm)}) \Gmmz (1+\Gmm{}^* V^{(mm)})\Big) .
\label{eq:ImG-2}
\end{eqnarray}%
Thus, \eref{eq:ImG-1} can be rewritten as
\begin{eqnarray}
{\rm Im}\, \Gcc 
&=&{\rm Im}\,\Big(\Gcc^* U^{(mc)}{}^\dag (1+\Gmm V^{(mm)}) \Gmmz (1+\Gmm{}^* V^{(mm)}) U^{(mc)} \Gcc\Big) .
\nonumber\\
\label{eq:ImG-3}
\end{eqnarray}%
When we apply the plain wave expansion for the
$\Gmmz$ in \eref{eq:ImG-3}, we have 
\begin{eqnarray}
\lefteqn{{\rmd W(c\cbar \rightarrow f;E)\over \rmd E}}
\nonumber\\
&=&
-{1\over \pi}{\rm Im}\int{\rm d}{\bi k}\bra f{\bi k}|\Gmmz(E)|f{\bi k}\ket
\Big|\sum_{f'}\int \rmd{{\bi k}'}
\bra\!\bra f'{\bi k}'(f{\bi k})| U^{(mc)}({\bi k}') \Gcc(E)|c \bar c\ket
\Big|^2
\nonumber\\
&=&{2\over \pi}\mu_f\int {k^2{\rm d}k\;\mu_f\Gamma_f\over (k_f^2-k^2)^2+(\mu_f\Gamma_f)^2}
\Big|\sum_{f'}\int \rmd{{\bi k}'}
\bra\!\bra f'{\bi k}'(f{\bi k})|U^{(mc)}({\bi k}')\Gcc(E)|c \bar c\ket\Big|^2,
\nonumber\\
\label{eq:X3872-width}
\end{eqnarray}%
where $k_f$ and $\mu_f$ stand for the three-momentum and 
the reduced mass of the final two-meson state where
$E=m_{1f}+m_{2f}+k_f^2/(2\mu_f)$.
$\Gamma_f$ is the the decay width of mesons in the final state $f$, {\it i.e.}\
0 if $f$ is $D\Dbar{}^*$, $\Gamma_\omega$ or $\Gamma_\rho$ when $f$ is $J/\psi \omega$ or $J/\psi \rho$.
$|f{\bi k}\ket$ stands for the plain wave of the channel $f$ with the momentum ${\bi k}$.
$|f'{\bi k}'(f{\bi k})\ket\!\ket$ stands for the distorted wave function 
of the channel $f'$ with the momentum ${\bi k}'$ which is generated from $|f{\bi k}\ket$.
This can be obtained by the Lippmann Schwinger equation as
\begin{eqnarray}
|f'{\bi k}'(f{\bi k})\ket\!\ket&=&(1+\Gmm V^{(mm)})|f{\bi k}\ket .
\label{eq:wfLS-eq}
\end{eqnarray}%
In the present model, only the $D\bar D{}^*$ channels couple directly to $c\bar c$.
The summation over $f'$ in \eref{eq:X3872-width} means summation over 
$D^{0}\Dbar{}^{*0}$ and $D^{+}D^{-*}$.
The final two-meson fraction expressed by $f$ in the above equations
can be  
$\Jpsi \omega$ or $\Jpsi\rho$ as well as $D\bar D{}^*$.
For the channels where $\Gamma_f$ is small, the transfer strength becomes 
\begin{eqnarray}
{\rmd W(c\cbar \rightarrow f;E)\over \rmd E}&\rightarrow&\mu_f k_f
\Big|\sum_{f'}\int \rmd{{\bi k}'}\bra\!\bra f'{\bi k}' (fk_f)|U^{(mc)}({\bi k}')\Gcc(E)|c \bar c\ket\Big|^2
~~~(\Gamma_f\rightarrow 0).
\nonumber\\
\end{eqnarray}%

\begin{figure}[htbp]
   \begin{center}
    \includegraphics[height=7cm,clip]{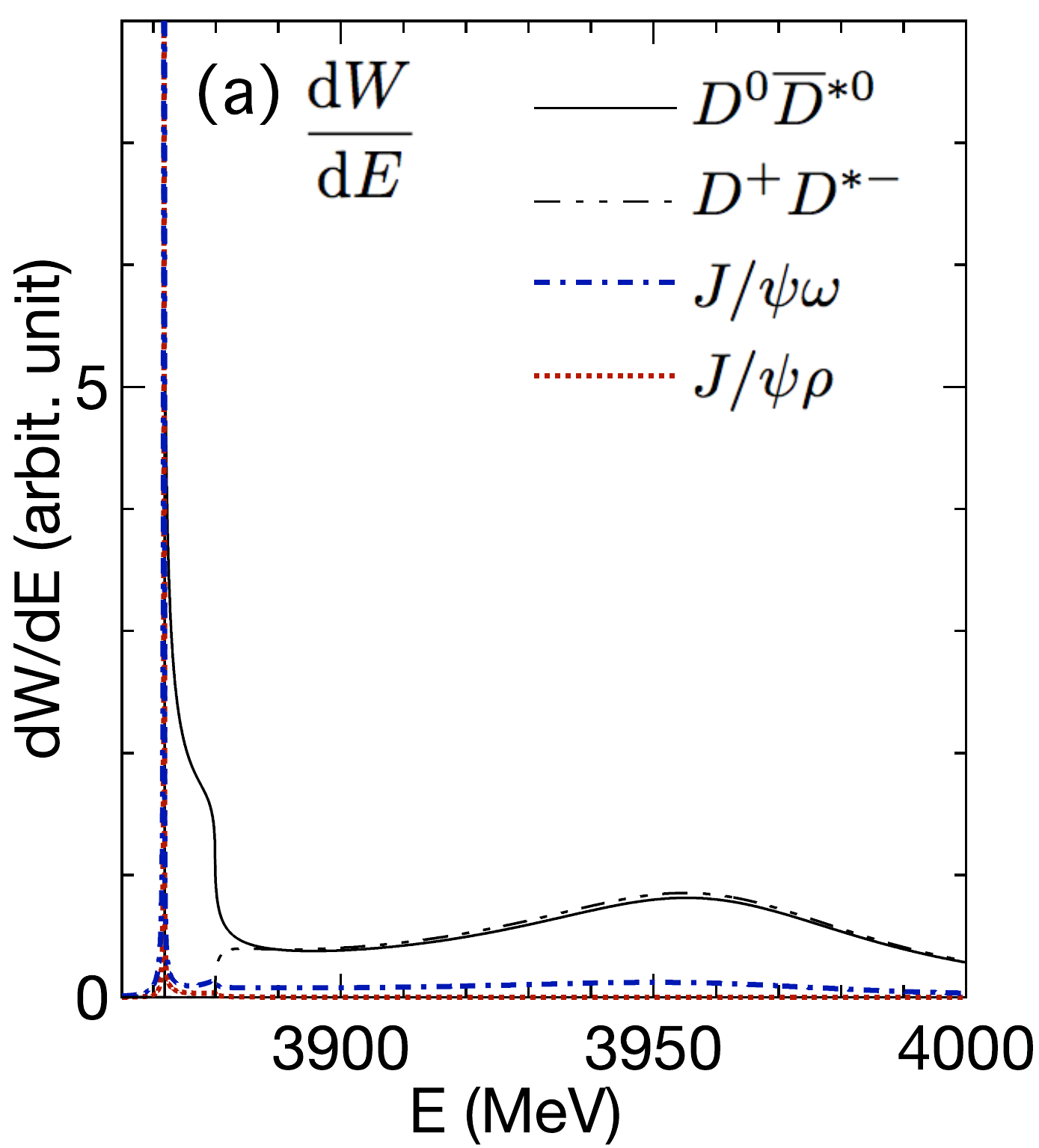}%
    \includegraphics[height=7cm,clip]{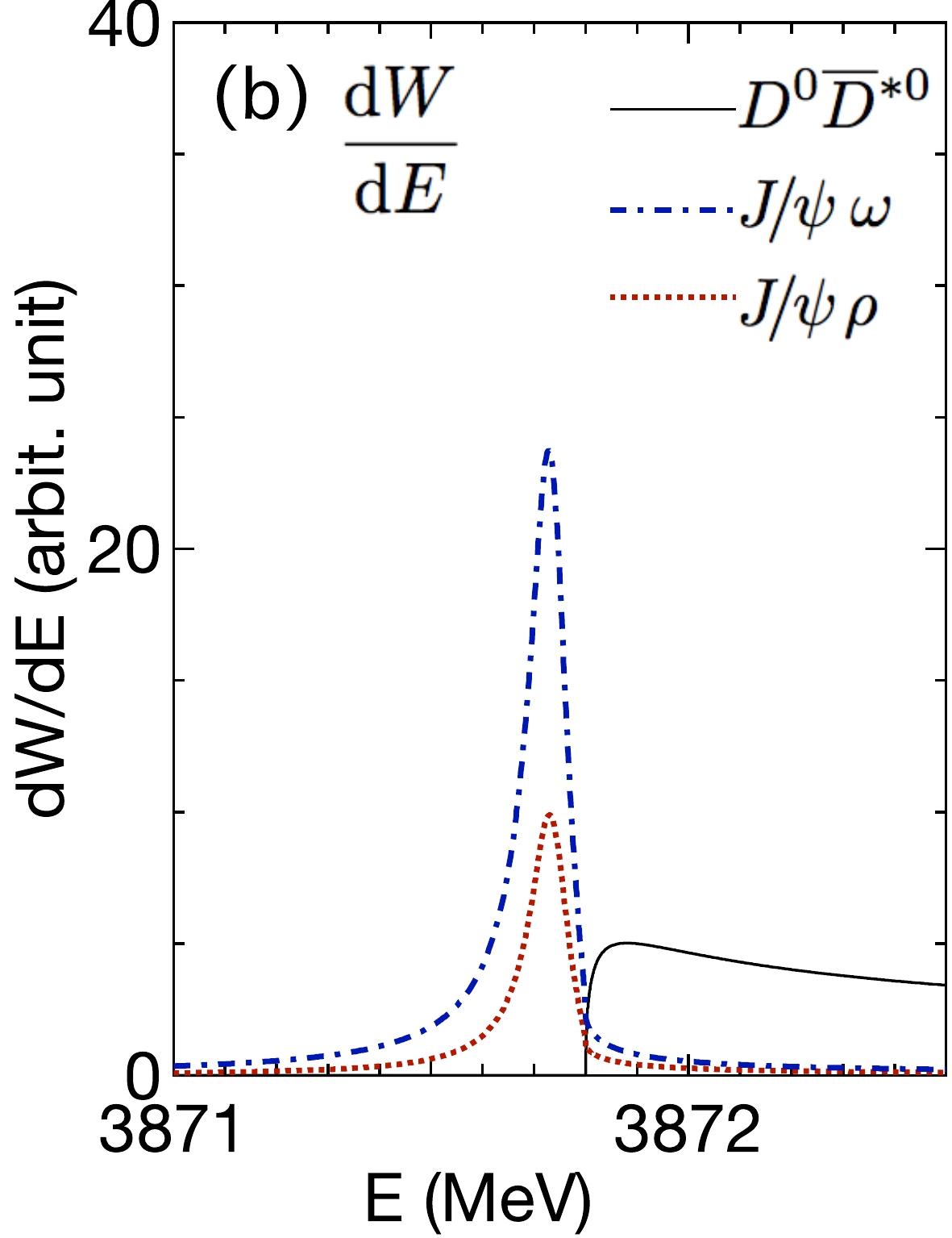}
   \end{center}
   \caption{The transfer strength from the $c\cbar$ state to the final two-meson states. 
   The energy $E$ is the center of mass energy of the two-meson states.
   Figure (b) is the same as (a) but magnified at around the $D^0\Dbar{}^{*0}$ threshold.
   Taken from \cite{Takeuchi:2014rsa}. }
   \label{fig:X3872decay}
  \end{figure}

The calculated $W$ spectrum is shown for each final state in \fref{fig:X3872decay}. 
The narrow peak at around the
$D^0\Dbar{}^{*0}$ threshold as well as 
the large isospin mixing are successfully reproduced by this picture of the
 $c\cbar$, $D^{0}\Dbar{}^{*0}$, $D^{+}D^{-*}$, $\Jpsi\omega$ and $\Jpsi\rho$.
The spectrum shape depends on the $X$(3872) binding energy,
but the qualitative feature does not.
The $\Jpsi\omega$ and $\Jpsi\rho$ peaks in the decay spectrum remain sharp
when the $X$(3872) is around the threshold\cite{Takeuchi:2014rsa}.

Let us discuss the mechanism to have the small width of the peak and the isospin mixing
in the final fractions, 
the two notable features of
the final $\Jpsi\omega(\rho)$ spectrum mentioned before.
The small width of $X$(3872) means that the corresponding pole of $\Gcc$ 
is close to the real axis.
The imaginary part in the denominator of $\Gcc$ 
comes from the imaginary part of $\Sigma_{c\bar c}$
as shown \eref{eq:X-Gcc}, which 
can be expanded by using \eref{eq:X-Sigcc} as
\begin{eqnarray}
\Sigma_{c\bar c}
&=&u^\dag G^{(D\bar D{}^*)} u+u^\dag G^{(D\bar D{}^*)} v G^{(J/\psi\omega,\rho)}_0 v G^{(D\bar D{}^*)}u
+\cdots,\nonumber\\
&=&
u^\dag G^{(D\bar D{}^*)} u
\Big(1+ 
{g_{c\bar c}}^{-1} v_0^2 (\Lambda_q^{-2} 
L_{\Lambda_q} 
G^{(J/\psi\omega,\rho)}_0 
L_{\Lambda_q} )(
\Lambda_q^{-3/2}
L_{\Lambda_q} G^{(D\bar D{}^*)} u)
+\cdots\Big),\nonumber\\
\label{eq:Sigcc-1}
\end{eqnarray}%
where
$G^{(D\bar D{}^*)}$ stands for the full propagator obtained within the $D\Dbar{}^*$ space,
and the second equation can be obtained for the separable potential,
\eref{eq:X3872-W} and \eref{eq:X-ccDD-coupling}.

On the right-hand side of \eref{eq:Sigcc-1}, 
the first term is real for the energy below the threshold.
It is because there is no bound state without the $c\bar c$ core,
and also because the decay width of $D^*$ is very small,
$83.4\pm 1.8 $ keV for $D^{*\pm}$ and $< 2.1$ MeV for $D^{*0}$ \cite{Tanabashi:2018oca},
and can be neglected.
The first term gives most of the size of the real part of the self energy, $\Sigma_{c\bar c}$,
which gives the mass difference between the $c\bar c$ core and the $X$(3872).

The second term has an imaginary part 
which comes from the width of the $\omega$ and $\rho$ 
in the propagator $G^{(J/\psi\omega,\rho)}_0$.
This term corresponds to the decay with the rearrangement,
$D\bar D{}^*\to \Jpsi\omega$, $\Jpsi\rho$.
The second term is very much suppressed.
One suppression comes from the color factor.
Probability to find a color-singlet $c\bar c$ pair in the $D\bar D{}^*$ channel
has a factor 1/9 in the color space.
In the present model this effect is taken into account by employing 
a small coupling constant $v_0$ in the transfer potential $v$. 
Another suppression comes from the range of the rearrangement.
It occurs with charm quark exchange, 
which means that the reaction occurs in the hadron size region
of the very large object, $X$(3872).
The factor $G^{(D\bar D{}^*)}u|c\bar c\ket$
can be regarded as the ${D\bar D{}^*}$ wave function
generated from $|c\bar c\ket$. 
So, one of the factors of the second term becomes a overlap:
\begin{eqnarray}
(
\Lambda_q^{-3/2}
L_{\Lambda_q} G^{(D\bar D{}^*)} u)
&\propto
4\pi \sqrt{2} \Lambda_q^{-3/2}\int{q^2\rmd q\over (2\pi)^3} L_{\Lambda_q}(q)
\psi_{X(3872)}(q)|_{D\bar D{}^*}
\label{eq:overlap-X}
\end{eqnarray}%
which is 0.109 for $I=0$, and 0.017 for $I=1$ when the wave function plotted 
in \fref{fig:X3872-DDwf-isospin} is used.
These suppressions, 
with kinematical factor for the final $J/\psi\omega$ channel discussed below, 
bring about the small imaginary part for the $X$(3872).

The isospin ratio of \eref{eq:overlap-X}, 0.15 (=0.017/0.109), 
is a rough size of the isospin
symmetry breaking in the short range part of the $X$(3872) wave function.
This ratio has been calculated in other literatures,  
$(11.5\pm 5.7)^{-1}$\cite{Braaten:2005ai},
or 
$0.27^2$ in \eref{TT_eq:1_2} \cite{Suzuki:2005ha}.
In order to compare this ratio to the experimental branching fractions \eref{TT_eq:1_1},
the kinematical factor, another factor in the second term, should be considered.
For the final $J/\psi\omega$ channel, the kinematical suppression 
appears 
because the charged and the neutral $D\bar D{}^*$ threshold difference 
is comparable in size to the $\omega$ decay width.
\begin{eqnarray}
{\rm Im}\int\rmd q\; \Lambda_q^{-2} L_{\Lambda_q}^2(q)G^{(J/\psi\omega)}_0(E=m_{X(3872)}) \Big/
(\omega \leftrightarrow \rho)= 0.15~ .
\label{eq:185}
\end{eqnarray}%
The corresponding value with 
a careful treatment for the off-shell 
$\omega$ meson is found to be
 0.0870 \cite{Braaten:2005ai}, or 1/13.3 \cite{Suzuki:2005ha}.
When both of the factors, \eref{eq:overlap-X} and \eref{eq:185}, are considered,
the ratio of the branching fractions becomes of the order of 1,
that is consistent with the observed one as discussed in \sref{sec:X3872}.
The above estimate gives
a rough size of the isospin violation in the $X(3872)$.
The new experiment, 
the $X(3872) \to \chi_{c1}\pi^0$ decay,
which we have not discussed here,
will also support this value 
because that decay fraction has a similar size of that of the $\pi\pi J/\psi$ \cite{Ablikim:2019soz}.

\begin{figure}[tbp]
   \begin{center}
    \includegraphics[height=7cm]{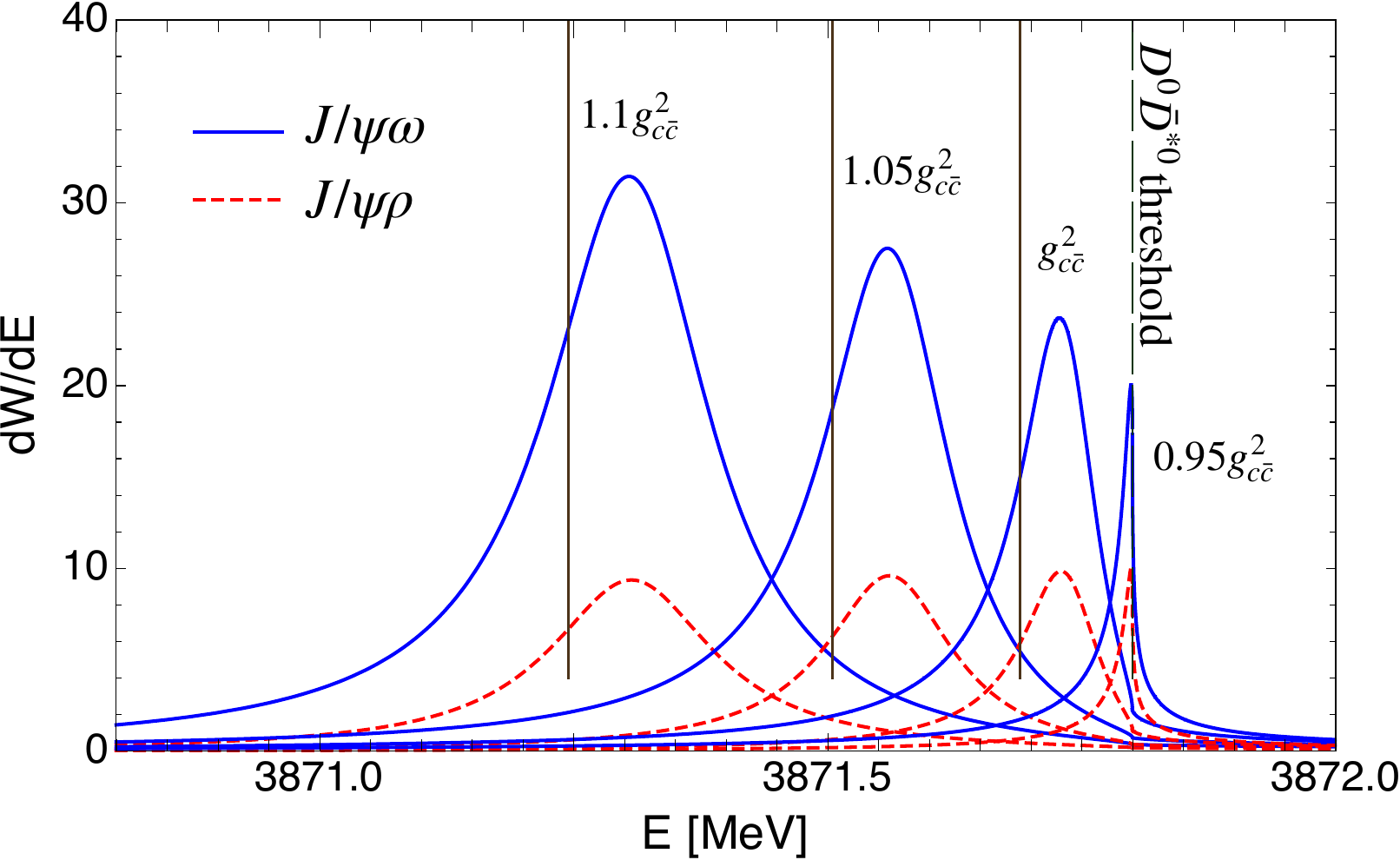}%
   \end{center}
   \caption{The transfer strength from the $c\cbar$ state to the final $\Jpsi\omega$ and $\Jpsi\rho$ states. Note that the energy abscissa is taken to be from 3870.8 to 3872 MeV.
   Final states are $\Jpsi\omega$ (Solid lines), and $\Jpsi\rho$ (dashed lines).
   The coupling $g_{c\bar c}^2$ is factored by 1.1, 1.05, 1, 0.95. 
   The corresponding bound state energies calculated without the decay widths 
   are shown by the left three vertical lines: 
   3871.25, 3871.51, and 3871.69 MeV, respectively. 
   The most right vertical line corresponds to the $D^0\Dbar{}^{*0}$ threshold.
   The one with the weakest coupling, 0.95$g_{c\bar c}^2$, does not have a bound state. 
   }
   \label{fig:X3872rhodecay}
  \end{figure}

In \fref{fig:X3872rhodecay},
the $\Jpsi\omega$ and $\Jpsi\rho$ peaks are plotted for the $c\bar c$ coupling with four different values of the strength $g_{c\bar c}$.
The corresponding binding energies are shown by the arrows,
which, as mentioned before, are calculated without the $\rho$ or $\omega$ meson widths.
The left two peaks correspond to the ones with stronger coupling than that of the reference \cite{Takeuchi:2014rsa}.
The third one is that of the original value with a binding energy of 0.16 MeV.
The most right one corresponds to the one where the coupling $g_{c\bar c}^2$ is reduced by a factor   0.95;
there is no bound state but a virtual state appears. 
As the coupling $g_{c\bar c}^2$ is weakened, the peak moves as the bound state moves to the threshold, but stops at the threshold when the pole moves to the
second Riemann sheet.
Namely, the peak energy of the final $\Jpsi \pi^n$ is kept
lower than or equal to the $D^0\Dbar{}^{*0}$ threshold energy.
On the other hand, the peak energy of the final $D^0\Dbar{}^{*0}$ fraction 
is higher than threshold by definition.
This means that the $X$(3872) mass is higher when measured by the
final $D\Dbar{}^*$ fraction, which is consistent with the observation.
Experimentally the $X$(3872) mass 
observed by the $J/\psi$ and anything is 3871.69$\pm$0.17 MeV with a width of $<$1.2 MeV \cite{Tanabashi:2018oca}, while that of the final $D\bar D{}^*$ is 3872.9${+0.6\atop -0.4}{+0.4\atop -0.5}$ MeV \cite{Adachi:2008sua}
or 3875.1${+0.7\atop -0.5}\pm0.5$ MeV\cite{Aubert:2007rva}.
The observed threshold mass is 3871.68$\pm$0.07 MeV  \cite{Tanabashi:2018oca}.
Thus the mass observed by the $J/\psi$ and anything 
is consistent with the heavier two peaks in \fref{fig:X3872rhodecay}. 
This suggests that 
the $X$(3872) is either a virtual state or a very shallow bound state.
These two states are very hard to distinguish from each other experimentally
when one considers the width of the component, such as the $\rho$ meson.
In literature, 
line shapes were studied by using amplitudes parametrized 
by effective range method~\cite{Braaten:2007dw}
or by the Flatte parametrization~\cite{Hanhart:2007yq,Voloshin:2007hh,Kalashnikova:2009gt}.  
It, however, seems difficult to determine the position of the resonance poles
just from the shape of the decay spectrum.
To discuss this subject, 
it will probably be necessary to treat the $\rho$ meson as
a resonance of the $\pi\pi$ continuum and perform dynamical analyses. 

\begin{figure}[tbp]
   \begin{center}
    \includegraphics[height=7cm]{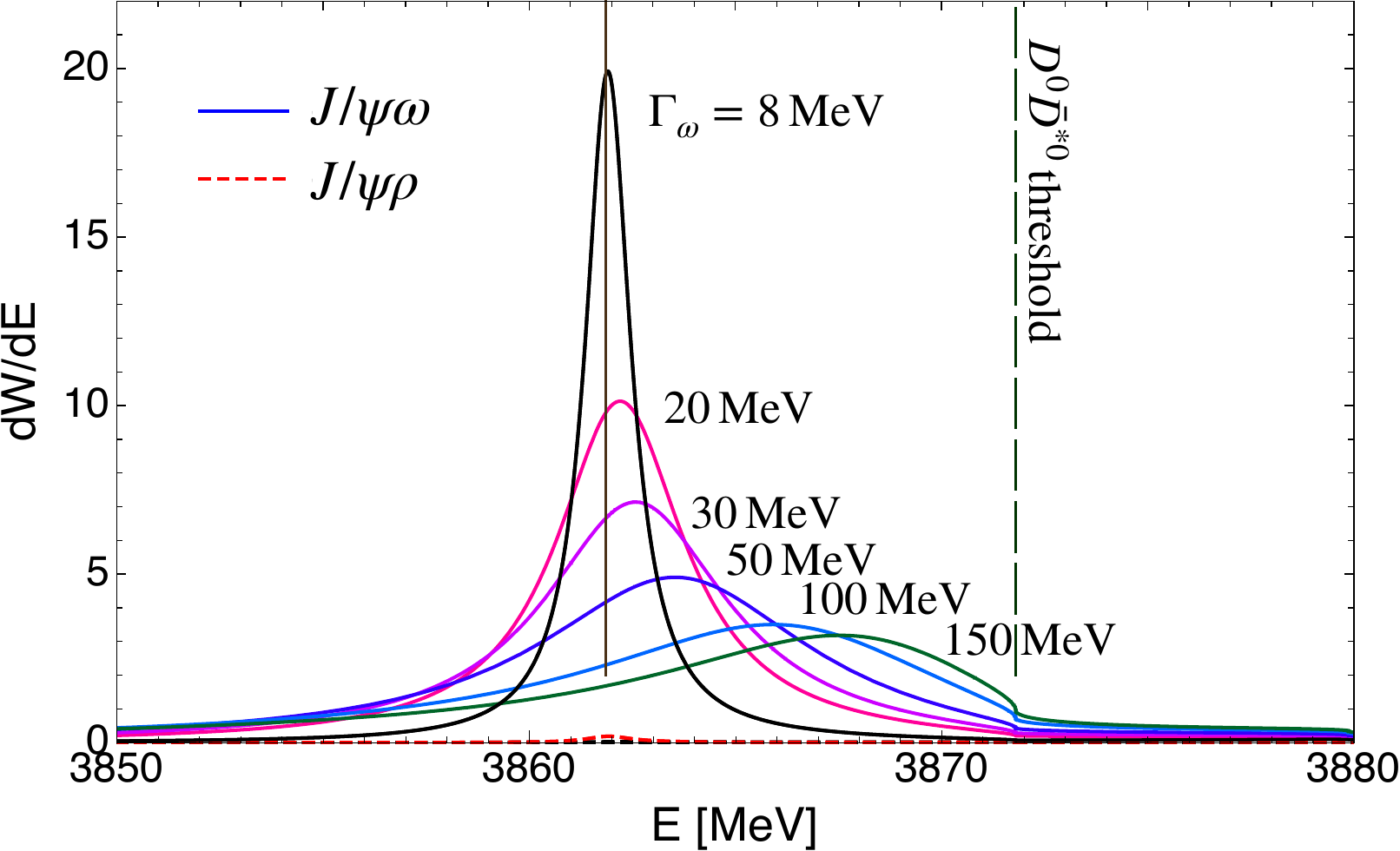}%
   \end{center}
   \caption{Trial calculation to estimate the effects of constructing mesons' width.  
   The $\Jpsi\omega$ channel is assumed to couple to the $c\bar c$ core, 
   and the width is enhanced by hand. (See text.) 
   The solid line is for the final $\Jpsi\omega$ fraction, the dashed line is for the 
   final $\Jpsi\rho$ fraction. 
   The latter is shown only for $\Gamma_\omega$ = 8 MeV. 
   The vertical lines correspond to the bound state energy 
   and the $\DDbarz$ threshold.}
   \label{fig:X3872omegadecay}
  \end{figure}

In order to discuss possibilities of the other mechanism for the $X$(3872) peak
as well as the effects of the meson decay width,
let us ignore the OZI rule just for now.
In strong decay,
where the isospin is conserved, 
$\Jpsi\rho$ does not directly decay from the $c\bar c$ core,
but $\Jpsi\omega$ may.
We estimate the effects of existence of such a process
by introducing a direct coupling between $\Jpsi\omega$-$c\bar c$. 
Suppose the $\Jpsi\omega$-$c\bar c$ coupling occurs by the same potential, $u$,
there is a bound state with a binding energy of 10 MeV
 below the $D^0\Dbar{}^{*0}$ threshold.
But as seen in 
 \fref{fig:X3872omegadecay} and as expected, 
almost all the decay fraction is $\Jpsi \omega$ and not  $\Jpsi \rho$  in that case,
which is excluded by experiment.
As far as $X$(3872) is concerned, no direct coupling seems to occur between
the $c\bar c$ core and the $J/\psi\omega$ channel. 
One interesting point of this trial calculation is that
the peak energy approaches the threshold 
as a less pronounced peak
when the width of the $\omega$ meson is enlarged by hand.
Therefore, exotic hadrons which appears at around a two-meson threshold 
and which contain meson(s) with a large decay width
should be examined carefully.
Let us make one more comment on the direct decay from the $c\bar c$ core.
The $\chi_{c1}(2P)$ peak may not be seen 
in the $D\Dbar{}^*$ decay spectrum, 
but it may be seen from the radiative decay.
There is a selection rule of $E1$ transition 
that reduces the fraction $\chi_{c1}(2P) \to J/\psi \gamma $ but not 
$\to \psi(2S)\gamma$, which may show clearly the existence of $\chi_{c1}(2P)$ \cite{Takizawa:2014nma,Takeuchi:2016hat},
and if so, support the $c\bar c$ admixture of $X$(3872)
discussed in this section.

Let us summarize the features of $X$(3872) and the $B$-decay to the final two-meson spectrum obtained in \sref{sec:X3872}.
The two-meson and the $c\bar c$ hadron model which consists of 
the $c\bar c(1^{++})$ core, $D^{(*)0}\Dbar{}^{(*)0}$ $({}^3S,{}^3D,{}^5D)$, $D^{(*)\pm} D^{(*)\mp}({}^3S,{}^3D,{}^5D)$, $\Jpsi\omega$ and $\Jpsi\rho^0$,
the following features are obtained: 
\begin{itemize}
\item 
$X$(3872) is a very shallow bound state or a virtual state which is close to the $D^0\bar D{}^{*0}$ threshold, which are very difficult to distinguish each other.
\item 
The state is molecular, mostly $D^0\bar D{}^{*0}$ in the long range region,
but has a considerable $D^\pm\bar D{}^{*\mp}$ component at the short distance. 
\item 
Two kinds of channel couplings provides attraction for $X$(3872):
one is OPEP tensor, which mixes $D^{(*)}\Dbar^{(*)}$ $S$-$D$-waves,
the other is the $c\bar c$-$D\bar D{}^*$ coupling.
These two effects are comparable in size.
\item
The amount of the $c\bar c$ component is found to be about 6 \%
in the model which contains both of the OPEP and the $c\bar c$-couplings,
which meets the requirement from the production rate in the $p\bar p$ experiments.
\item 
When considering the whole energy spectrum of the $B$ weak decay,
there is one very narrow peak at the $D^0\bar D{}^{*0}$ threshold,
but not around the $D^\pm\bar D{}^{*\mp}$ threshold, nor around the $c\bar c(1^{++})$ bare mass.
\item 
Among the final products, amounts of $D\Dbar^*$'s is the largest, which are produced directly from $c\cbar$ core.
There are small amount of  $\Jpsi\omega$ and  $\Jpsi\rho$ final product,
which are comparable to each other.
\item
The spectrum of the $\Jpsi\omega$ and 
the $\Jpsi\rho$ final products makes a very narrow peak at the bound state energy, if a bound state exists,
or at the $D^0\bar D{}^{*0}$ threshold, if not.
For instance, according to \fref{fig:X3872rhodecay}, the width is around 0.1 MeV when the binding energy 
is 0.5 MeV.
\end{itemize}

\section{Charged heavy mesons}
\label{sec:charged}

\subsection{Brief overview in experimental status}
\label{sec:charged_overview}

Here we overview current experimental status about charged exotic mesons.
We have summarized major experimental results of the charged charmonium-like and bottomonium-like exotic candidates in Table \ref{tab:charged_exotics}.
\begin{table*}[tbp]
\begin{center}
\begin{tabular}{lllllll}
\hline
State & Mass [MeV] & Width [MeV] &  Reaction  & Exp. (Year) \\
\hline
$Z_c(3900)^\pm$ & $3899.0 \pm 3.6 \pm 4.9$ & $46 \pm 10 \pm 20$ &  $Y(4260) \to \pi^+ \pi^- J/\psi$ &  BESIII(2013) \\
 & $3894.5 \pm 6.6 \pm 4.5$ & $63 \pm 24 \pm 26$ & $Y(4260) \to \pi^+ \pi^- J/\psi$ &  Belle(2013) \\
 & $3886 \pm 4 \pm 2$ & $37 \pm 4 \pm 8$ & $\psi(4160) \to \pi^+ \pi^- J/\psi$ & CLEO-c(2013) \\
 & $3883.9 \pm 1.5 \pm 4.2$ & $24.8 \pm 3.3 \pm 11.0$ & $Y(4260) \to (D \bar D^\ast)^\pm \pi^\mp$ &  BESIII(2014) \\
 & $3895.0 \pm 5.2 \pm ^{+4.0}_{-2.7}$ & & $H_b \to J/\psi \pi^+ \pi^- + X$ & D0(2018) \\
$Z_c(3900)^0$ & $3894.8 \pm 2.3 \pm 3.2$ & $29.6 \pm 8.2 \pm 8.2$ &  $Y(4260) \to \pi^0 \pi^0 J/\psi$ &  BESIII(2015) \\ 
 & $3885.7^{+4.3}_{-5.7} \pm 8.4$ & $35^{+11}_{-12} \pm 15$ & $Y(4260) \to (D \bar D^\ast)^0 \pi^0$ &  BESIII(2015) \\
$Z_c(4020)^\pm$ & $4022.9 \pm 0.8 \pm 2.7$ & $7.9 \pm 2.7 \pm 2.6$ & $e^+ e^- \to \pi^+ \pi^- h_c$ &  BESIII(2013) \\
$Z_c(4025)^\pm$ & $4026.3 \pm 2.6 \pm 3.7$ & $24.8 \pm 2.6 \pm 7.7$ &  $Y(4260) \to (D^\ast \bar D^\ast)^\pm \pi^\mp$ &  BESIII(2014) \\
$Z_c(4025)^0$ & $4025.5^{+2.0}_{-4.7} \pm 3.1$ & $23.0 \pm 6.0 \pm 1.0$ &  $Y(4260) \to (D^\ast \bar D^\ast)^0 \pi^0$ &  BESIII(2015) \\
$Z_c(4050)^+$ & $4051 \pm 14^{+20}_{-41}$ & $82^{+21}_{-17}{}^{+47}_{-22}$ &  $\bar B^0 \to \chi_{c1}(1P) K^- \pi^+$ &  Belle(2008) \\
$Z_c(4200)^+$ & $4196^{+31}_{-29}{}^{+17}_{-13}$ & $370^{+70}_{-70}{}^{+70}_{-132}$ &  $\bar B^0 \to J/\psi K^- \pi^+$ &  Belle(2014) \\  
$Z_c(4250)^+$ & $4248^{+44}_{-29}{}^{+180}_{-35}$ & $177^{+54}_{-39}{}^{+316}_{-61}$ & $\bar B^0 \to \chi_{c1}(1P) K^- \pi^+$ & Belle(2008) \\
$Z_c(4430)^+$ & $4433 \pm 4 \pm 2$ & $45^{+18}_{-13}{}^{+30}_{-13}$ & $B \to K \pi^+ \psi(2S)$ & Belle(2008) \\
$Z_c(4430)^-$ & $4485^{+22+28}_{-22-11}$ & $200^{+41+26}_{-46-35}$ & $B^0 \to K^+ \pi^- \psi(2S)$ & Belle(2013) \\
$Z_c(4430)^-$ & $4475 \pm 7^{+15}_{-25} $ & $172 \pm 13^{+37}_{-34}$ & $B^0 \to K^+ \pi^- \psi(2S)$ & LHCb(2014) \\
$Z_b(10610)^\pm$ & $10607.2 \pm 2.0$ & $18.4 \pm 2.4$ & $\Upsilon(5S) \to \pi^+ \pi^- \Upsilon(1,2,3S)$ &  Belle(2012) \\
 & & & $\Upsilon(5S) \to \pi^+ \pi^- h_b(1,2P)$ &   \\
$Z_b(10610)^0$ & $10609 \pm 4 \pm 4$ & 18.4 (input) & $\Upsilon(5S) \to \pi^0 \pi^0 \Upsilon(2,3S)$ & Belle(2013) \\
$Z_b(10650)^\pm$ & $10652.2 \pm 1.5$ & $11.5 \pm 2.2$ & $\Upsilon(5S) \to \pi^+ \pi^- \Upsilon(1,2,3S)$ &  Belle(2012) \\
 & & & $\Upsilon(5S) \to \pi^+ \pi^- h_b(1,2P)$ &   \\
\hline
\end{tabular}
\end{center}
\caption{
Experimental status of the charged charmonium-like and bottomonium-like exotic candidates with the neutral isospin partners.
}
\label{tab:charged_exotics}
\end{table*}
Below, we pickup $Z_{c}(3900)$, $Z_{c}(4200)$, $Z_{c}(4430)$, $Z_b(10610)$ and $Z_b(10650)$, as well-known candidates.
All these states have the quantum numbers $I^{G}(J^{PC})=1^{+}(1^{+-})$.

$Z_{c}(3900)$ with mass $3887.2 \pm 2.3$ MeV and decay width $28.2 \pm 2.6$ MeV was found in the decay of $Y(4260)$:
$Y(4260)\rightarrow \pi^{\mp}Z_{c}(3900)^{\pm}$, 
$Z_{c}(3900)^{\pm} \rightarrow \pi^{\pm} J/\psi$ in BESIII~\cite{Ablikim:2013mio} and was also confirmed by Belle~\cite{Liu:2013dau} and CLEO~\cite{Xiao:2013iha}.
The mass lies about 11 MeV above the $D^{+}\bar{D}^{\ast0}$ (or 12 MeV above the $D^{\ast+}\bar{D}^{0}$) threshold.
The process $e^+ e^- \to \pi^\pm (D \bar D^\ast)^\mp$ at $\sqrt{s} = 4.26$ GeV was studied and the strong threshold enhancement, 
$Z_c(3885)$ with mass $3883.9 \pm 1.5 \pm 4.2$ MeV and width $24.8 \pm 3.3 \pm 11.0$ MeV was observed in~\cite{Ablikim:2013xfr}.
In PDG~\cite{Tanabashi:2018oca}, $Z_c(3885)$ is assumed to be related to $Z_c(3900)$.
Its quantum numbers, $I^{G}(J^{P})=1^{+}(1^{+})$, have been determined in a partial wave analysis of the process $e^+ e^- \to \pi^+ \pi^- J/\psi$ by BESIII~\cite{Collaboration:2017njt}.
The decay $Z_{c}(3900) \rightarrow \pi^{\pm}h_{c}$ was not seen in BESIII~\cite{Ablikim:2013wzq}.

The $\pm1$ electric charge of the $Z_{c}(3900)^{\pm}$, indicating the quark contents $u\bar{d}c\bar{c}$ and $d\bar{u}c\bar{c}$, suggests the existence of the isospin partner, $(u\bar{u}-d\bar{d})c\bar{c}$.
Indeed, a 3.5 $\sigma$ level significance evidence of a neutral $Z_{c}(3900)^{0}$ was reported in~\cite{Xiao:2013iha}.
Later, $Z_{c}(3900)^{0}$ was also observed in $e^{+}e^{-} \rightarrow \pi^{0}Z_{c}(3900)^{0} \rightarrow \pi^{0}\pi^{0}J/\psi$ (10.4 $\sigma$)~\cite{Ablikim:2015tbp} and in $e^{+}e^{-} \rightarrow \pi^{0}Z_{c}(3900)^{0} \rightarrow \pi^{0} (D \bar D^\ast)^0$ 
(greater than 10$\sigma$)~\cite{Ablikim:2015gda}.

Belle observed $Z_{c}(4200)$ via $Z_{c}(4200) \rightarrow J/\psi \pi^{+}$ in $\bar{B}^{0} \rightarrow K^{-}\pi^{+}J/\psi$, and it has mass $4196^{+31}_{-29}{}^{+17}_{-13}$ MeV and decay width $370^{+70}_{-70}{}^{+70}_{-132}$ MeV with the quantum numbers $I^{G}(J^{P})=1^{+}(1^{+-})$~\cite{Chilikin:2014bkk}.  
LHCb found the evidence for $Z_{c}(4200)$ contributions to $\Lambda_b \to J/\psi p \pi^-$ decays~\cite{Aaij:2016ymb}, and 
found some structure where the invariant mass of $J/\psi \pi^-$ is near 4200 MeV in $B^0 \to J/\psi K^+ \pi^-$ decays~\cite{Aaij:2019ipm}.
 
$Z_{c}(4430)$ was first observed in Belle with the mass of $4433 \pm 4 \pm 2$ MeV~\cite{Choi:2007wga}. The present world average of the mass, $4478^{+15}_{-18}$ MeV~\cite{Tanabashi:2018oca}, is about 50 MeV higher than the original value, though the name $Z_c(4430)$ is still used. 
An interesting point is that $Z_c(4430)$ is first observed in the channel including the radially excited state of the charmonium $\psi(2S)$,
not the ground state charmonium $J/\psi$.
Dalitz analysis of $B \rightarrow K \pi \psi(2S)$ decays was performed in~\cite{Mizuk:2009da} and the full amplitude analysis of $B^{0} \rightarrow K^{+}\pi^{-}\psi(2S)$ decays was done in~\cite{Chilikin:2013tch}. As for the quantum numbers of the $Z_c(4430)$, $J^{P}=1^{+}$ were 
favored in~\cite{Chilikin:2013tch} and confirmed by LHCb~\cite{Aaij:2014jqa}. 
LHCb also performed Argand diagram analysis and showed its resonance character.

The decay patterns of $Z_{c}(4430)$ exhibit interesting features:
\begin{eqnarray}
\fl {\cal B}(B^{0}\rightarrow K^{+}Z_c^{-}(4430)) \times {\cal B}(Z_c^{-}(4430)\rightarrow \pi^{-}\psi(2S)) =  
(6.0^{+1.7+2.5}_{-2.0-1.4}) \times 10^{-5}
\end{eqnarray}
from Ref.~\cite{Chilikin:2013tch}, and
\begin{eqnarray}
\fl {\cal B}(\bar{B}^{0}\rightarrow K^{-}Z_c^{+}(4430)) \times {\cal B}(Z_c^{+}(4430)\rightarrow \pi^{+}J/\psi) = 
(5.4^{+4.0+1.1}_{-1.0-0.9}) \times 10^{-6}
\end{eqnarray}
from Ref.~\cite{Chilikin:2014bkk}.
From the above two results, we obtain the branching ratio:
\begin{eqnarray}
   \frac{{\cal B}(Z_c^{-}(4430)\rightarrow \pi^{-}\psi(2S))}{{\cal B}(Z_c^{+}(4430)\rightarrow \pi^{+}J/\psi)}
\simeq 10,
\end{eqnarray}
which indicates that the decay to $\pi\psi(2S)$ is enhanced relative to the decay to $\pi J/\psi$.
This ordering is against the naive intuition; the decay rate to $\pi\psi(2S)$ should be smaller than the decay to $\pi J/\psi$ if only their phase space is considered.
Further investigations will be useful for understanding the internal structure of the $Z_{c}(4430)$.

There are charged bottomonium-like states.
The $Z_{b}^{\pm}(10610)$ and $Z_{b}^{\pm}(10650)$ were found in $\Upsilon(5S) \rightarrow \pi^{+}\pi^{-}\Upsilon(nS)$ ($n=1,2,3$) and $\Upsilon(5S) \rightarrow \pi^{+}\pi^{-}h_{b}(mP)$ ($m=1,2$) by Belle~\cite{Belle:2011aa,Garmash:2014dhx}\footnote{$\Upsilon(10860)$ was regarded as $\Upsilon(5S)$.}.
$Z_{b}^{\pm}(10610)$ has the mass $10607.2\pm2.0$ MeV and the decay width $18.4\pm2.4$ MeV,
and $Z_{b}^{\pm}(10650)$ has the mass $10652.2\pm1.5$ MeV and the decay width $11.5\pm2.2$ MeV.
It is noted that their masses are very close to the $B\bar{B}^{\ast}$ and $B^{\ast}\bar{B}^{\ast}$ thresholds.
The fact that $Z_{b}^{\pm}(10610)$ and $Z_{b}^{\pm}(10650)$ were found in $e^{+}e^{-}\rightarrow\Upsilon(nS)\pi^{+}\pi^{-}$ favors $I^{G}(J^{P})=1^{+}(1^{+})$, as it is expected from the final state when $S$-waves are assumed~\cite{Garmash:2014dhx}.

It is interesting that $Z_{b}^{\pm}(10610)$ and $Z_{b}^{\pm}(10650)$ have a larger probability in the decay to open heavy mesons rather than 
the decay to bottomonia; $Z_{b}^{\pm}(10610) \rightarrow B\bar{B}^{\ast}$ and
$Z_{b}^{\pm}(10650) \rightarrow B^{\ast}\bar{B}^{\ast}$
are the dominant channels with the fractions about 86 \% and 74 \%, respectively~\cite{Garmash:2015rfd}.
For a charged state, $Z_{b}^{\pm}(10610)$ decays to $\pi \Upsilon(nS)$ ($n=1,2,3$) and to 
$\pi h_{b}(mP)$ ($n=1,2$) with the fractions whose values are of the same order.
Situation is the same for the $Z_{b}^{\pm}(10650)$.
This indicates that the heavy quark spin symmetry is violated largely 
for $Z_{b}^{\pm}(10610)$ and $Z_{b}^{\pm}(10650)$~\cite{Bondar:2011ev}.
$Z_{b}^{\pm}(10610)$ and $Z_{b}^{\pm}(10650)$ were also found in $\Upsilon(11020)\rightarrow h_{b}(nP)\pi^{+}\pi^{-}$~\cite{Abdesselam:2015zza}, where the mass of $\Upsilon(11020)$ lies above $\Upsilon(5S)$.
The neutral partner, $Z_{b}^{0}(10610)$, was found in $\Upsilon(5S)\rightarrow \pi^{0}\pi^{0}\Upsilon(nS)$ ($n=1,2,3$) by 
Belle\cite{Krokovny:2013mgx}.
However, no significance signal was obtained for $Z_{b}^{0}(10650)$.

The structures of the charged charmonium-like and bottomoniumu-like states discussed above have been studied theoretically 
in the hadronic-molecular approaches~\cite{Liu:2007bf,Liu:2008xz,Ding:2008mp,Braaten:2007xw,Cleven:2013sq,Yamaguchi:2013ty,Voloshin:2013dpa,
Wilbring:2013cha,Zhao:2014gqa,Aceti:2014uea,He:2014nya,Dias:2014pva,Karliner:2015ina,He:2015mja,Zhou:2015jta,Albaladejo:2015lob,Voloshin:2016cgm,
Kang:2016ezb,Coito:2016ads,Goerke:2016hxf,Gong:2016hlt,He:2017lhy,Gong:2016jzb,Wang:2018jlv}, 
in the hadrocharmonium approach~\cite{Voloshin:2018vym}, 
in the QCD sum rule approaches~\cite{Lee:2007gs,Bracco:2008jj,Zhang:2013aoa,Wang:2013vex,Cui:2013yva,Wang:2013daa,Wang:2013zra,Chen:2015ata,
Agaev:2016dev,Agaev:2017tzv,Ozdem:2017jqh,Wang:2017lot,Wang:2017dce}, 
in the tetraquark approaches~\cite{Ali:2011ug,Faccini:2013lda,Braaten:2013boa,Braaten:2014qka,Deng:2014gqa,Zhao:2014qva,Deng:2015lca,Deng:2017xlb,
Anwar:2018sol,Wu:2018xdi}, 
in the heavy quark spin symmetry approaches~\cite{Bondar:2011ev, Guo:2013sya,Cleven:2015era,Baru:2017gwo} 
and in the lattice QCD approaches~\cite{Meng:2009qt,Prelovsek:2013xba,Chen:2014afa,Prelovsek:2014swa,Albaladejo:2016jsg,Ikeda:2016zwx,Ikeda:2017mee}.
The production and decay processes of the charged charmonium-like and bottomoniumu-like states have been studied 
in~\cite{Liu:2008qx,Liu:2008yy,Matsuki:2008gz,Branz:2010sh,Li:2012uc,Ohkoda:2012rj,Dong:2012hc,Li:2012as,Voloshin:2013ez,Wang:2013hga,Liu:2013vfa,
Dong:2013iqa,Chen:2013bha,Chen:2013coa,Chen:2013wca,Dong:2013kta,Liu:2014eka,Gutsche:2014zda,Wang:2015pfa,Swanson:2014tra,Guo:2014iya,Wang:2015lwa,
Szczepaniak:2015eza,Pakhlov:2014qva,Bediaga:2015tga,Chen:2015fsa,Liu:2015taa,Chen:2015igx,Huang:2015xud,Chen:2015jgl,Abreu:2016xlr,Huo:2015uka,
Goerke:2017svb,Voloshin:2017gnc,Yang:2017nde,Agaev:2017lmc,Voloshin:2018pqn,Nakamura:2019emd,Nakamura:2019btl,Wu:2019vbk,Wang:2018pwi}.

\subsection{Isovector $P^{(\ast)}\bar{P}^{(\ast)}$ molecule with OPEP} 
\label{sec:ZcZbOPEP}
In this subsection, 
we study 
how the $Z_c$ and $Z_b$ states are generated as
isovector $P^{(\ast)}\bar{P}^{(\ast)}$ 
molecular state with the OPEP.
We focus on the states with $J^{PC}=0^{++},$ $1^{+-}$ and $1^{++}$, where
the $S$-wave $P^{(\ast)}\bar{P}^{(\ast)}$ component is included.
Among them, $J^{PC}=1^{+-}$ is assigned as the quantum number of $Z_c(3900)$, $Z_c(4200)$, and $Z_c(4430)$.
The $J^{PC}=1^{++}$ state has not been reported, but it is the isospin partner of $X(3872)$.

The components of the isovector $P^{(\ast)}\bar{P}^{(\ast)}$ states for 
$J^{PC}=0^{++},1^{++}$, and $1^{+-}$ are given by~\cite{Tornqvist:1993ng,Ohkoda:2011vj}
\begin{eqnarray}
 0^{++}: && P\bar{P}(^1S_0), P^{\ast}\bar{P}^{\ast} (^1S_0,  {^5D_0}),
  \label{eq:Isov0++} \\
 1^{++}: && 
  \frac{1}{\sqrt{2}}\left(P\bar{P}^\ast-P^\ast\bar{P}\right)
  (^3S_1, {^3D_1}), P^\ast\bar{P}^\ast (^5D_1), \label{eq:Isov1++}  \\
 1^{+-}: && 
  \frac{1}{\sqrt{2}}\left(P\bar{P}^\ast+P^\ast\bar{P}\right)
  (^3S_1, {^3D_1}), P^\ast\bar{P}^\ast (^3S_1, {^3D_1}). \label{eq:Isov1+-}
\end{eqnarray}
The lowest threshold of the $J^{PC}=0^{++}$ state is $P\bar{P}$, while 
$P\bar{P}^\ast$ and $P^{\ast}\bar{P}$ are the lowest thresholds for $J^{PC}=1^{++}$ and $1^{+-}$.
 We note that the phase convention in~\eref{eq:Isov0++}-\eref{eq:Isov1+-}
 is different from the one in the literatures~\cite{Tornqvist:1993ng,Ohkoda:2011vj} 
 as discussed in \sref{sec:quark-hadronic-models}.

In the basis \eref{eq:Isov0++}-\eref{eq:Isov1+-},
the matrix elements of the OPEP in~\eref{eq_VPPtoP*P}-\eref{eq_VP*P*toP*P*}
are given by~\cite{Tornqvist:1993ng,Ohkoda:2011vj}
\begin{eqnarray}
\fl V^{0^{++}}_{\pi}(r) &=& \left(\frac{g_A}{2f_\pi}\right)^2\frac{1}{3}\left(
		     \begin{array}{ccc}
		      0& -\sqrt{3}{\cal C}_\pi&\sqrt{6}{\cal T}_\pi \\
		      -\sqrt{3}{\cal C}_\pi&2C_\pi &\sqrt{2}T_\pi \\
		      \sqrt{6}{\cal T}_\pi&\sqrt{2}T_\pi &-C_\pi+2T_\pi \\
		     \end{array}
		    \right),   \label{eq:OPEP-Zc0++}  \\
\fl V^{1^{++}}_\pi(r)&=&\left(\frac{g_A}{2f_\pi}\right)^2\frac{1}{3}\left(
	\begin{array}{ccc}
	 -{\cal C}_\pi&\sqrt{2}{\cal T}_\pi &\sqrt{6}T_\pi \\
	 \sqrt{2}{\cal T}_\pi& -{\cal C}_\pi-{\cal T}_\pi& \sqrt{3}T_\pi\\
	 \sqrt{6}T_\pi& \sqrt{3}T_\pi& -C_\pi+T_\pi\\			 
	\end{array}
   \right),   \label{eq:OPEP-Zc1++}  \\
\fl V^{1^{+-}}_\pi(r)&=&\left(\frac{g_A}{2f_\pi}\right)^2\frac{1}{3} \left(
		      \begin{array}{cccc}
		       {\cal C}_\pi& -\sqrt{2}{\cal T}_\pi& -2C_\pi& -\sqrt{2}T_\pi\\
		       -\sqrt{2}{\cal T}_\pi&{\cal C}_\pi+{\cal T}_\pi &-\sqrt{2}T_\pi& -2C_\pi+T_\pi \\
		       -2C_\pi& -\sqrt{2}T_\pi& C_\pi& -\sqrt{2}T_\pi\\
		       -\sqrt{2}T_\pi& -2C_\pi+T_\pi& -\sqrt{2}T_\pi&C_\pi+T_\pi \\
		      \end{array}
						\right), \label{eq:OPEP-Zc1++}
\end{eqnarray}
where
the functions ${\cal C}_\pi$, ${\cal T}_\pi$, $C_\pi$ and $T_\pi$ are given in~\eref{eq:OPEP-X3872}.
The potential form in the isovector channel is almost 
the same as the one in the isoscalar channel,
while the isospin factors are different,
${\btau}_1\cdot{\btau}_2=+1$ for $I=1$ and ${\btau}_1\cdot{\btau}_2=-3$ for $I=0$, 
compare \eref{eq:OPEP-Zc1++} and \eref{eq:OPEP-X3872}.

As in~\sref{sec:DDbar_molecule_with_OPEP}, the $P^{(\ast)}\bar{P}^{(\ast)}$ 
state is studied by solving the coupled channel Schr\"odinger equation with 
the Hamiltonian
\begin{eqnarray}
 H^{J^{PC}} &=& K^{J^{PC}} + V^{J^{PC}}_\pi .
\end{eqnarray}
The kinetic term is given by
\begin{eqnarray}
\fl K^{0^{++}} &=& {\rm diag}\left(
			   -\frac{1}{2\mu_{P\bar{P}}}\Delta_0, 
			   -\frac{1}{2\mu_{P^\ast\bar{P}^\ast}}\Delta_0 + 2\Delta m_{PP^{\ast}}, 
			   -\frac{1}{2\mu_{P^\ast\bar{P}^\ast}}\Delta_2 + 2\Delta m_{PP^{\ast}}, 
			  \right) \\
\fl K^{1^{++}} &=& {\rm diag}\left(
			   -\frac{1}{2\mu_{P\bar{P}^\ast}}\Delta_0, 
			   -\frac{1}{2\mu_{P\bar{P}^\ast}}\Delta_2, 
			   -\frac{1}{2\mu_{P^\ast\bar{P}^\ast}}\Delta_2 + \Delta m_{PP^{\ast}}, 
			  \right) \\
\fl K^{1^{+-}} &=& {\rm diag}\left(
			   -\frac{1}{2\mu_{P\bar{P}^\ast}}\Delta_0, 
			   -\frac{1}{2\mu_{P\bar{P}^\ast}}\Delta_2, 
			   -\frac{1}{2\mu_{P^\ast\bar{P}^\ast}}\Delta_0 + \Delta m_{PP^{\ast}}, 			    
			  \right.  \nonumber \\
\fl  && \left. -\frac{1}{2\mu_{P^\ast\bar{P}^\ast}}\Delta_2 + \Delta m_{PP^{\ast}} \right),
\end{eqnarray}
where $\mu_{P^{(\ast)}\bar{P}^{(\ast)}}$, $\Delta_{\ell}$ and $\Delta m_{P^{(\ast)}\bar{P}^{(\ast)}}$ are defined 
in \eref{eq:Delta-ell}-\eref{eq:delta-m}, and 
the masses of $P^{(\ast)}=D^{(\ast)}, B^{(\ast)}$ are 
given in~\tref{table:Hadronmass}.

\begin{figure}[t]  
   \begin{center}
    $J^{PC}=0^{++}$\\
    \includegraphics[width=7cm,clip]{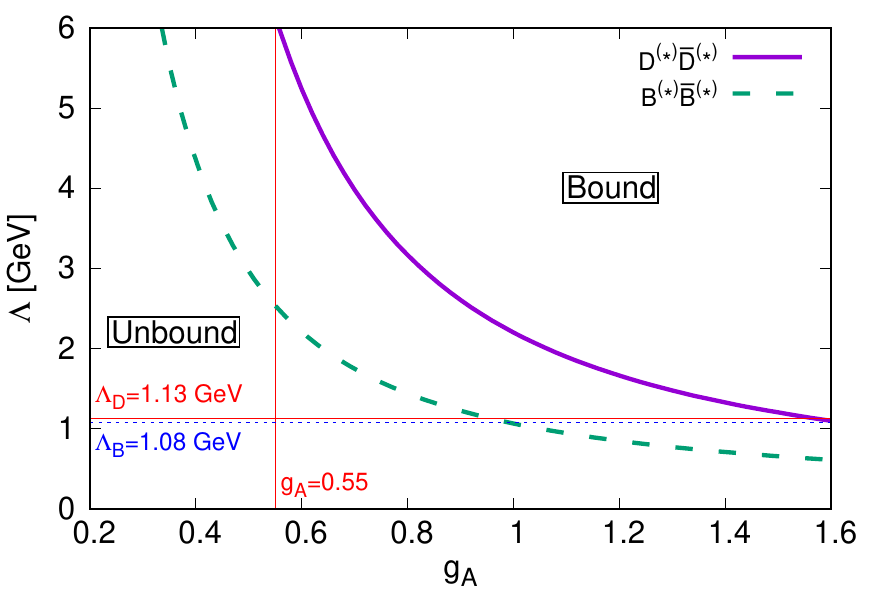}\\
     \begin{tabular}{cc}
     $J^{PC}=1^{++}$ & $J^{PC}=1^{+-}$ \\
     \includegraphics[width=7cm,clip]{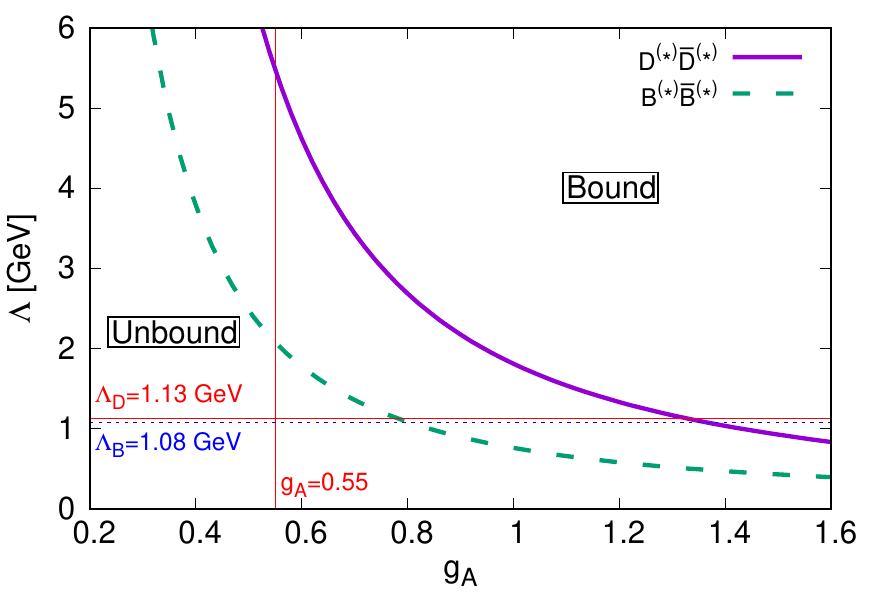} & 
	  \includegraphics[width=7cm,clip]{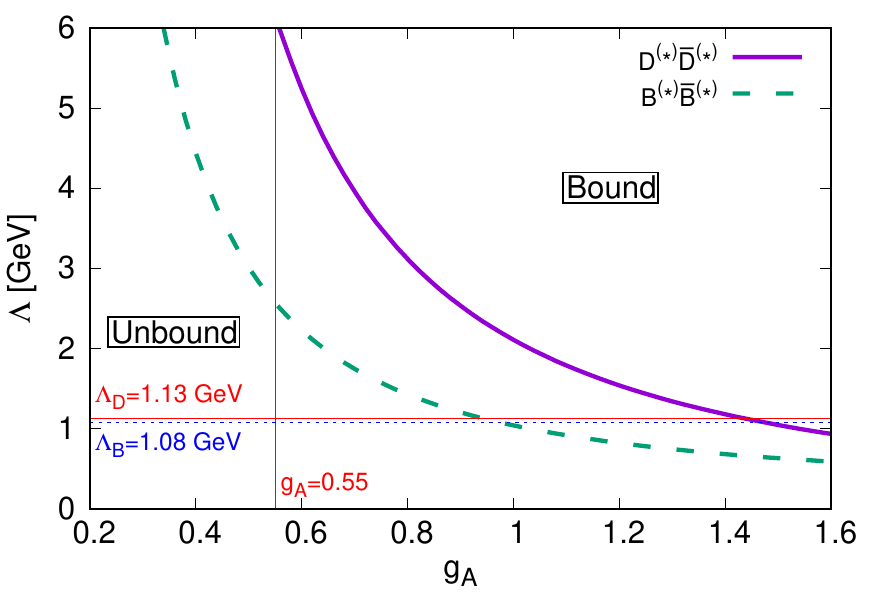} \\
     \end{tabular}
   \end{center}
   \caption{\label{fig:Zc_0++1++1+-}
   The boundary lines of the isovector 
   $P^{(\ast)}\bar{P}^{(\ast)}$ 
   bound states in the $(g_A,\Lambda)$ 
   plane for $J^{PC}=0^{++}$, $1^{++}$ and $1^{+-}$.
   The results of the $D^{(\ast)}\bar{D}^{(\ast)}$ and $B^{(\ast)}\bar{B}^{(\ast)}$ states are shown 
   by the solid and dashed lines, respectively.
   The right side beyond the line 
   is the bound region, 
   while the left side is the unbound region. 
   The vertical solid lines is the values at $g_A=0.55$,
   while the horizontal solid and dashed lines are the values 
   at $\Lambda=\Lambda_D=1.13$ GeV and $\Lambda=\Lambda_B=1.08$ GeV, respectively.   
   }
  \end{figure}

As studied in~\sref{sec:DDbar_molecule_with_OPEP}, 
we search the parameter region which gives a bound state
by varying 
the parameters $g_A$ and $\Lambda$.
As a result, the boundary of the isovector $D^{(\ast)}\bar{D}^{(\ast)}$  and $B^{(\ast)}\bar{B}^{(\ast)}$ 
bound states in the $(g_A, \Lambda)$ plane is shown in~\fref{fig:Zc_0++1++1+-}.
The results for $J^{PC}=0^{++}$ and $1^{+-}$ are similar, while
we note that the lowest thresholds are different, $P\bar{P}$ for $J^{PC}=0^{++}$, and 
$P\bar{P}^{\ast}$ $(P^{\ast}\bar{P})$ for $J^{PC}=1^{+-}$.
The bound region of the $J^{PC}=1^{++}$ is slightly larger than the others,
and hence the attraction for $J^{PC}=1^{++}$ 
is larger than that for $J^{PC}=0^{++}$ and $1^{+-}$.
Comparing the results of the $D^{(\ast)}\bar{D}^{(\ast)}$ and $B^{(\ast)}\bar{B}^{(\ast)}$ states,
the $B^{(\ast)}\bar{B}^{(\ast)}$ bound region is 
wider
than the $D^{(\ast)}\bar{D}^{(\ast)}$ one.
This is because 
the heavier mass suppresses the kinetic energy, and 
because the the small $BB^{\ast}$ mass splitting enhances the attraction from the coupled channel effect.
For the reference point 
$(g_A,\Lambda)=(0.55,1.13\,{\rm GeV})$,
no bound state is found for the isovector channels of both the charm and bottom sectors.
To accommodate bound states, we need larger $g_A$ and/or $\Lambda$. 
In fact, our previous choice of the overestimated coupling strength corresponds to the vertical line $g_A \sim 0.83 = \sqrt{2} \times 0.59$~\cite{Ohkoda:2011vj} 
(see the footnote\footref{note:sqrt2} in page \pageref{note:sqrt2})
which allowed a shallow bound state for the isovector $B^{(\ast)} \bar B^{(\ast)}$ channel. 
When we 
have only the OPEP, 
larger $g_A$ or $\Lambda$ is 
needed to produce a isovector 
$P^{(\ast)}\bar{P}^\ast$ bound state.

\begin{figure}[t]  
   \begin{center}
    \includegraphics[width=11cm,clip]{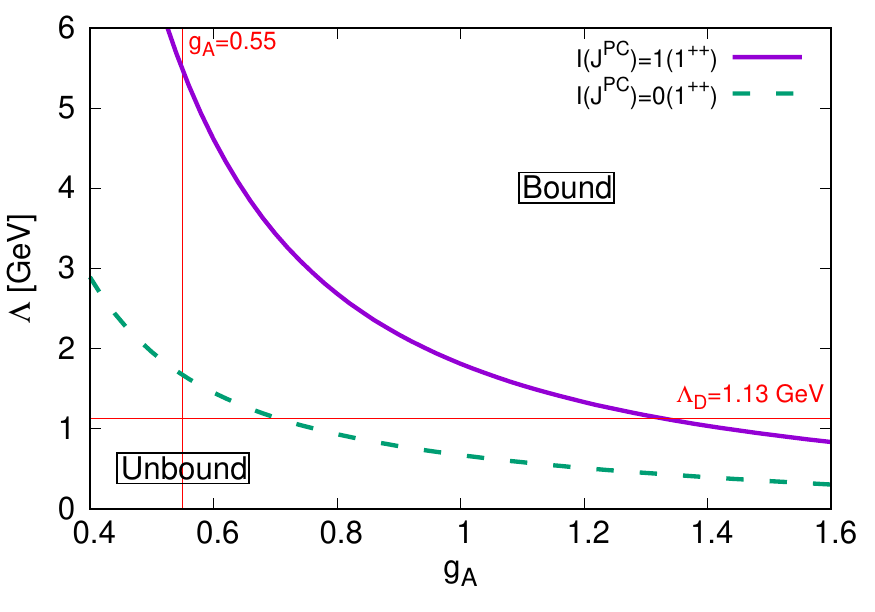}    
   \end{center}
   \caption{\label{fig:XandZc_1++}
   The boundary lines of the $D^{(\ast)}\bar{D}^{(\ast)}$ bound states $J^{PC}=1^{++}$ 
   in the $(g_A,\Lambda)$ plane for the isospin $I=1$ and $0$.
   The solid and dashed lines show the results for $I=1$ 
   and for $I=0$, 
   respectively, which are obtained in \fref{fig:Zc_0++1++1+-} and \fref{fig:X3872-ulim-1}.
   The right side beyond the line 
   is the bound region, 
   while the left side is the unbound region. 
   The vertical and horizontal solid lines 
   show the values at $g_A=0.55$  and
   at $\Lambda=\Lambda_D=1.13$ GeV, respectively.   
   }
  \end{figure}

  Finally, we compare the results for the isovector and isoscalar channels.
  In \fref{fig:XandZc_1++}, 
  the boundary lines of the $D^{(\ast)}\bar{D}^{(\ast)}$ bound states for $I(J^{PC})=I(1^{++})$  ($I=0,1$) are shown, 
  which were obtained in~\fref{fig:Zc_0++1++1+-} for $I=1$ and in \fref{fig:X3872-ulim-1} for $I=0$.
  As seen in \fref{fig:XandZc_1++},
  the bound region for $I=0$ is obviously larger than that for $I=1$.
  This fact indicates that the attraction in the $I=0$ channel is stronger than that in the $I=1$ channel.
  The difference between them
  comes from the isospin dependence of the OPEP,
  which is given by 
  the isospin factor $\btau_1\cdot\btau_2$ in~\eref{eq_VPPtoP*P}-\eref{eq_VP*P*toP*P*};
  ${\btau}_1\cdot{\btau}_2=-3$
  for the isoscalar channel, while ${\btau}_1\cdot{\btau}_2=+1$ for the isovector channel.
  For the OPEP, the tensor force in the off-diagonal term has the dominant role to produce an attraction 
  rather than the diagonal term.
  For the off-diagonal term, the sign of the potential is not important, but the magnitude is important because 
  the off-diagonal term contributes as the second order of the perturbation.
  Thus, the attraction in the isoscalar channel with $|{\btau}_1\cdot{\btau}_2|=3$ is larger than 
  that in the isovector one with $|{\btau}_1\cdot{\btau}_2|=1$ by about a factor $9$.

\subsection{$P^{(\ast)}\bar{P}^{(\ast)}$ molecule with OPEP and $\sigma$ exchange potential}
\label{sec:ZcZbOPEP-sigma}
In the previous subsection, we have seen that the OPEP contribution is rather small in the isovector channel.
In such a situation, the short and middle range interactions may become important.
Such interaction includes $\sigma$, $\rho$ and $\omega$ exchange interactions~\cite{Chen:2017vai}. 
In the isoscalar $P^{(\ast)}\bar{P}^{(\ast)}$ channel for $X(3872)$, 
we have not considered them because of 
a reason discussed 
below. 
In the isovector channel, there is another reason that 
we expect that 
the vector meson exchanges 
are not important; 
$\rho$ and $\omega$ exchange interactions have opposite signs and 
mostly cancel. 

The $\sigma$ exchange potential may be effective not only for the isovector channel but also for the $X(3872)$ isoscalar channel. 
One of the reasons that we have considered only OPEP 
in the previous sections 
is that the effect of the short range interaction 
including the $\sigma$ exchange has been effectively taken care of by the suitable choice of the cutoff parameter $\Lambda$.
To determine the reference value of $\Lambda$, we have used the binding energy of the deuteron. 
The OPEP thus determined for the nucleon-nucleon interaction is extrapolated to that of $D\bar{D}^\ast$ by assuming 
hadron structures by constituent quarks and the pion coupling to the light quarks. 
Strictly, however, we do not know to what extent such extrapolation works.
Therefore, in this subsection we consider the role of $\sigma$-exchange interaction in some detail.
Analysis here also provides an estimate of ambiguities coming from short range interactions.

The $\sigma$ exchange potential is derived by the 
effective Lagrangian of the heavy and 
$\sigma$ mesons~\cite{Bardeen:2003kt}, 
\begin{eqnarray}
 {\cal L}_{\sigma HH}=g_\sigma {\rm Tr}\left[H \sigma \bar{H}\right].
  \label{eq:LsigmaHH}
\end{eqnarray}
From 
this Lagrangian, the $\sigma$ exchange potential 
is obtained as
\begin{eqnarray}
 V^\sigma_{P\bar{P}{\mathchar`-}P\bar{P}}(r) = -\left(\frac{g_\sigma}{m_\sigma}\right)^2 C(r;m_\sigma,\Lambda), 
\label{eq:sigma-PPbar}\\
 V^\sigma_{P\bar{P}^{\ast}{\mathchar`-}P\bar{P}^\ast}(r) = -\left(\frac{g_\sigma}{m_\sigma}\right)^2
  C(r;m_\sigma,\Lambda), \label{eq:sigma-PPbar*}\\
 V^\sigma_{P^\ast\bar{P}^{\ast}{\mathchar`-}P^\ast\bar{P}^\ast}(r) = -\left(\frac{g_\sigma}{m_\sigma}\right)^2
  C(r;m_\sigma,\Lambda), \label{eq:sigma-P*Pbar*}
\end{eqnarray}
where the $\sigma$ mass $m_\sigma=550$ MeV is used.
The factor $1/m^2_\sigma$ is multiplied because the function $C(r;m_\sigma,\Lambda)$ 
includes $m^2_\sigma$ (see \eref{eq:poteC}). 

There is ambiguity in the value of the coupling constant $g_\sigma$. 
In~\cite{Chen:2017vai}, $g_\sigma=3.65$ is taken, which
is determined by 
a quark model estimation.
This value is one-third of the value of the $\sigma NN$ coupling $g_{\sigma NN}$ according to 
the quark number counting, 
because the $\sigma$ meson couples to the scalar charge of hadrons which is additive. 
Another way to estimate $g_\sigma$ is to use a chiral theory for quarks such 
as the Nambu-Jona-Lasinio model~\cite{Nambu:1961tp,Nambu:1961fr,Hatsuda:1994pi}.
By using the equality of the $\sigma qq$ and $\pi qq$ couplings, and the Goldberger-Treiman relation 
for the constituent quark, we have the relation 
\be
\frac{g_\sigma}{m_q} = \frac{g_A^q}{f_\pi} .
\label{eq_GTquark}
\ee
By using the parameter values 
$m_q \sim 300$ MeV, $f_\pi = 93$ MeV, $g_A^q \sim 0.55$, we find 
$g_\sigma \sim 1.8$.  
Yet, 
in~\cite{Bardeen:2003kt}, 
an even smaller value $g_\sigma=g_\pi/2\sqrt{6}\sim 0.76$ is obtained, where 
$g_\pi=\Delta M/f_\pi$ and $\Delta M$ is the mass difference between
the $0^+$ and $0^-$ heavy-light mesons.
$\Delta M=349$ MeV is used in~\cite{Bardeen:2003kt}, which is 
the mass difference between $D^{\ast +}_{s0}$ and $D^+_s$.
In this subsection, we 
present the results for $g_\sigma=0.76$ and $3.65$, which are regarded as the 
lower and upper limits of the attractive contribution due to the sigma meson exchange potential.  

Using $\Lambda_N=681$ MeV for the $\pi\sigma$ potential 
for the nucleon as shown in~\tref{table:cutoffNN},
$\Lambda_D$ and $\Lambda_B$ are obtained by $919$ MeV and $878$ MeV, respectively.
In the basis of~\eref{eq:Isov0++}-\eref{eq:Isov1++}, 
the matrix elements of the $\sigma$ exchange potential for the given $J^{PC}$ 
are 
obtained by
\begin{eqnarray}
V^{0^{++}}_{\sigma}(r) &=& -\left(\frac{g_\sigma}{m_\sigma}\right)^2\left(
		     \begin{array}{ccc}
		      C_\sigma&0 &0 \\
		      0&C_\sigma &0 \\
		      0&0 &C_\sigma \\
		     \end{array}
		    \right),   \label{eq:sigma-Zc0++}  \\
V^{1^{++}}_\sigma(r)&=&-\left(\frac{g_\sigma}{m_\sigma}\right)^2\left(
	\begin{array}{ccc}
	 C_\sigma&0 &0 \\
	 0&C_\sigma &0 \\
	 0&0 &C_\sigma \\	 
	\end{array}
   \right),   \label{eq:sigma-Zc1++}  \\
V^{1^{+-}}_\sigma(r)&=&-\left(\frac{g_\sigma}{m_\sigma}\right)^2\left(
		      \begin{array}{cccc}
		       C_\sigma&0 &0 &0 \\
		       0&C_\sigma &0 &0 \\
		       0&0 &C_\sigma &0 \\	 
		       0&0 &0 &C_\sigma \\	 		       
		      \end{array}
			\right), \label{eq:sigma-Zc1++}
\end{eqnarray}
with the function $C_\sigma=C(r;m_\sigma,\Lambda)$.

  \begin{figure}[t]  
   \begin{center}
    (i) $J^{PC}=0^{++}$ \\
    \includegraphics[width=7cm,clip]{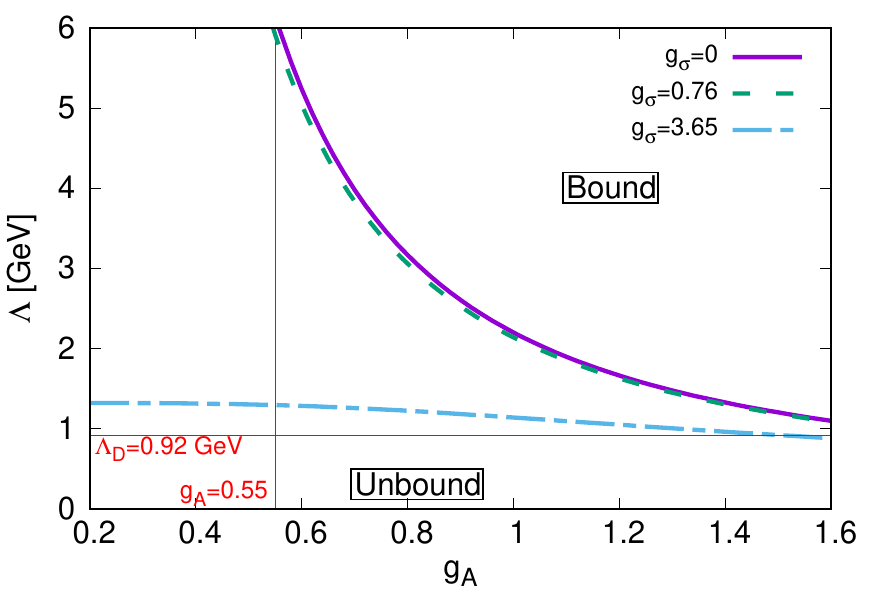} \\
    \begin{tabular}{cc}
     (ii) $J^{PC}=1^{++}$ & (iii) $J^{PC}=1^{+-}$ \\
     \includegraphics[width=7cm,clip]{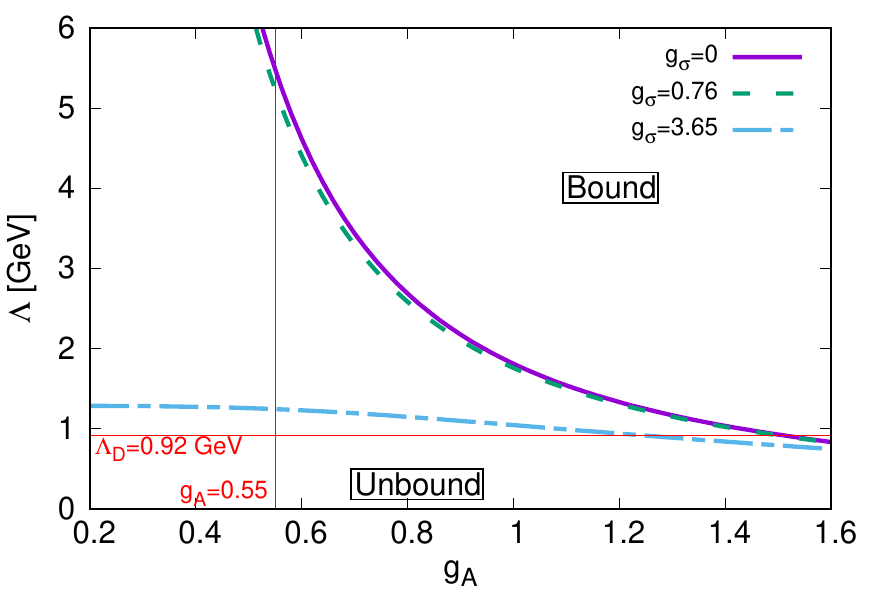} & 
	 \includegraphics[width=7cm,clip]{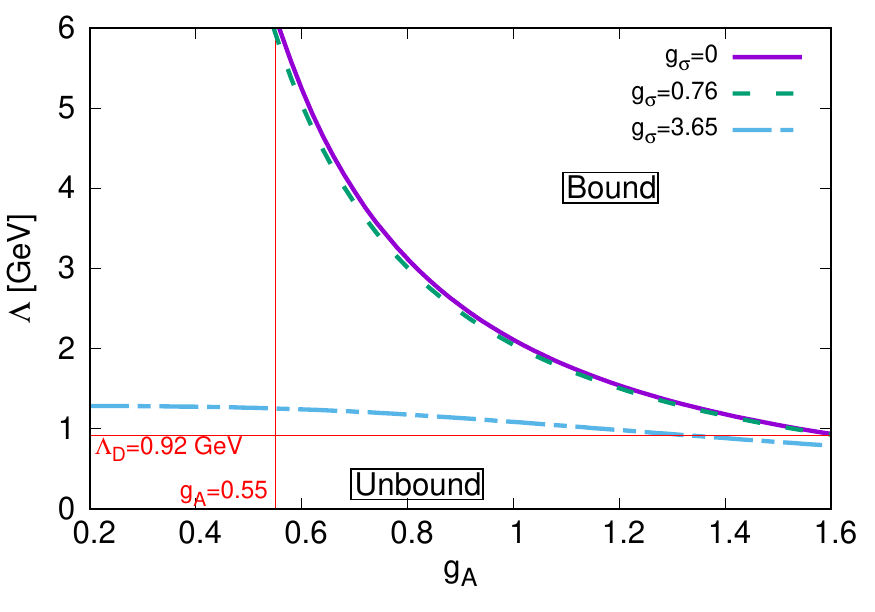} \\
    \end{tabular}    
   \end{center}
   \caption{\label{fig:pi_vs_pi+sigma_Zc}
   The boundary of the isovector $D^{(\ast)}\bar{D}^{(\ast)}$
   bound state with $V_\pi$ and $V_\sigma$ for (i) $J^{PC}=0^{++}$, (ii) $1^{++}$ and (iii) $1^{+-}$ in the $(g_A,\Lambda)$ plane.
   The solid line shows the result 
   for $g_\sigma=0$, namely only with $V_\pi$, while 
   the dashed and dashed-dot lines are the results 
   for $g_\sigma=0.76$ and $3.65$, respectively.
   The right side beyond the boundary line is the bound region, 
   while the left side is the unbound region. 
   The vertical and horizontal solid lines show the values at $g_A=0.55$  and 
   at $\Lambda=\Lambda_D=0.92$ GeV, respectively.   
   }
  \end{figure} 

In~\fref{fig:pi_vs_pi+sigma_Zc}, 
the boundaries of the isovector $D^{(\ast)}\bar{D}^{(\ast)}$ bound states 
are shown 
for the case only with OPEP ($V_\pi$) is used and for the case of the $\pi \sigma$ exchange potential ($V_\sigma$) used in the $(g_A, \Lambda)$ plane.
The result with $g_\sigma=0$ is 
corresponds to 
the one only with the OPEP as
shown in~\fref{fig:Zc_0++1++1+-}.
Since $V_\sigma$
is attractive, the bound region for $g_\sigma\neq 0$
is larger than that for $g_\sigma=0$.
For the small coupling $g_\sigma=0.76$, the boundary is close to the one for $g_\sigma=0$.
Thus the $V_\sigma$ contribution is small, and the OPEP plays the dominant role.
For $g_\sigma=3.65$, however, the bound region is much larger than that for $g_\sigma=0$ and $0.76$.

  \begin{figure}[t]  
   \begin{center}
    (i) $J^{PC}=0^{++}$ \\
    \includegraphics[width=7cm,clip]{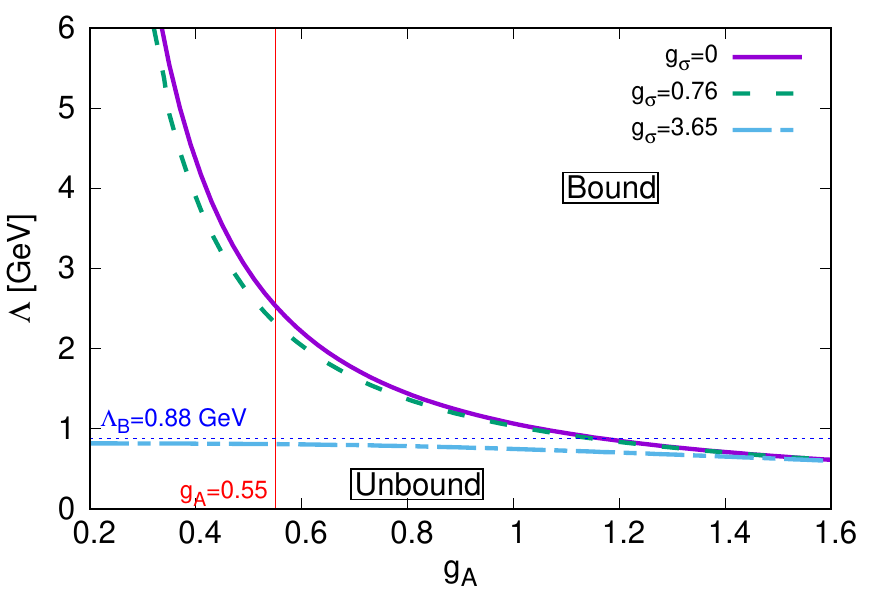}
    \begin{tabular}{ccc}
     (ii) $J^{PC}=1^{++}$ & (iii) $J^{PC}=1^{+-}$ \\
     \includegraphics[width=7cm,clip]{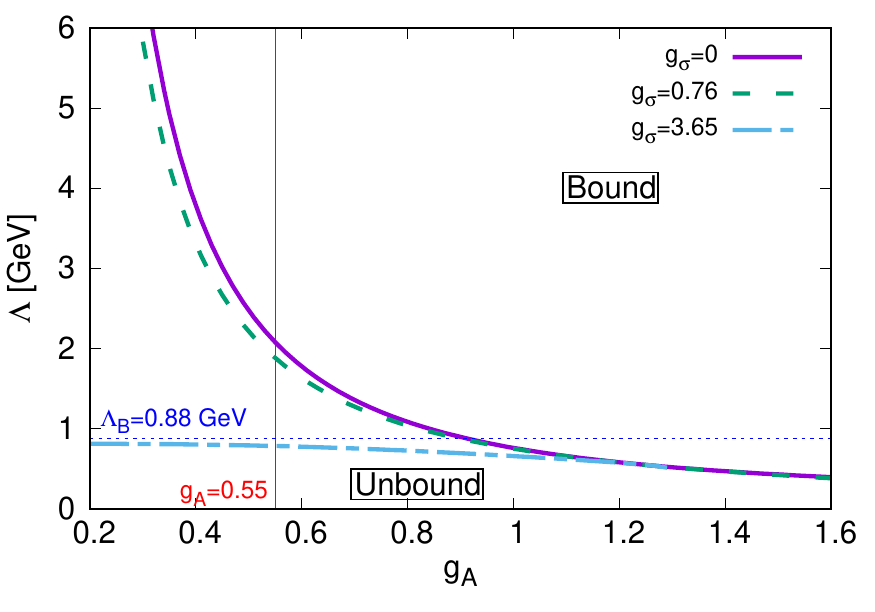} & 
	 \includegraphics[width=7cm,clip]{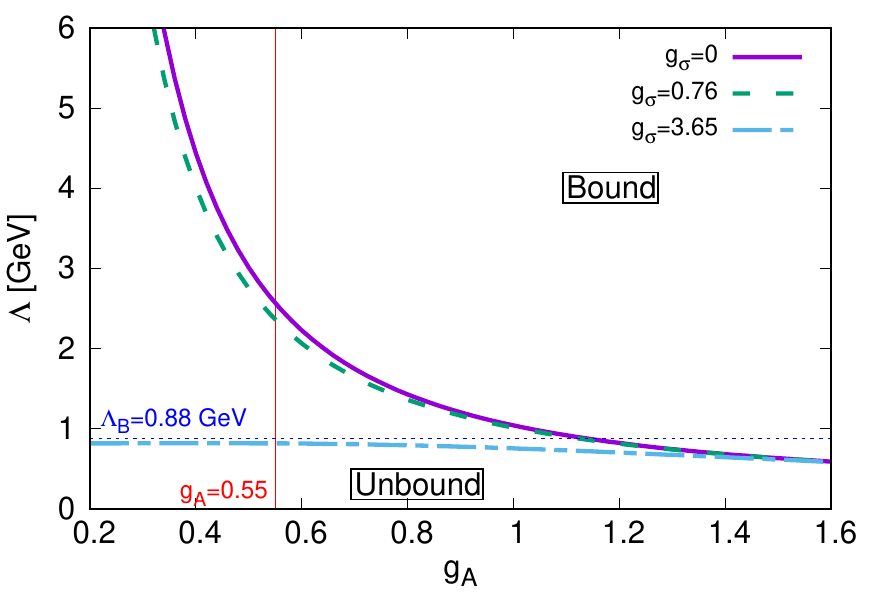} \\
    \end{tabular}    
   \end{center}
   \caption{\label{fig:pi_vs_pi+sigma_Zb}
   The boundary of the isovector $B^{(\ast)}\bar{B}^{(\ast)}$
   bound state for (i) $J^{PC}=0^{++}$, (ii) $1^{++}$ and (iii) $1^{+-}$ in the $(g_A,\Lambda)$ plane.
   The same convention is used as~\fref{fig:pi_vs_pi+sigma_Zc}, while 
   the vertical and horizontal solid lines show the values at $g_A=0.55$  and 
   $\Lambda=\Lambda_B=0.88$ GeV, respectively.   
   }
  \end{figure} 

The results 
for the isovector $B^{(\ast)}\bar{B}^{(\ast)}$ state are summarized in~\fref{fig:pi_vs_pi+sigma_Zb}
for $J^{PC}=0^{++}, 1^{++}$ and $1^{+-}$.
As seen in the $D^{(\ast)}\bar{D}^{(\ast)}$ state,
the bound region of the $B^{(\ast)}\bar{B}^{(\ast)}$ state becomes large as the coupling $g_\sigma$ increases.
The result for $g_\sigma=0.76$ is the similar to the one for $g_\sigma=0$. 
For $g_\sigma=3.65$, the bound region is much larger than those for $g_\sigma=0$ and $0.76$.
For 
$(g_A,\Lambda)=(0.55, 0.88\,{\rm GeV})$, there are bound states for $J^{PC}=0^{++}, 1^{++}$ and $1^{+-}$.
The binding energies are 
$0.92$ MeV for $0^{++}$, 
$1.51$ MeV for $1^{++}$, 
and $0.76$ MeV for $1^{+-}$.  
In experiments, however, 
the charged $Z^{(\prime)}_b$ states 
have been found above the $B^{(\ast)}\bar{B}^{(\ast)}$ threshold
so far.
The attraction generated by the interaction parameters 
$(g_A,\Lambda)=(0.55, 0.88\,{\rm GeV})$
could be overestimated.
In the study of the hadronic molecules,
the uncertainty of the parameters remains a problem, which 
should be addressed.

  \begin{figure}[t]  
   \begin{center}
    \begin{tabular}{ccc}
     (i) $D^{(\ast)}\bar{D}^{(\ast)}$ & (ii) $B^{(\ast)}\bar{B}^{(\ast)}$ \\
     \includegraphics[width=7cm,clip]{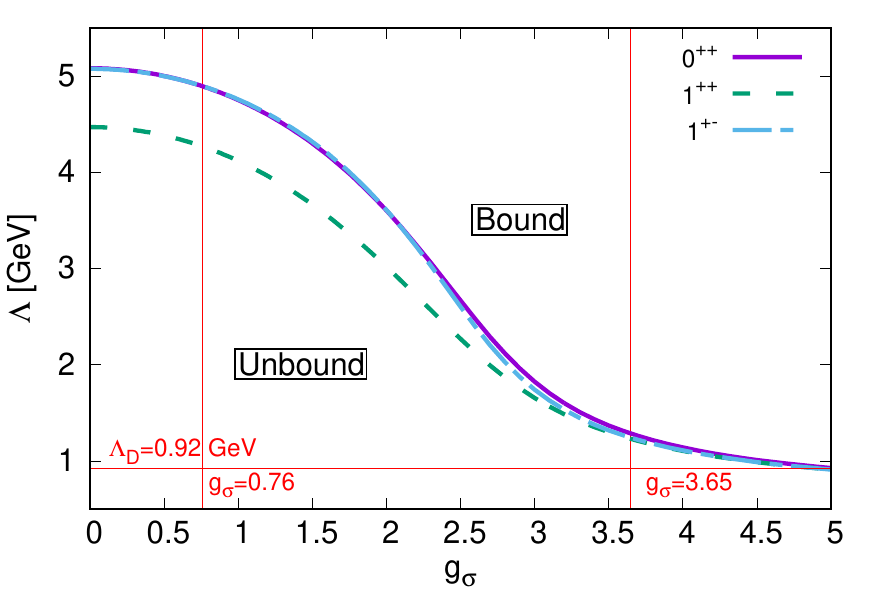} & 
	 \includegraphics[width=7cm,clip]{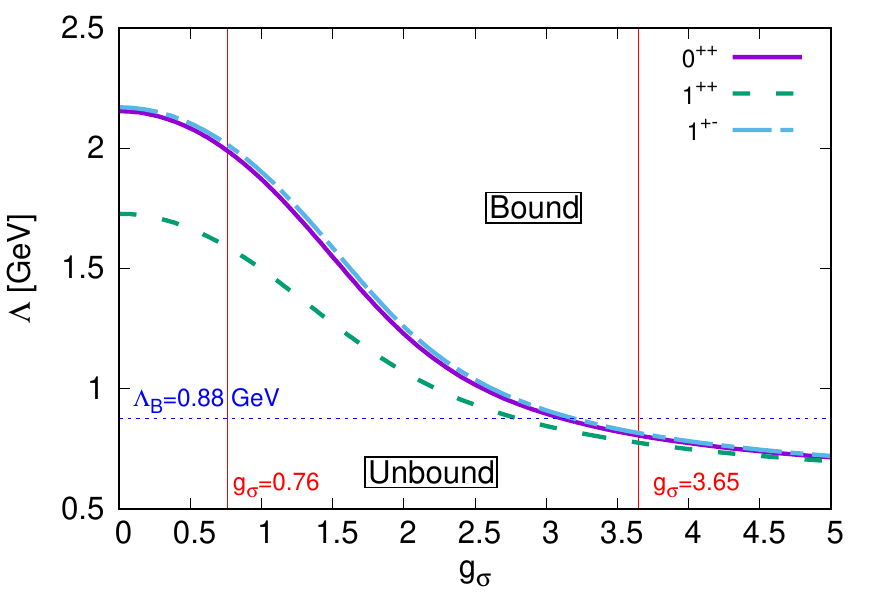} \\
    \end{tabular}    
   \end{center}
   \caption{\label{fig:gs_vs_Lambda_ZcZb}
   The boundary of the isovector (i) $D^{(\ast)}\bar{D}^{(\ast)}$ 
   and (ii) $B^{(\ast)}\bar{B}^{(\ast)}$ bound states
   in the $(g_\sigma,\Lambda)$ plane.
   The vertical lines show the values at $g_s=0.76$ 
   and $g_s=3.65$.
   The horizontal lines show the values 
   at $\Lambda=\Lambda_D=0.92$ GeV 
   for (i) $D^{(\ast)}\bar{D}^{(\ast)}$ and 
   at $\Lambda=\Lambda_B=0.88$ GeV 
   for (ii) $B^{(\ast)}\bar{B}^{(\ast)}$.
   }
  \end{figure} 

In the end, we show 
$g_\sigma$-$\Lambda$ plots to see continuously the change in the role
of $\sigma$ exchange as $g_\sigma$ is varied~\fref{fig:gs_vs_Lambda_ZcZb}.  
For the charm sector, the left figure indicates an unlikely situation for the molecular states to be generated
at the mean value $g_\sigma \lesssim$ 1.8, where a very large cutoff  is needed, $\Lambda \sim 4$ GeV.  
For bottom sector, the molecular states are not yet generated there, but only slight increase of 
$g_\sigma$ will do.

\section{Pentaquark baryons}
\label{sec:summary}

The discussion of exotic hadrons was activated when two new observations were reported in 2003; 
one is the $X(3872)$  which we have discussed in detail in this review, 
and the other the pentaquark $\Theta^+$~\cite{Nakano:2003qx}.  
While the observation of the $X(3872)$  was rather accidental in the  study of 
charmonium spectrum, that of $\Theta^+$ was  motivated
by the theoretical prediction by Diakonov et al~\cite{Diakonov:1997mm}.  
The expected flavor quantum number $S = + 1$ of the $\Theta^+$ requires 
the minimal quark content $uudd\bar s$, 
and hence the genuine multiquark exotic state.  
Various theoretical models have shown very different features, which was thought as an indication 
that we understood hadrons beyond the ground states only poorly.  
As in Ref.~\cite{Diakonov:1997mm}, a model of chiral symmetry with strong pionic soliton correlations predicts
a positive parity state with a relatively low mass at around 1.5 GeV. 
The positive parity is also explained due to the strong pionic spin and isospin correlation~\cite{Hosaka:2003jv}.
One unique feature of this model  is that the decay width is very narrow around 
10 MeV or less.
A model with strong diquark correlations can also make similar predictions, though their mechanisms 
are very much different~\cite{Jaffe:2003sg}.  
Contrary, conventional quark model predicts a negative parity state at relatively large mass around 
1.7 GeV or more and with a wide decay width~\cite{Hosaka:2004bn}.
Therefore, the experimental signals of mass at around 1.5 GeV with a narrow width seemed to support 
the chiral or diquark model.  

After the first observation many data appeared supporting the $\Theta^+$, 
which are, however, followed by a number of experiments with no evidence~\cite{Amsler:2008zzb}.  
There, more experimental data were taken and higher statistics analyses were done.  
In Ref.~\cite{Hicks:2007vg} how statistical fluctuations would have lead no evidence is discussed.  
One cannot, however, definitely conclude that these results have proven that 
the $\Theta^+$ does not exist.  
There are also discussions how multiquark exotics can be seen in some processes 
while not in the others~\cite{Azimov:2007hq}.  
It is particularly so when many experiments utilizes indirect processes induced 
by photons, leptons and hadrons (protons or pions) because 
the direct formation experiments, for instance, $K^+ + n \to \Theta^+$ with a free neutron, is not possible.  
Therefore, we may expect further analyses with improved signal to noise ratio or direct experiments~\cite{WWW}.  

Aside from the $\Theta^+$,  $\Lambda(1405)$ has long been  an exotic baryon candidate 
that is described as a $\bar K N$ molecule~\cite{Oset:1997it,Jido:2003cb,Hyodo:2011ur,Tanabashi:2018oca}.
Its existence was first inferred by the analysis of $\bar K N$ scattering~\cite{Dalitz:1959dn}.  
It is the lowest negative parity hyperon with an excitation energy of about 300 MeV.  
This amount is significantly small as compared to the others.
For instance, the lowest negative parity nucleon $N(1535)$ is about 600 MeV above the ground state nucleon.  
This feature is not easy to be explained by the conventional quark model where baryons are made from three valence quarks.
In Ref.~\cite{Takeuchi:2007tv}, by introducing five quark states corresponding to the molecular channels 
such as $\bar K N$ coupled to the three-quark states, they have shown that the resonance structure at around 1405 MeV was generated.

Whether the picture of hadronic molecule $\bar K N$ crucially depends on the interaction between them.  
In the low energy theorem of chiral symmetry, 
the $\bar K N$ interaction appears attractive as given by the Weinberg-Tomozawa theorem.  
Physically, 
much of attraction strength is provided by  $\omega$ meson exchange.
At the quark level, it originates from the interaction between the antiquark $\bar q$ in the antikaon $\bar K$ 
and the quarks $q$ in the nucleon $N$.  
Due to charge conjugation, the sign of the vector type interaction flips from the one between quarks.  
Such a picture has been shown in the Skyrme model, which is one of successful chiral models 
for baryons~\cite{Callan:1985hy,Callan:1987xt,Ezoe:2017dnp,Ezoe:2016mkp}.  
As a result, a $\bar KN$ bound state appears near the $\bar K N$ threshold.
By coupling to the lower $\pi \Sigma$ channel, the bound state turns into a resonance, which is 
a typical mechanism of a Feshbach resonance.  
An experimental study of the $\bar KN$ molecule is 
going
at J-PARC and analysis is 
underway~\cite{JPARCproposal,Inoue:2017kwd,Kawasaki:2017vxj}.  

Another important topic is the hidden charm $P_c$ baryons observed by LHCb in the weak decay  $\Lambda_b \to J/\! \psi p \bar K$.  
Generally speaking, heavier constituents are more likely to be bound or resonate due to the suppression of their kinetic energies.  
Thus we expect more chances for exotic baryons.  
The first observation was reported in 2015 with a prominent peak structure in the invariant mass plot 
of $J/\! \psi p $~\cite{Aaij:2015tga}.  
In their detailed analysis, they claimed that the peak was generated by two resonant states, one at 4380 MeV and the other at 4450 MeV.  
The analysis has been further performed with more statistics data in 2019, and 
they have reported three narrow peaks~\cite{Aaij:2019vzc}; 
two at 4440 MeV and 4457 MeV that seem to split the strength of the former prominent peak, 
and  one at 4312 MeV that was not seen in the former analysis.
Thus they are denoted as $P_c(4312)^+, P_c(4440)^+$ and $P_c(4457)^+$.  
Interestingly these three peaks are just below the thresholds; 
the higher two below $\Sigma_c \bar D^*$ threshold, and 
the lower one below the $\Sigma_c \bar D$ threshold.  

The new observation has  lead to a number of theory discussions of heavy quark multiplets formed by the four combinations of 
$\Sigma_c,\Sigma_c^{*}$ and  $\bar D,\bar D^{*}$~\cite{Chen:2019bip,Xiao:2019aya,Liu:2019tjn,He:2019ify}.
In the heavy quark limit, the pair of $\bar D$ and $\bar D^{*}$, and the pair of $\Sigma_c$ and $\Sigma_c^{*}$ 
are considered as a spin doublet of $J = 0$ and 1, and the one of $J = 1/2$ and 3/2, respectively.  
Their hadronic molecules also form multiplets of heavy quark spin symmetry.  
In connection with the present discussions, in the formation of these multiplets  
the tensor interaction of OPEP seem to play an important role~\cite{Yamaguchi:2019seo}.  
It not only binds the constituent hadrons of these states but also splits 
the above two $P_c(4440)^+$, $P_c(4457)^+$ as a spin doublet.  

Because the LHCb observed these states in the $J/\! \psi p$ final state, the isospin of these $P_c$ states is $I = 1/2$.  
Therefore their masses are, respectively, 23 MeV and 6 MeV below 
the isospin averaged threshold of $\Sigma_c \bar D^*$, 4463 MeV.  
A relevant question is the origin of the masses, decay widths and quantum numbers of these states.  
Assuming that the orbital motion of the molecules is dominated by $S$-waves, possible total spin values are $J = 1/2$ and 3/2.  
Now the crucial observation is that the tensor interaction can be effective for both states because the $S$-$D$ couplings survive 
for both channels. 
This is understood by the fact that the sum of $S(\Sigma_c) = 1/2, S(\bar D^*) = 1, L = 0$ 
and the sum of $S(\Sigma_c) = 1/2, S(\bar D^*) = 1, L = 2$
can both make the total spin $J = 1/2$ and 3/2.  
This contrasts with the two nucleon system, where two states of $J = 1$ and 0 are possible
while the tensor force is effective only for the $J = 1$ state (corresponding to the deuteron).  
Therefore, the two $P_c$'s provide an interesting opportunity to study the role of the tensor force in the OPEP.

Having said this much, 
in~\cite{Yamaguchi:2019seo} the role of the tensor force in the OPEP 
has been discussed in a coupled channel model of $\Sigma_c^{(*)} \bar D^{(*)}$, $\Lambda_c \bar D^{(*)}$ 
with OPEP supplemented by the short range interaction that is brought 
about by the coupling of the molecular states with compact five quark states~\cite{Yamaguchi:2017zmn}.  
An interesting observation there is that by adjusting one most important model parameter for the short range interaction, 
they have made predictions of ten states.  
Three of them correspond to the $P_c$ states of the LHCb with good agreement with the observed masses and decay widths.
The quantum numbers for the would be doublet $P_c(4440)$ and $P_c(4457)$ are then identified with  $J = 3/2$ and 1/2, respectively.
It turns out that the mechanism of lowering the $J = 3/2$ state is mostly due to the tensor force.
The assignment of these quantum numbers is unique because the tensor force acts in the second order and 
the sign of the interaction does not matter.  
In general spin-spin interaction with heavy quark symmetry is employed to explain the splitting.  
However, the quantum numbers are not uniquely determined.  
Therefore, in Ref.~\cite{Liu:2019tjn}, two options were investigated for the spin assignments.
The spins of the $P_c$ states are not yet known, and hence the determination of them is very 
important for the further understanding of these states.

\section{Summary and complements}
\label{sec:summary}

\subsection{Brief summary}

In this article we have discussed the hadronic molecule as one of the exotic structures of hadrons.  
It has become possible experimentally to observe 
various exotic phenomena long after the predictions made more than half century ago, which
have stimulated a diverse body of theoretical work.  
The ingredients of hadronic molecules are constituent hadrons and their interactions.  
The constituent hadrons also couple to compact structures. 
Therefore, we have discussed in detail  how the admixture model has been applied to the $X(3872)$.  

From a first-principle point of view, such a picture should effectively and conveniently 
replace the direct but complicated approach of QCD.
In other words, the model should be economized~\cite{Weinberg:1962hj,Weinberg:1963zza,Weinberg:1965zz}, 
such that its  work region, where and how, is under control.  
As emphasized in the introduction, hadronic molecules are expected to appear near threshold regions.  
Their formation is a consequence of finely tuned hadron dynamics, as their binding or resonant energies 
are of order MeV which is much smaller than the  scale of low energy QCD, $\Lambda_{QCD} \sim$ some hundreds MeV.  
We have discussed why such conditions are likely to be realized for heavy hadrons.  
Their kinetic energies are suppressed and relatively weak force is sufficient 
to generate hadronic molecules.  

For the interaction, we have emphasized the role of the one-pion exchange interaction.
Pion dynamics are well established because the pion is the Nambu-Goldstone boson 
of spontaneous breaking of chiral symmetry, where the pion interaction is  
dictated by the low energy theorems.  
In the constituent quark picture the pion interacts with the light $u,d$ quarks by the 
pseudoscalar Yukawa coupling of $\bsigma \cdot \bi q$ type, 
whose strength is extracted from the empirically known axial coupling constants of hadrons.

The long range part of hadron interactions is provided by the OPEP,
which we have discussed in detail in this paper.  
Because of the spin structure of the Yukawa coupling, the OPEP contributes to the transitions 
$D \to D^*$ and $D^* \to D^*$.  
It turns out that they are effective in the formation of a $D\bar D^*$ molecule for the  $X(3872)$ channel .  
Therefore, an emphasis has been put on the role of the tensor component of the OPEP 
that causes  mixing of states (channels) differing in angular momenta by 2$\hbar$. 
Because this transition leads to second order process, the resulting interaction  
for the relevant low lying channel
must be attractive, with more attraction with more channel couplings.  

There still remains a question for the short-range part, because the bare OPEP is singular.  
To avoid this we introduce a form factor.
Such a prescription is known to work well for low energy properties of the deuteron~\cite{Ericson:1988gk}
by employing a cutoff parameter around $\Lambda \sim 800$ MeV (see Table~\ref{table:cutoffNN}).  
To extend such a prescription to $D D^*$ molecule, in particular to determine the coupling strength and cutoff values 
for $D D^*$, 
we have employed a counting of quarks and 
the sizes of the nucleon and $D^{(*)}$ using the quark model.

It is interesting to note that the 
employed cutoff, $\Lambda \sim 800$ MeV for the nucleon, is consistent 
with the size of the nucleon core  $\sqrt{6}/\Lambda \sim 0.5$ fm.
Discussions of the nucleon core have a long history. 
It is recognized as the repulsive core of the $NN$ interaction, 
which has been explicitly shown in the study by the quark cluster model~\cite{Oka:2000wj,Shimizu:2000wm}.  
It was also discussed in the chiral bag model where the nucleon is expressed 
as a quark core with pion clouds~\cite{Hosaka:1996ee,Hosaka:2001ux}.

Having this construction of the interactions, the OPEP provides a non-negligible amount of attraction 
particularly for hidden heavy hadrons such as $X(3872)$ with isospin 0, 
where the tensor force plays the dominant role, while 
the central component plays little.  
Therefore, the inclusion of the channel coupling of $SD$ waves  (generally, states that differ in angular momenta 
by 2$\hbar$) is very important.  
The resulting strength of the attraction, however, turns out not to be sufficient to generate the $X(3872)$ 
as a molecular state of $D \bar D^*$.
The coupling/mixing with a compact state of $c \bar c$ supplies additional attraction, 
if the compact state has a larger mass 
such as that theoretically expected for $\chi_{c1}(2P)$ charmonium meson.  
The mixing is also required to explain the large production rates of the $X(3872)$ in high energy hadron processes.  
Quantitative estimates of the production rate, however, 
have 
to be done carefully~\cite{Bignamini:2009sk,Albaladejo:2015lob,Esposito:2016noz,Braaten:2004fk,Braaten:2004ai}.
In the present analysis of the $X(3872)$, the OPEP and the short-range coupling play roughly equal roles.  
In general, however,  their relative importance depends on the system under study.

Another possible molecule that we have discussed is the $Z_c(3900)$.
However, the strength of the OPEP for the $Z_c(3900)$ of isospin one  is smaller than that for the $X(3872)$ 
of isospin zero by factor three. 
As a result, the attraction is reduced and the formation of molecular state is less likely. 
In the remaining part of this article, we have also discussed the above features for the bottom sector.

\subsection{Resonances or cusps}

Here, we briefly mention a question which one would yet like to ask; 
whether the observed exotic phenomena imply physical resonant states 
or cusps of virtual states.  
At this moment, we have no decisive answer to this question for the observed signals, while there are many 
articles discussing the nature of the signals theoretically.
Here we just refer only to a few of them in relation with 
the $X(3872)$~\cite{Hanhart:2007yq,Braaten:2007dw,Voloshin:2007hh,Kalashnikova:2009gt,Zhang:2009bv}.
There, amplitude analyses are performed by using parametrizations of Flatt\` e or effective range expansion types.  
Then an observation was made that by suitably choosing parameters, the line shape $X(3872)$ was 
shown to emerge as a virtual state cusp at the threshold~\cite{Braaten:2007dw}.  

To reproduce a very narrow or sharp peak at the $D^0 \bar D^{*0}$ threshold, however, 
there must be sufficient amount of attraction between them.  
If the attraction is larger than the critical value, the peak appears as a resonant state, 
otherwise as a virtual state cusp.   
The difference between them is subtle because only a small change in the interaction strength may change 
the nature of the peak.  
Moreover, in such a situation it is difficult to differentiate them experimentally.  
But then the important question is; what would be the  mechanism to provide that suitable amount of the attraction?
In this paper, we have tried to offer an option that a model with the pion exchange interaction does it, supplemented
by a coupling to a short distance structure.  
This is a dynamical approach for the construction of amplitudes that we discuss shortly below.

\subsection{Hadron interactions and exotics}

The last issue that we would like to mention is 
the dynamical approach for the construction of amplitudes from reliable hadron interactions.
For heavy hadrons including charm or bottom quarks, it is formidably difficult to derive
interactions from experiments.  
This is the reason that we have resorted to a model for the study of the the $X(3872)$ in this paper.  

Yet another powerful and promising method is lattice QCD, which is, in principle, 
the first principle method for the strong interaction.  
In the so called HAL QCD method, hadron interactions are obtained by using 
the Nambu-Bethe-Salpeter amplitude~\cite{Ishii:2006ec,Aoki:2012tk}.  
To obtain hadron interactions, this method is practically more powerful than 
the widely used Luscher's method~\cite{Iritani:2017rlk,Iritani:2018vfn}.
An attempt was made for  the $Z_c(3900)$ with coupled channels of $D \bar D^*$, $\eta_c\rho$, $J/\! \psi \pi$, where they have derived 
the interactions between these channels and solved the coupled channel problem~\cite{Ikeda:2016zwx,Ikeda:2017mee}. 
Unexpectedly it was found that there is a rather strong coupling between $J/\! \psi \pi$ and 
$D \bar D^*$ channels, which effectively causes an attraction in the $J/\! \psi \pi$ channel.  
As a consequence, they have found rather than a resonance,  a virtual state pole that contributes to an enhancement near the $D\bar D^*$ threshold corresponding to $Z_c(3900)$.  

For the study of exotic hadrons,  an approach based on the coupled channel method with 
suitable hadron interactions is highly desired.
It is a non-trivial program because many channels may couple, including those with more 
than two particles. 
With complementary approaches of experiments, effective theories and lattice simulations, 
such an approach can be further elaborated, thereby enabling elucidation of the nature of exotic hadrons.  

\section*{Acknowledgments}
The authors thank S. Yasui for fruitful discussions and useful comments.
This work is supported in part by the Special Postdoctoral Researcher Program (SPDR) of RIKEN (Y.Y.),
and by JSPS KAKENHI Grant Numbers JP16K05361 (S.T. and M.T.),
 JP17K05441, and Grants-in-Aid for Scientific Research on Innovative Areas (No.\ 18H05407) (A.H.).

\section*{References}
    \bibliographystyle{iopart-num} 
    \bibliography{./reference}

\end{document}